\documentclass[11pt]{iopart}

\expandafter\let\csname equation*\endcsname\relax
\expandafter\let\csname endequation*\endcsname\relax

\usepackage{amsmath,latexsym,amssymb,amsfonts}
\usepackage[dvips]{graphicx}
\usepackage{cite,color}
\usepackage{bm}
\usepackage{mathrsfs}
\usepackage{amsthm}
\usepackage{textcomp}
\usepackage{upgreek}
\usepackage{wasysym}
\usepackage{slashed}
\newcommand{\rqm}{{%
\declareslashed{}{\text{-}}{0.04}{0}{f}\slashed{f}}}

\begin{document}

\title[]{Separable wave
equations for gravitoelectromagnetic perturbations of 
rotating charged black strings}

\author{Alex S Miranda$^{a}$,\; Jaqueline Morgan$^{b}$, Alejandra 
Kandus$^{a}$, and Vilson T Zanchin$^{c}$}

\address{$^{a}\,$Departamento de Ci\^encias Exatas e Tecnol\'ogicas,
Universidade Estadual de Santa Cruz, Rodovia Jorge Amado, km 16, 45662-900 
Ilh\'eus, BA, Brazil\\
{$^{b}\,$Instituto Federal de Educa\c{c}\~ao, Ci\^encia e 
Tecnologia do Rio Grande do Sul, Rua Avelino Ant\^onio de Souza 1730,
95043-700 Caxias do Sul, RS, Brazil}\\
{$^{c}\,$Centro de Ci\^encias Naturais e Humanas, 
Universidade Federal do ABC, Avenida dos Estados 5001, 09210-580 Santo 
Andr\'e, SP, Brazil}}
\eads{\mailto{asmiranda@uesc.br}, \mailto{jaqueline.morgan@caxias.ifrs.edu.br},
\mailto{kandus@uesc.br}, \mailto{zanchin@ufabc.edu.br}}

\begin{abstract}

Rotating charged black strings are exact solutions of four-dimensional
Einstein-Maxwell equations with a negative cosmological constant and a non-trivial
spacetime topology. According to the AdS/CFT correspondence, these black strings
are dual to rotating thermal states of a strongly interacting quantum field theory
with nonzero chemical potential that lives in a cylinder. The dynamics of linear fluctuations
in the dual field theory can be studied from the perturbation equations
for classical fields in a black-string spacetime. With this motivation in mind,
we develop here a completely gauge and tetrad invariant perturbation approach to
deal with the gravitoelectromagnetic fluctuations of rotating charged black strings
in the presence of sources. As usual, for any charged black hole, a perturbation in
the background electromagnetic field induces a metric perturbation and vice versa.
In spite of this coupling and the non-vanishing angular momentum, we show that
linearization of equations of the Newman-Penrose formalism leads to four
separated second-order complex equations for suitable combinations of
the spin coefficients, the Weyl and the Maxwell scalars. Then, we generalize the
Chandrasekhar transformation theory by the inclusion of sources and apply
it to reduce the perturbation problem to four decoupled inhomogeneous wave
equations --- a pair for each sector of perturbations. The radial part of such
wave equations can be put into Schr\"odinger-like forms after Fourier transforming
them with respect to time. We find that the resulting effective potentials form two 
pairs of supersymmetric partner potentials and, as a consequence, the 
fundamental variables of one perturbation sector are related to the 
variables of the other sector. The relevance of such a symmetry in connection
to the AdS/CFT correspondence is discussed, and
future applications of the pertubation theory developed here are outlined.

\end{abstract}

\section{Introduction}
\label{introd} 

During the last two and a half decades, black holes in asymptotically anti-de Sitter
(AdS) spacetimes have been recognized as important objects for the study
of the foundations of the gravitational interaction and its connections to other
areas of physics. In particular, three-dimensional
AdS black holes \cite{Banados:1992wn,Martinez:1999qi,Zanchin:2003nu} have been explored to
study the emergence of quantum gravitational effects in a simpler setting than the 
four-dimensional case (see, e.g., refs. \cite{Carlip:1995qv,Carlip:2005zn,Witten:2007kt}), and
eternal Schwarzschild-AdS black holes have played an important role in the recently
discovered relation between the entanglement of quantum states and the Einstein-Rosen
non-traversable wormholes \cite{Maldacena:2013xja,Jensen:2013ora,Sonner:2013mba,
Gharibyan:2013aha,Chernicoff:2013iga,JohnCBaez:2014sra}.

Another important reason for the interest in anti-de Sitter black holes is the well-known
AdS/CFT correspondence \cite{Maldacena:1997re,Witten:1998qj,Gubser:1998bc,Aharony:1999ti},
which affirms that AdS black holes are dual to equilibrium thermal states of a
large $N$ strongly coupled conformal field theory (CFT) on the boundary of the AdS spacetime.
In such a context, perturbations of a black hole are associated to small deviations
from equilibrium of the CFT thermal system, and the dynamics of linear fluctuations
in the dual field theory can be directly studied from the perturbation equations
for classical fields in a black-hole spacetime.
The AdS/CFT duality at nonzero temperature has become now an
important tool to investigate in- and out-of-equilibrium properties of
quantum field theories at strong coupling. As a consequence of it, the study
of AdS black holes have encountered applications that range from QCD
to condensed-matter physics (for reviews, see refs.
\cite{Son:2007vk,Gubser:2007zz,Myers:2008fv,Berti:2009kk,Herzog:2009xv,
Hartnoll:2009sz,McGreevy:2009xe,Hubeny:2010ry}).

The rich structure of the anti-de Sitter spacetime has also been uncovered 
and has become manifest with the advance of applications of the AdS/CFT correspondence. By itself,
the existence of a cosmological constant changes the spacetime asymptotic 
behavior and also determines
the different topologies that a black hole may have. For the case of asymptotically flat
four-dimensional spacetimes, Hawking's topology theorem assures that, under certain
reasonable conditions, the horizon of a black hole must be topologically spherical
\cite{Hawking:1971vc,Hawking:1973uf}. However, the presence of a negative cosmological
constant renders possible the existence of a multiply connected spacetime with an event
horizon. A specific identification of points in a planar Reissner-Nordstr\"om-AdS black
hole \cite{Lemos:1994fn,cai:1996} generates a charged black string (also called cylindrical
charged black hole) or a charged black torus \cite{huang:1995}, which in turn can be put to
rotate through an improper coordinate transformation in the sense of Stachel \cite{Stachel:1981fg},
giving rise to a rotating charged black string or black torus \cite{Lemos:1994xp,Lemos:1995cm}.

The perturbations of static charged anti-de Sitter black holes have been 
the focus of great interest
in the last years, especially due to their applications
to holographic condensed matter systems (see refs. 
\cite{Edalati:2010hk,Edalati:2010pn,Brattan:2010bw,
Brattan:2010pq,Davison:2011uk,Ge:2010yc,Davison:2013bxa,Phukon:2013tda,Kim:2014bza,Blake:2014lva}
for a sample). In a similar way, rotating uncharged black holes (either in asymptotically flat or
AdS spacetimes) have attracted a lot of attention in connection with the Kerr/CFT
correspondence \cite{Guica:2008mu,Lu:2008jk,Dias:2009ex,Guica:2010ej,Mei:2012wd} (see also
\cite{Compere:2012jk} and references therein). On the other hand, while the geometric and thermodynamic
equilibrium properties of rotating charged black holes have been well studied since
the discovery of the Kerr-Newman solution \cite{Newman:1965my}, the decoupling of the
gravitational and electromagnetic perturbation equations for these black holes remains as a
long-standing open problem in general relativity theory \cite{Lee:1976,Chitre:1976bb,Chandrasekhar:1978ab}.
   
Motivated by the correspondence between the near-extremal Kerr-Newman solution
and conformal field theories \cite{Hartman:2008pb,Hartman:2009nz}, recently 
there has been a renewed interest in testing the stability of rotating 
charged black holes against
gravitoelectromagnetic fluctuations. This issue has been attacked in different ways,
which include: the study of the weak charge limit \cite{Mark:2014aja} and the slow rotation
limit \cite{Pani:2013ija,Pani:2013wsa} with perturbative expansions, respectively, around
the Kerr and Reissner-Nordstr\"om solutions; the computation of the black-hole quasinormal mode
spectrum of frequencies by solving a coupled system of perturbation
equations \cite{Dias:2015wqa}; and the performance of numerical simulations in the
full Einstein-Maxwell theory \cite{Zilhao:2014wqa}.

The aforementioned studies practically settled the question of the linear modal stability of the
Kerr-Newman black hole and gave strong evidence in favor of its stability beyond the linear level.
However, the computation of quantities like scattering amplitudes, greybody factors and
correlation functions in the dual CFT requires, in general, the separability of the
perturbation equations. In the case of charged black strings, the fact that rotation is
implemented via an `illegitimate' boost in the compact direction is
a key ingredient for the obtaining of decoupled equations. Hence, 
the study of rotating charged black strings offers a valuable opportunity
to investigate, in a unified way, how the parameters of a black hole (mass,
charge and angular momentum) and the global properties of a spacetime (cosmological
constant and topology) affect the dynamics of a gravitoeletromagnetic perturbation.

In this paper, we investigate the first-order coupled gravitational and electromagnetic
fluctuations of four-dimensional rotating charged black strings
in the presence of sources. We generalize here the Chandrasekhar transformation theory
\cite{Chandrasekhar:1985kt,Chandrasekhar:1975,Chandrasekhar:1979iz} with the inclusion of
source terms, and by combining it with the linearized Newman-Penrose (NP) equations
\cite{Newman:1961qr}, we are able to reduce the perturbation problem to four separated,
decoupled inhomogeneous equations of Schr\"odinger-like form.
The set of effective potentials appearing in such equations forms two pairs of supersymmetric partner
potentials. Among other things, this implies that the fundamental variables 
governing one sector of perturbations are related to the variables of the 
other sector.

The layout of the present article is the following. In the next section, the 
main geometric properties of a rotating charged black string, 
and its description in terms of the variables of the Newman-Penrose 
formalism are reviewed.
Section \ref{perturbations} contains the core of the perturbation 
theory for the charged rotating black string, with the basic set of 
equations in subsection \ref{Basic_set}, the
separation of variables in subsection \ref{Separation_var}, the system of 
equations for the tetrad invariant NP scalars in \ref{Tetrad_invariant}, and 
the decoupling of the perturbation equations in
\ref{Decoupling_eqs}. Section \ref{transformation-theory} is dedicated to
transform a pair of decoupled complex equations into four real Schr\"odinger-like
wave equations by means of a generalized Chandrasekhar transformation theory,
developed in \ref{Chandra_transformations}.
Section \ref{SUSY_QM} is devoted to investigate the emergence of
a quantum-mechanical supersymmetry in the fundamental equations of the
black-string perturbation theory. Finally, in section \ref{final} we
present a summary of the results and conclude.

\section{Rotating charged black strings}
\label{rot_charged_bs} 
\subsection{The background spacetime}
\label{properties}

We consider in this work a four-dimensional Einstein-Maxwell theory
with a negative cosmological constant $\Lambda_{c}=-3/\ell^2$, whose action
takes the form
\begin{equation}
I=-\frac{1}{16\uppi G}\int d^{4}x\sqrt{-g}\left(R
+\frac{6}{\ell^{2}}-\ell^{2}F_{\mu\nu}F^{\mu\nu}\right),
\label{acaocompleta}
\end{equation}
where $F_{\mu\nu}=\partial_{\mu}A_{\nu}- \partial_{\nu}A_{\mu}$ and the Greek
indices ($\mu,\,\nu,\, ...$) run over all the spacetime dimensions.
The resulting equations of motion admit a family of asymptotically AdS
stationary solutions \cite{Lemos:1995cm} given by the metric
\begin{equation}
ds^2=-\frac{r^2}{\ell^2}\upgamma^2 f\left(dt-ad\varphi\right)^2
+\frac{r^2}{\ell^2}\upgamma^2 \left(\ell d\varphi-\frac{a}{\ell}dt\right)^2
+\frac{r^2}{\ell^2}dz^2+\frac{\ell^2 dr^2}{r^2 f},
\label{background1}
\end{equation}
and by the gauge potential
\begin{equation}
A=-\upgamma\frac{c}{r}(dt-ad\varphi), 
\label{background2}
\end{equation}
where
\begin{equation}
f(r)=1-\frac{\ell^3 b}{r^3}+\frac{\ell^4 c^2}{r^4}\qquad\quad
\mbox{and}\qquad\quad\frac{1}{\upgamma}=\sqrt{1-\frac{a^2}{\ell^2}}.
\label{back-functions}
\end{equation}
  
If the parameters satisfy $b\geq b_{crit}=4(c^{2}/3)^{3/4}$ and the
$\varphi-z$ surfaces have cylindrical topology ($S^{1}\times \mathbb{R}$),
the metric \eqref{background1} describes the spacetime geometry of a rotating
charged black string with the inner and outer horizon radius, $r_{-}$ and
$r_{+}$, given by the real roots of $f(r)=0$. For $b=b_{crit}$, the horizons
coalesce and we have an extreme black string. The parameters $a$, $b$ and $c$
can be written in terms of the conserved mass $M$, angular momentum $J$ and
charge $Q$ per unit of black-string length as \cite{Lemos:1995cm,Dehghani:2002rr}
\begin{equation}
\begin{aligned}[2]
& a=-\frac{3\ell^2}{2 J}\left(\mathcal{Z}-M\right),\qquad\qquad &
b=2G\left(3\mathcal{Z}-M\right),\\
& c=2Q\ell\sqrt{\frac{3\mathcal{Z}-M}{\mathcal{Z}+M}},\qquad\qquad
&\mathcal{Z}=\sqrt{M^2-\frac{8J^2}{9\ell^2}}.
\end{aligned}
\end{equation}
Parameter $a$ is the analogous
of the Kerr rotation parameter, representing the angular momentum per unit
mass of the black string. The extremal rotation limit in which
$J^2/\ell^{2}M^2\rightarrow 1$ or, equivalently, the limit $\mathcal{Z}\rightarrow
M/3$, implies in $a=\ell$. On the other hand, the limit of zero angular 
momentum $J=0$, or, equivalently, the limit $\mathcal{Z}\rightarrow M$, 
implies in $a=0$.

\subsection{Description in the Newman-Penrose formalism}
\label{NP_formalism}

For the sequence of the work, it is important to describe the rotating 
charged black strings in the Newman-Penrose formalism 
\cite{Newman:1961qr}, see also \ref{apenA}. The spacetime
\eqref{background1} is algebraically type D in the Petrov classification.
So it is convenient to define the Newman-Penrose (NP) quantities  in terms of a
Kinnersley-like null frame \cite{Kinnersley:1969zza}:
\begin{align}
D&=l^{\mu}\partial_{\mu}=\frac{1}{r^2 f}\left(\ell^2\upgamma\partial_{t}
+a\upgamma\partial_{\varphi}+ r^2 f\partial_{r}\right), \label{vec-l}\\
\Delta&=n^{\mu}\partial_{\mu}=\frac{1}{2\ell^2}(\ell^2\upgamma\partial_{t}
+a\upgamma\partial_{\varphi}-r^2f\partial_{r}), \label{vec-n}\\
\delta&=m^{\mu}\partial_{\mu}=\frac{1}{r\sqrt{2}}
(\upgamma\partial_{\varphi}+a\upgamma\partial_{t}+i\ell\partial_{z}),
\label{vec-m}
\end{align}
where $l^{\mu}$ and $n^{\mu}$ are the double principal null directions
of the Weyl tensor. With this choice of basis vectors,
the only non-vanishing spin coefficients are
\begin{equation}
\rho=-\frac{1}{r},\qquad\quad\mu=-\frac{rf}{2\ell^2},\qquad\quad\gamma=
\frac{1}{4\ell^{2}}\frac{d}{dr}(r^2 f).
\label{spin_coefficients}
\end{equation}
The fact that in the chosen null frame all of the other spin coefficients
are identically zero, and particularly $\kappa=\sigma=\lambda=\nu=0$,
confirms the type-D character of the spacetime \eqref{background1}.
In accordance with the Goldberg-Sachs theorem \cite{goldberg1962},
the Weyl scalars $\Psi_{0}$, $\Psi_{1}$, $\Psi_{3}$ and $\Psi_{4}$ vanish,
and a direct computation shows that
\begin{equation}
\Psi_{2}=C_{\mu\nu\rho\sigma}l^{\mu} m^{\nu} m^{\ast\rho} n^{\sigma} 
=-\frac{\ell(br-2\ell c^{2})}{2r^{4}}, \label{weyl}
\end{equation}
where the asterisk denotes complex conjugation.

In the Newman-Penrose formalism, the electromagnetic field
is described by three complex Maxwell scalars: $\phi_0$, $\phi_1$ and
$\phi_2$. For a rotating charged black string,
$\phi_0=\phi_2=0$ and
\begin{equation}
\phi_{1}=\frac{1}{2}F_{\mu\nu}(l^{\mu} n^{\nu} + m^{\ast \mu} m^{\nu})=\frac{c}{2r^{2}}.
\label{maxwell}
\end{equation}

\section{Gravitational and electromagnetic perturbations}
\label{perturbations}

For the present analysis of the gravitoelectromagnetic perturbations of
rotating charged black strings via NP formalism we follow a similar procedure
as that presented in Ref.~\cite{Chandrasekhar:1985kt}
(see also \ref{apenA} for notation and sign conventions). In such an approach 
the metric and gauge-field linear perturbations are related to infinitesimal changes in the
null tetrad vectors and Maxwell scalars, which accordingly lead to
first-order perturbations in the spin coefficients and curvature scalars.
Among the complete set of NP quantities, we will be mainly interested in the
scalars that govern the evolution of the coupled gravitational and
electromagnetic waves in the background spacetime \eqref{background1}.

\subsection{The set of basic equations}
\label{Basic_set}

In the study of gravitational perturbations of electrically neutral black holes
\cite{Bardeen:1973xb,Teukolsky:1973ha},
an important fact to simplify the problem is the linearity and homogeneity 
of a set of NP equations (four of the Bianchi identities
and two of the Ricci identities) in quantities that vanish in the background spacetime.
However, when investigating charged black hole perturbations, we must also
consider the Maxwell equations with sources:
\begin{align}
&(\delta^{\ast}+\pi-2\alpha)\phi_{0}-(D-2\rho)\phi_{1}-\kappa\phi_{2}=2\uppi J_l,
\label{Maxwell1}\\
&(\Delta+2\mu)\phi_{1}-(\delta-\tau+2\beta)\phi_{2}-\nu\phi_{0}=2\uppi J_n,
\label{Maxwell2}\\
&(\Delta+\mu-2\gamma)\phi_{0}-(\delta-2\tau)\phi_{1}-\sigma\phi_{2}=2\uppi J_m,
\label{Maxwell3}\\
&(\delta^{\ast}+2\pi)\phi_{1}-(D-\rho+2\varepsilon)\phi_{2}
-\lambda\phi_{0}=2\uppi J_{m^{\ast}}.
\label{Maxwell4}
\end{align}
Note that these equations contain terms that involve directional derivatives of the scalar
$\phi_1$, like $D \phi_1$, $\delta \phi_1$ and so on. Since $\phi_{1}$ is non-vanishing in
the background spacetime, a direct linearization of Maxwell equations would give rise to 
perturbations in the basis vectors (and also in the spin coefficients $\rho$, $\tau$,
$\pi$ and $\mu$), making difficult the decoupling of the resulting equations.

An alternative to overcome this problem, which was used with success in the Reissner-Nordstr\"om
black-hole case \cite{Lee:1976}, is looking for another set of equations which are already linearized,
in the sense that they are linear and homogeneous in quantities that vanish in
the background geometry. By taking into account the presence of source terms, we generalize
below the procedure employed by Chandrasekhar \cite{Chandrasekhar:1985kt} in the study
of the gravitoelectromagnetic perturbations of Reissner-Nordstr\"om black holes.

Initially we apply the operator $(\delta-2\tau-\alpha^{\ast}-
\beta+\pi^{\ast})$ to the Maxwell equation \eqref{Maxwell1} and the operator
$(D-\varepsilon+\varepsilon^{\ast}-2\rho-\rho^{\ast})$ to the Maxwell equation
\eqref{Maxwell3} and subtract one equation from the other to get
\begin{equation}
\begin{gathered}
\left[(\delta-2\tau-\alpha^{\ast}-\beta+\pi^{\ast})
(\delta^{\ast}+\pi-2\alpha)-(D-\varepsilon+\varepsilon^{\ast}
-2\rho-\rho^{\ast})(\Delta+\mu-2\gamma)\right]\phi_{0}=\\
\left[(\delta-2\tau-\alpha^{\ast}-\beta+\pi^{\ast})\kappa
-(D-\varepsilon+\varepsilon^{\ast}-2\rho-\rho^{\ast})\sigma
\right]\phi_{2}+\kappa\delta\phi_{2}-\sigma D\phi_{2}\\
+(\delta D-D\delta)\phi_{1}+
\left[(\varepsilon-\varepsilon^{\ast}+\rho^{\ast})
(\delta-2\tau)-(\alpha^{\ast}+\beta-\pi^{\ast})(D-2\rho)\right]\phi_{1}\\
-2(\delta\rho)\phi_{1}+2(D\tau)\phi_{1}
+2\uppi\left[(\delta-2\tau-\alpha^{\ast}-\beta+\pi^{\ast})J_{l}
-(D-\varepsilon+\varepsilon^{\ast}-2\rho-\rho^{\ast})J_{m}\right].
\label{pgrande1}
\end{gathered}
\end{equation}
To simplify the resulting equation, we utilize the commutation relation
\eqref{NP-comut2} and the Ricci identities \eqref{Ricci-3} and \eqref{Ricci-5}
to eliminate, respectively, the operator $(\delta D-D\delta)$ and the quantities
$D\tau$ and $\delta\rho$ from \eqref{pgrande1}. After some manipulations,
it results in the following equation:
\begin{equation}
\begin{gathered}
\left[(\delta-2\tau-\alpha^{\ast}-\beta+\pi^{\ast})
(\delta^{\ast}+\pi-2\alpha)-(D-\varepsilon+\varepsilon^{\ast}
-2\rho-\rho^{\ast})(\Delta+\mu-2\gamma)\right]\phi_{0}=\\
\left[(\delta-2\tau-\alpha^{\ast}-\beta+\pi^{\ast})\kappa
-(D-\varepsilon+\varepsilon^{\ast}-2\rho-\rho^{\ast})\sigma
\right]\phi_{2}+\kappa\delta\phi_{2}-\sigma D\phi_{2}\\
+2\phi_{1}\left[(\Delta-3\gamma-\gamma^{\ast}-\mu+\mu^{\ast})\kappa-
(\delta^{\ast}-3\alpha+\beta^{\ast}-\tau^{\ast}-\pi)\sigma\right]
+4\Psi_{1}\phi_{1}+\kappa\Delta\phi_{1}\\
-\sigma\delta^{\ast}\phi_{1}+2\uppi\left[(\delta-2\tau-\alpha^{\ast}
-\beta+\pi^{\ast})J_{l}-(D-\varepsilon+\varepsilon^{\ast}-2\rho-\rho^{\ast})J_{m}\right].
\end{gathered}
\label{pgrande2}
\end{equation}
In a similar way, applying the operator $(\delta^{\ast}-\tau^{\ast}+
\alpha+\beta^{\ast}+2\pi)$ to equation \eqref{Maxwell2}
and the operator $(\Delta+\mu^{\ast}-\gamma^{\ast}+\gamma+2\mu)$
to equation \eqref{Maxwell4} and subtracting one of the resulting equations
from the other, we obtain
\begin{equation}
\begin{gathered}
\left[(\delta^{\ast}-\tau^{\ast}+\alpha+\beta^{\ast}+2\pi)
(\delta-\tau+2\beta)-(\Delta+\mu^{\ast}-\gamma^{\ast}+
\gamma+2\mu)(D-\rho+2\varepsilon)\right]\phi_{2}=\\
-\left[(\delta^{\ast}-\tau^{\ast}+\alpha+\beta^{\ast}+2\pi)
\nu-(\Delta+\mu^{\ast}-\gamma^{\ast}+\gamma+2\mu)\lambda\right]\phi_{0}
-\nu\delta^{\ast}\phi_{0}+\lambda\Delta\phi_{0}\\
+(\delta^{\ast}\Delta-\Delta\delta^{\ast})\phi_{1}
-\left[(\mu^{\ast}-\gamma^{\ast}+\gamma)
(\delta^{\ast}+2\pi)+(\tau^{\ast}-\alpha-\beta^{\ast})
(\Delta+2\mu)\right]\phi_{1}+2(\delta^{\ast}\mu)\phi_{1}\\
-2(\Delta\pi)\phi_{1}
-2\uppi\left[(\delta^{\ast}-\tau^{\ast}+\alpha+\beta^{\ast}+2\pi)J_{n}
-(\Delta+\mu^{\ast}-\gamma^{\ast}+\gamma+2\mu)J_{m^{\ast}}\right].
\end{gathered}
\label{pgrande3}
\end{equation}
Then, using the complex conjugate of the commutation relation \eqref{NP-comut3}
and the Ricci identities \eqref{Ricci-4} and \eqref{Ricci-6} to eliminate, 
respectively, 
the operator $(\delta^{\ast}\Delta-\Delta\delta^{\ast})$ and the quantities
$\Delta\pi$ and $\delta^{\ast}\mu$ from \eqref{pgrande3}, we find
\begin{equation}
\begin{gathered}
\left[(\delta^{\ast}-\tau^{\ast}+\alpha+\beta^{\ast}+2\pi)
(\delta-\tau+2\beta)-(\Delta+\mu^{\ast}-\gamma^{\ast}+
\gamma+2\mu)(D-\rho+2\varepsilon)\right]\phi_{2}=\\
-\left[(\delta^{\ast}-\tau^{\ast}+\alpha+\beta^{\ast}+2\pi)
\nu-(\Delta+\mu^{\ast}-\gamma^{\ast}+\gamma+2\mu)\lambda\right]\phi_{0}
+\lambda\Delta\phi_{0}-\nu\delta^{\ast}\phi_{0}\\
-2\phi_{1}\left[(D+3\varepsilon+\varepsilon^{\ast}+\rho-
\rho^{\ast})\nu-(\delta+\pi^{\ast}+\tau-\alpha^{\ast}+
3\beta)\lambda\right]+4\Psi_{3}\phi_{1}-\nu D\phi_{1}\\
+\lambda\delta\phi_{1}-2\uppi\left[(\delta^{\ast}-\tau^{\ast}+
\alpha+\beta^{\ast}+2\pi)J_{n}-(\Delta+\mu^{\ast}
-\gamma^{\ast}+\gamma+2\mu)J_{m^{\ast}}\right].
\end{gathered}
\label{pgrande4}
\end{equation}

Equations \eqref{pgrande2} and \eqref{pgrande4} are already
linearized in the sense that they give rise to equations
which are linear and homogeneous in the quantities that vanish in the background.
In fact, the terms involving $\phi_{2}$ in \eqref{pgrande2} and the terms
involving $\phi_{0}$ in \eqref{pgrande4} consist of quantities of second order,
and so they can be ignored in a linear perturbation theory. On basis of the
Maxwell equations \eqref{Maxwell1}-\eqref{Maxwell4} for the background spacetime,
the terms $\Delta\phi_{1}$, $\delta^{\ast}\phi_{1}$,
$D\phi_{1}$ and $\delta\phi_{1}$ can be replaced, respectively,
by $-2\mu\phi_{1}$, $-2\pi\phi_{1}$, $2\rho\phi_{1}$ and $2\tau\phi_{1}$.
As a result, the linearized versions of equations \eqref{pgrande2}
and \eqref{pgrande4} become
\begin{equation}
 \begin{split}
[\delta\delta^{\ast}-(D-2\rho-\rho^{\ast})(\Delta+\mu & -2\gamma)]
\phi_{0}^{\scriptscriptstyle{(1)}}-2\phi_{1}\left[(\Delta-3\gamma-
\gamma^{\ast}-2\mu+\mu^{\ast})\kappa^{\scriptscriptstyle{(1)}}\right.\\
&\left. -\delta^{\ast}\sigma^{\scriptscriptstyle{(1)}} 
+2\Psi_{1}^{\scriptscriptstyle{(1)}}\right]=
2\uppi \left[\delta J_l-\left(D-2\rho-\rho^{\ast}\right)J_m\right],
\label{pmedia1}
\end{split}
\end{equation}
\begin{equation}
\begin{split}
[\delta^{\ast}\delta-(\Delta+\mu^{\ast}-\gamma^{\ast}+
\gamma+2\mu)(D & -\rho)]\phi_{2}^{\scriptscriptstyle{(1)}}-
2\phi_{1}[\delta\lambda^{\scriptscriptstyle{(1)}}-(D+2\rho-\rho^{\ast})
\nu^{\scriptscriptstyle{(1)}}\\
&+2\Psi_{3}^{\scriptscriptstyle{(1)}}]=2\uppi\left[\left(\Delta+3\mu\right)
J_{m^{\ast}}-\delta^{\ast}J_{n}\right],
\label{pmedia2}
\end{split}
\end{equation}
where the superscript $(1)$ is used to distinguish the first-order
perturbation of a NP quantity from its value in the stationary background 
state,
and the vanishing of some of the spin coefficients in the background
spacetime was used to simplify the above equations.

In addition to the linearized equations \eqref{pmedia1} and \eqref{pmedia2}, it is
important to consider here the pair of Ricci equations \eqref{Ricci-1} and
\eqref{Ricci-2} and the set of Bianchi identities \eqref{Bianchi-full1}-\eqref{Bianchi-full4}.
The main difference of these equations in relation to the electrically 
neutral black hole case is the appearance of Ricci (source) terms in the 
Bianchi identities, which
depend on the square of the Maxwell scalars and arise due to the coupling 
between the metric and gauge field.

The linearized Bianchi identities \eqref{Bianchi-full1}-\eqref{Bianchi-full4},
governing the radiative (nontrivial) parts of the black-string perturbations,
take respectively the forms
\begin{equation}
\begin{split}
\delta^{\ast}\Psi_{0}^{\scriptscriptstyle{(1)}}-(D-4\rho)\Psi_{1}^{\scriptscriptstyle{(1)}}
+2\ell^2\phi_{1}^{\ast}D\phi_{0}^{\scriptscriptstyle{(1)}}-(3\Psi_{2}-
4\ell^2\phi_{1}\phi_{1}^{\ast})\kappa^{\scriptscriptstyle{(1)}}=&\\
4\uppi G\left[-(D-2\rho)T^{\mbox{\tiny{(MAT)}}}_{lm}+\delta T^{\mbox{\tiny{(MAT)}}}_{ll}\right]&, 
\label{Bianchi1}
\end{split}
\end{equation}
\begin{equation}
\begin{split}
(\Delta-4\gamma+\mu)\Psi_{0}^{\scriptscriptstyle{(1)}}-\delta\Psi_{1}^{\scriptscriptstyle{(1)}}
-2\ell^2 \phi_{1}^{\ast}\delta \phi_{0}^{\scriptscriptstyle{(1)}}
-(3\Psi_{2}+4\ell^2\phi_{1}\phi_{1}^{\ast})\sigma^{\scriptscriptstyle{(1)}}=&\\
4\uppi G\left[-(D+\rho^{\ast})T^{\mbox{\tiny{(MAT)}}}_{mm}+\delta T^{\mbox{\tiny{(MAT)}}}_{lm}\right]&,
\label{Bianchi2}
\end{split}
\end{equation}
\begin{equation}
\begin{split}
\delta\Psi_{4}^{\scriptscriptstyle{(1)}}-(\Delta+2\gamma+4\mu)
\Psi_{3}^{\scriptscriptstyle{(1)}}+2\ell^2\phi_{1}^{\ast}
(\Delta+2\gamma)\phi_{2}^{\scriptscriptstyle{(1)}}
+(3\Psi_{2}-4\ell^2\phi_{1}\phi_{1}^{\ast})\nu^{\scriptscriptstyle{(1)}}=&\\
4\uppi G\left[(\Delta+2\mu^{\ast}+2\gamma)T^{\mbox{\tiny{(MAT)}}}_{nm^{\ast}}-\delta^{\ast}
T^{\mbox{\tiny{(MAT)}}}_{nn}\right]&, 
\label{Bianchi3}
\end{split}
\end{equation}
\begin{equation}
 \begin{split}
(D-\rho)\Psi_{4}^{\scriptscriptstyle{(1)}}-\delta^{\ast}\Psi_{3}^{\scriptscriptstyle{(1)}}
-2\ell^2\phi_{1}^{\ast}\delta^{\ast}\phi_{2}^{\scriptscriptstyle{(1)}}
+(3\Psi_{2}+4\ell^2\phi_{1}\phi_{1}^{\ast})\lambda^{\scriptscriptstyle{(1)}}=&\\
4\uppi G\left[-(\Delta+\mu^{\ast})T^{\mbox{\tiny{(MAT)}}}_{m^{\ast}m^{\ast}}
+\delta^{\ast}T^{\mbox{\tiny{(MAT)}}}_{nm^{\ast}}\right]&,
\label{Bianchi4}
\end{split}
\end{equation}
where $T_{ab}^{\mbox{\tiny{(MAT)}}}$ ($a,b=l,n,m,m^{\ast}$) are the tetrad compoments of
the energy-momentum tensor of all forms of matter and all nonelectromagnetic and
nongravitational fields. The above equations are supplemented by the following
linearized Ricci identities:
\begin{equation}
(D-\rho-\rho^{\ast})\sigma^{\scriptscriptstyle{(1)}}-
\delta\kappa^{\scriptscriptstyle{(1)}}=\Psi_{0}^{\scriptscriptstyle{(1)}},
\label{Ricci-new1}
\end{equation}
\begin{equation}
(\Delta+\mu+\mu^{\ast}+3\gamma-\gamma^{\ast})\lambda^{\scriptscriptstyle{(1)}}
-\delta^{\ast}\nu^{\scriptscriptstyle{(1)}}=\Psi_{4}^{\scriptscriptstyle{(1)}}.
\label{Ricci-new2}
\end{equation}

The set of equations \eqref{pmedia1}-\eqref{Ricci-new2} forms the basic set of
equations for the study of the gravitoelectromagnetic perturbations of the
rotating charged black strings.

\subsection{Separation of variables}
\label{Separation_var}

As usual we take the Fourier transform of the perturbation functions, 
i.e.,
we assume a dependence on the coordinates $t$, $\varphi$ and $z$
of the form $\exp[-i\omega t+im\varphi+ikz)]$,
where $m$ is an integer number. So the action of the
directional derivatives $D$, $\Delta$, $\delta$ and $\delta^\ast$
on functions with such an exponential dependence become
\begin{equation}
D=\mathscr{D}_{0},\qquad\Delta=-\frac{r^2 f}{2\ell^2}\mathscr{D}_{0}^{\dagger},
\qquad\delta=\frac{i}{r\sqrt{2}}p,\qquad
\delta^{\ast}=\frac{i}{r\sqrt{2}}p^{\ast},
\label{operadores}
\end{equation}
where
\begin{equation}
\mathscr{D}_{n}=\frac{d}{d r}-\frac{i\ell^2 }{r^2 f}\varpi
+n\frac{d}{dr}\ln\left(\frac{r^4}{\ell^2}f\right),\qquad\quad
\mathscr{D}_{n}^{\dagger}=\frac{d}{dr}+\frac{i\ell^2 }{r^2 f}\varpi
+n\frac{d}{dr}\ln\left(\frac{r^4}{\ell^2}f\right),
\end{equation}
and the constants $p$ and $\varpi$ are given by
\begin{equation}
p=\frac{\upgamma}{\ell}(m-a\omega)+ik,\quad\qquad\varpi=\upgamma\left(\omega-\frac{a
m}{\ell^2}\right). 
\end{equation}
The differential operators $\mathscr{D}_{n}$ and $\mathscr{D}_{n}^{\dagger}$
satisfy some identities that will be useful in the sequence of this work:
\begin{equation}
(\mathscr{D}_{n})^{\ast}=\mathscr{D}_{n}^{\dagger},\qquad\quad
\mathscr{D}_{n}\left[r^{N'}\left(\frac{r^4}{\ell^2}f\right)^{N}\right]=r^{N'}
\left(\frac{r^4}{\ell^2}f\right)^{N}\left[\mathscr{D}_{(n+N)}+\frac{N'}{r}\right],
\label{identidades}
\end{equation}
where $n$, $N$ and $N'$ are integer numbers.

We work from now on with the Fourier transformed versions of equations 
\eqref{pmedia1}-\eqref{Ricci-new2}, but keep the same symbols for the
Fourier-transforms as the original NP quantities. Substituting the background values
\eqref{spin_coefficients}-\eqref{maxwell} for the spin coefficients $\rho$,
$\gamma$ and $\mu$, for the Weyl scalar $\Psi_2$ and the Maxwell
scalar $\phi_1$, equations \eqref{pmedia1}-\eqref{Ricci-new2} become
\begin{equation}
\begin{gathered}
-\frac{ip^{\ast}}{\sqrt{2}r}\Psi_{0}^{\scriptscriptstyle{(1)}}+\left(\mathscr{D}_{0}+
\frac{4}{r}\right)\Psi_{1}^{\scriptscriptstyle{(1)}}-\frac{c\,\ell^2}{r^2}\,\mathscr{D}_{0}
\,\phi_{0}^{\scriptscriptstyle{(1)}}
-\frac{(3\ell br-4\ell^2 c^{2})}{2r^{4}}\kappa^{\scriptscriptstyle{(1)}}=
\frac{\mathcal{T}_{\scriptscriptstyle{\cal{A}}}}{\sqrt{2}r},\\
\frac{r^2 f}{2\ell^2}\left(\mathscr{D}_{2}^{\dagger}-\frac{3}{r}\right)
\Psi_{0}^{\scriptscriptstyle{(1)}}+\frac{ip}{\sqrt{2}r}
\left(\Psi_{1}^{\scriptscriptstyle{(1)}}+\frac{c\,\ell^2}{r^2}
\phi_{0}^{\scriptscriptstyle{(1)}}\right)-\frac{(3\ell br-8\ell^2 c^{2})}{2r^{4}}
\sigma^{\scriptscriptstyle{(1)}}=\frac{\mathcal{T}_{\scriptscriptstyle{\cal{B}}}}{2r^2},\\
\left(\mathscr{D}_{0}+\frac{2}{r}\right)\sigma^{\scriptscriptstyle{(1)}}
-\frac{ip}{\sqrt{2}r}\kappa^{\scriptscriptstyle{(1)}}-\Psi_{0}^{\scriptscriptstyle{(1)}}=0,\\
\frac{r^2 f}{2\ell^2}\left(\mathscr{D}_{2}^{\dagger}-\frac{5}{r}\right)
\kappa^{\scriptscriptstyle{(1)}}+\frac{ip^{\ast}}{\sqrt{2}r}\,
\sigma^{\scriptscriptstyle{(1)}}+\left[\frac{r^4 f}{\ell^2}
\left(\mathscr{D}_{1}+\frac{1}{r}\right)\left(\mathscr{D}_{1}^{\dagger}-\frac{1}{r}\right)
-p^2\right]\frac{\phi_{0}^{\scriptscriptstyle{(1)}}}{2c}-2\Psi_{1}^{\scriptscriptstyle{(1)}}=
\frac{\mathcal{J}_{\scriptscriptstyle{\cal{B}}}}{\sqrt{2}},
\end{gathered}
\label{eqbasica1}
\end{equation}
and
\begin{equation}
\begin{gathered}
\frac{ip}{\sqrt{2}r}\Psi_{4}^{\scriptscriptstyle{(1)}}+
\frac{r^2 f}{2\ell^2}\left(\mathscr{D}_{-1}^{\dagger}+\frac{6}{r}\right)
\Psi_{3}^{\scriptscriptstyle{(1)}}-\frac{cf}{2}\left(\mathscr{D}_{-1}^{\dagger}+\frac{2}{r}\right)
\phi_{2}^{\scriptscriptstyle{(1)}}-\frac{(3\ell br-4\ell^2 c^{2})}{2r^{4}}
\nu^{\scriptscriptstyle{(1)}}=\frac{\mathcal{T}_{\scriptscriptstyle{\cal{C}}}}{\sqrt{2}r^5},\\
\left(\mathscr{D}_{0}+\frac{1}{r}\right)\Psi_{4}^{\scriptscriptstyle{(1)}}
-\frac{ip^{\ast}}{\sqrt{2}r}\left(\Psi_{3}^{\scriptscriptstyle{(1)}}+
\frac{c\ell^2}{r^2}\phi_{2}^{\scriptscriptstyle{(1)}}\right)-
\frac{(3\ell br-8\ell^2 c^{2})}{2r^{4}}\lambda^{\scriptscriptstyle{(1)}}=
\frac{\mathcal{T}_{\scriptscriptstyle{\cal{D}}}}{r^4},\\
\frac{r^2}{2\ell^2}f\left(\mathscr{D}_{-1}^{\dagger}+\frac{4}{r}\right)
\lambda^{\scriptscriptstyle{(1)}}+\frac{ip^{\ast}}{\sqrt{2}r}
\nu^{\scriptscriptstyle{(1)}}-\Psi_{4}^{\scriptscriptstyle{(1)}}=0,\\
-\left(\mathscr{D}_{0}-\frac{1}{r}\right)\nu^{\scriptscriptstyle{(1)}}+
\frac{ip}{\sqrt{2}r}\lambda^{\scriptscriptstyle{(1)}}-\left[\frac{r^4 f}{\ell^2}
\left(\mathscr{D}_{0}^{\dagger}+\frac{3}{r}\right)\left(\mathscr{D}_{0}+\frac{1}{r}\right)
-p^2\right]\frac{\phi_{2}^{\scriptscriptstyle{(1)}}}{2c}+2\Psi_{3}^{\scriptscriptstyle{(1)}}=
\frac{\sqrt{2}}{r^2}\mathcal{J}_{\scriptscriptstyle{\cal{D}}},
\end{gathered}
\label{eqbasica2}
\end{equation}
where the source terms in the above equations are given by
\begin{align}
\mathcal{T}_{\scriptscriptstyle{\cal{A}}} & =4\uppi G\left[\sqrt{2}r\left(\mathscr{D}_{0}+
\frac{2}{r}\right)T^{\mbox{\tiny{(MAT)}}}_{lm}-ipT^{\mbox{\tiny{(MAT)}}}_{ll}\right],\\ 
\mathcal{T}_{\scriptscriptstyle{\cal{B}}} & =4\sqrt{2}\uppi G r\left[\sqrt{2}r\left(\mathscr{D}_{0}+\frac{1}{r}
\right)T^{\mbox{\tiny{(MAT)}}}_{mm}-ipT^{\mbox{\tiny{(MAT)}}}_{lm}\right],\\
\mathcal{J}_{\scriptscriptstyle{\cal{B}}} & =\frac{2\uppi r}{c}\left[ipJ_{l}-\sqrt{2}r\left(\mathscr{D}_{0}+\frac{3}{r}
\right)J_{m}\right],\\
\mathcal{T}_{\scriptscriptstyle{\cal{C}}} & =-4\uppi Gr^4\left[\frac{r^3 f}{\sqrt{2}\ell^2}\left(\mathscr{D}_{-1}^{\dagger}+
\frac{4}{r}\right)T^{\mbox{\tiny{(MAT)}}}_{nm^{\ast}}+ip^{\ast}T^{\mbox{\tiny{(MAT)}}}_{nn}\right],\\ 
\mathcal{T}_{\scriptscriptstyle{\cal{D}}} & =2\sqrt{2}\uppi G r^3\left[\frac{r^3 f}{\sqrt{2}\ell^2}\left(\mathscr{D}_{0}^{\dagger}
+\frac{1}{r}\right)T^{\mbox{\tiny{(MAT)}}}_{m^{\ast}m^{\ast}}+ipT^{\mbox{\tiny{(MAT)}}}_{nm^{\ast}}\right],\\
\mathcal{J}_{\scriptscriptstyle{\cal{D}}} & =\frac{\uppi r^3}{c}\left[ip^{\ast}J_{n}+\frac{r^3 f}{\sqrt{2}\ell^2}
\left(\mathscr{D}_{0}^{\dagger}+\frac{3}{r}\right)J_{m^{\ast}}\right],
\end{align}
and the subscripts $\cal{A}$, $\cal{B}$, $\cal{C}$ and $\cal{D}$ are used only
to distinguish the different source terms.

\subsection{Equations for tetrad-invariant scalars}
\label{Tetrad_invariant}

An important issue in a perturbation theory is the invariance of the
basic quantities under general coordinate transformations and infinitesimal
gauge transformations respectively in the coordinates and null basis 
vectors. Being scalar functions
with vanishing background values, the variables $\Psi_0$, $\Psi_1$, $\Psi_3$, $\Psi_4$,
$\phi_0$, $\phi_2$ and the spin coefficients $\sigma$, $\kappa$, $\lambda$ and $\nu$
are invariant under {\it{gauge transformations of the first and second kind}}
\cite{Sachs:1964}, i.e., they are independent of the choice of the coordinate
system on the physical manifold and independent of the identification map
between points of the fictitious background spacetime and the physical spacetime.

As discussed in detail in \ref{gauge_invariance}, there are
also gauge degrees of freedom associated to infinitesimal transformations on the
tetrad vectors. The basic NP variables appearing in equations \eqref{eqbasica1} and
\eqref{eqbasica2} are invariant under transformations of class III, but change under
null rotations of class I and II, as shown in equations \eqref{spin-I2},
\eqref{Weyl-Maxwell-I2} and \eqref{Spin-Weyl-Maxwell-II}. However, it is easy
to verify that the following combinations of NP quantities
\begin{equation}
\begin{aligned}
&\Psi_{0}^{\scriptscriptstyle{(1)}},\quad\Psi_{1}^{\scriptscriptstyle{(1)}}
-\frac{3}{2}\left(\frac{\Psi_{2}}{\phi_{1}}\right)
\phi_{0}^{\scriptscriptstyle{(1)}},\quad \kappa^{\scriptscriptstyle{(1)}}
+\frac{1}{2}\left(D-\rho\right)
\frac{\phi_{0}^{\scriptscriptstyle{(1)}}}{\phi_{1}},\quad
\sigma^{\scriptscriptstyle{(1)}}+\frac{1}{2}\,\delta\,
\frac{\phi_{0}^{\scriptscriptstyle{(1)}}}{\phi_{1}},\\
&\Psi_{3}^{\scriptscriptstyle{(1)}}-\frac{3}{2}
\left(\frac{\Psi_{2}}{\phi_{1}}\right)\phi_{2}^{\scriptscriptstyle{(1)}},
\quad\Psi_{4}^{\scriptscriptstyle{(1)}},\quad
\nu^{\scriptscriptstyle{(1)}}-\frac{1}{2}\left(\Delta+\mu+2\gamma\right)
\frac{\phi_{2}^{\scriptscriptstyle{(1)}}}{\phi_{1}},\quad
\lambda^{\scriptscriptstyle{(1)}}-\frac{1}{2}\,\delta^{\ast}\,
\frac{\phi_{2}^{\scriptscriptstyle{(1)}}}{\phi_{1}},
\end{aligned}
\label{new_variables}
\end{equation}
are invariant under infinitesimal rotations of the null basis.
The foregoing variables constitute a set of tetrad and
coordinate-gauge independent quantities governing the
gravitoelectromagnetic perturbations of rotating charged black strings.

Equations \eqref{eqbasica1} and \eqref{eqbasica2} take particularly simple
and symmetrical forms with the introduction of the following new variables
\begin{equation}
\begin{gathered}
\Phi_{0}=\Psi_{0}^{\scriptscriptstyle{(1)}},\qquad
\Phi_{1}=r\sqrt{2}\left[\Psi_{1}^{\scriptscriptstyle{(1)}}+\frac{3}{cr}
\left(\frac{b\ell}{2}-\frac{c^2\ell^2}{r}\right)\phi_{0}^{\scriptscriptstyle{(1)}}\right],\\
\mathcal{K}=\frac{1}{\sqrt{2}}\left[\frac{\kappa^{\scriptscriptstyle{(1)}}}{r^{2}}+
\frac{1}{c}\left(\mathscr{D}_{0}+\frac{3}{r}\right)\phi_{0}^{\scriptscriptstyle{(1)}}\right],\qquad
\mathcal{S}=\frac{1}{r}\left(\sigma^{\scriptscriptstyle{(1)}}+\frac{r}{\sqrt{2}c}\,ip\,
\phi_{0}^{\scriptscriptstyle{(1)}}\right),
\end{gathered}
\end{equation}
and
\begin{equation}
\begin{gathered}
\Phi_{3}=\frac{r^{3}}{\sqrt{2}}\left[\Psi_{3}^{\scriptscriptstyle{(1)}}+\frac{3}{cr}
\left(\frac{b\ell}{2}-\frac{c^2\ell^2}{r}\right)\phi_{2}^{\scriptscriptstyle{(1)}}\right],\qquad
\Phi_{4}=r^{4}\Psi_{4}^{\scriptscriptstyle{(1)}},\\
\mathcal{N}=\frac{r^{2}}{\sqrt{2}}\left[\nu^{\scriptscriptstyle{(1)}}+
\frac{r^4 f}{2c\ell^2}\left(\mathscr{D}_{-1}^{\dagger}+\frac{5}{r}\right)
\phi_{2}^{\scriptscriptstyle{(1)}}\right],\qquad
\mathcal{L}=\frac{1}{2}r\left(\lambda^{\scriptscriptstyle{(1)}}-\frac{r}{\sqrt{2}c}\,ip^{\ast}\,
\phi_{2}^{\scriptscriptstyle{(1)}}\right),
\end{gathered}
\end{equation}
which are proportional to the Fourier transforms of the quantities 
\eqref{new_variables}. This replacement of variables leads to the
following set of perturbation equations:
\begin{align}
&-ip^{\ast}\Phi_{0}+\left(\mathscr{D}_{0}+\frac{3}{r}\right)\Phi_{1}
-\left(3\ell b-\frac{4\ell^2 c^{2}}{r}\right)\mathcal{K}= 
\mathcal{T}_{\scriptscriptstyle{\cal{A}}},\label{bcharge1}\\
&\frac{r^4 f}{\ell^2}\left(\mathscr{D}_{2}^{\dagger}-\frac{3}{r}\right)
\Phi_{0}+ip\,\Phi_{1}-\left(3\ell b-\frac{8\ell^2 c^{2}}{r}\right)\mathcal{S}=
\mathcal{T}_{\scriptscriptstyle{\cal{B}}},\label{bcharge2}\\
&\left(\mathscr{D}_{0}+\frac{3}{r}\right)\mathcal{S}
-ip\,\mathcal{K}-\frac{\Phi_{0}}{r}=0,\label{bcharge3}\\
&\frac{r^4 f}{\ell^2}\left(\mathscr{D}_{2}^{\dagger}-\frac{3}{r}\right)
\mathcal{K}+ip^{\ast}\mathcal{S}-2\frac{\Phi_{1}}{r}=
\mathcal{J}_{\scriptscriptstyle{\cal{B}}},\label{bcharge4}
\end{align}
and
\begin{align}
&ip\,\Phi_{4}+\frac{r^4 f}{\ell^2}\left(\mathscr{D}_{-1}^{\dagger}+\frac{3}{r}\right)\Phi_{3}
-\left(3\ell b-\frac{4\ell^2 c^{2}}{r}\right)\mathcal{N}=\mathcal{T}_{\scriptscriptstyle{\cal{C}}},
\label{bcharge5}\\
&\left(\mathscr{D}_{0}-\frac{3}{r}\right)\Phi_{4}-ip^{\ast}\Phi_{3}
-\left(3\ell b-\frac{8\ell^2 c^{2}}{r}\right)\mathcal{L}=
\mathcal{T}_{\scriptscriptstyle{\cal{D}}},\label{bcharge6}\\
&\frac{r^4 f}{\ell^2}\left(\mathscr{D}_{-1}^{\dagger}+\frac{3}{r}\right)\mathcal{L}
+ip^{\ast}\mathcal{N}-\frac{\Phi_{4}}{r}=0,\label{bcharge7}\\
&-\left(\mathscr{D}_{0}-\frac{3}{r}\right)\mathcal{N}+ip\,\mathcal{L}
+2\frac{\Phi_{3}}{r}=\mathcal{J}_{\scriptscriptstyle{\cal{D}}},\label{bcharge8}
\end{align}
where, to obtain these final relations, we have made use of the second one 
of the identities given in equation \eqref{identidades}.

\subsection{Decoupling of the perturbation equations}
\label{Decoupling_eqs}

The set of perturbation equations \eqref{bcharge1}-\eqref{bcharge8} include
two decoupled systems of equations: \eqref{bcharge1}-\eqref{bcharge4} govern
the evolution of $\Phi_0$, $\Phi_1$, $\mathcal{K}$ and $\mathcal{S}$, and
\eqref{bcharge5}-\eqref{bcharge8} govern the scalars $\Phi_3$, $\Phi_4$,
$\mathcal{L}$ and $\mathcal{N}$. In the following, we study first the set of
equations \eqref{bcharge1}-\eqref{bcharge4}. The decoupling of these equations
and their subsequent reduction to a pair of second-order differential
equations can be carried out with the introduction of the functions
\begin{equation}
\begin{aligned}[2]
&\mathcal{F}_{+1}=\Phi_{0}-i\frac{q_{1}}{p^{\ast}}\mathcal{K},\qquad\quad
&\mathcal{G}_{+1}=\Phi_{1}+i\frac{q_{1}}{p}\mathcal{S},\\
&\mathcal{F}_{+2}=\Phi_{0}-i\frac{q_{2}}{p^{\ast}}\mathcal{K},\qquad\quad
&\mathcal{G}_{+2}=\Phi_{1}+i\frac{q_{2}}{p}\mathcal{S},
\end{aligned}
\end{equation}
where $q_{1}$ and $q_{2}$ are defined as 
\begin{equation}
q_{1}=\frac{1}{2}\ell\left[3b+\sqrt{9b^{2}+16c^{2}p^{2}}\right],\qquad
q_{2}=\frac{1}{2}\ell\left[3b-\sqrt{9b^{2}+16c^{2}p^{2}}\right],
\end{equation}
with
\begin{equation}
p^{2}=pp^{*}=\frac{\upgamma^{2}}{\ell^{2}}(m-a\omega)^{2}+k^{2}.
\end{equation}
The quantities $q_1$ and $q_2$ obey the following relations:
\begin{equation}
q_{1}+q_{2}=3\ell b\qquad\mbox{and}\qquad -q_{1}q_{2}=4\ell^2 c^{2}p^{2}.
\label{q1q2_relations}
\end{equation}

A set of equations governing the evolution of variables $\mathcal{F}_{+i}$ and
$\mathcal{G}_{+i}$ ($i=1,2$) can be found from suitable combinations of
equations \eqref{bcharge1}-\eqref{bcharge4}.
For instance, the addition of equation \eqref{bcharge2} to equation 
\eqref{bcharge4} multiplied by the factor $-iq_{1}/p^{\ast}$ furnishes
\begin{equation}
\frac{r^4}{\ell^2}f\left(\mathscr{D}_{2}^{\dagger}-\frac{3}{r}\right)\mathcal{F}_{+1}
+ip\left(1+\frac{2q_{1}}{p^{2}r}\right)\mathcal{G}_{+2}=\mathcal{T}_{\scriptscriptstyle{\cal{B}}}-
\frac{iq_1}{p^{\ast}}\mathcal{J}_{\scriptscriptstyle{\cal{B}}},
\label{mcharge1}
\end{equation}
where, to simplify the above equation, it was made use of relations \eqref{q1q2_relations}.
An analogous combination of equations \eqref{bcharge2} and \eqref{bcharge4}, with
the term $-iq_{2}/p^{\ast}$ in place of $-iq_{1}/p^{\ast}$, results in a relation
of the same form as equation \eqref{mcharge1}, except by the interchange of 
the indices $1$ and $2$. Both formulas can be compacted into the single expression
\begin{equation}
\frac{r^4}{\ell^2}f\left(\mathscr{D}_{2}^{\dagger}-\frac{3}{r}\right)\mathcal{F}_{+i}+
ip\left(1+\frac{2q_{i}}{p^{2}r}\right)\mathcal{G}_{+j}=\mathcal{T}_{\scriptscriptstyle{\cal{B}}}
-\frac{iq_i}{p^{\ast}}\mathcal{J}_{\scriptscriptstyle{\cal{B}}},
\qquad(i,j=1,2;\; i\neq j).
\label{mcharge2}
\end{equation}
A similar procedure involving equations \eqref{bcharge1} and \eqref{bcharge3}
leads to
\begin{equation}
\left(\mathscr{D}_{0}+\frac{3}{r}\right)\mathcal{G}_{+i}-
ip^{\ast}\left(1+\frac{q_{i}}{p^{2}r}\right)\mathcal{F}_{+j}=\mathcal{T}_{\scriptscriptstyle{\cal{A}}},
\qquad(i,j=1,2;\; i\neq j).
\label{mcharge3}
\end{equation}
The convention that $i$ and $j$ assume values 1 and 2 with $i\neq j$ will be used
from now on in various expressions of this work without explicit indication.

A further simplification of equations \eqref{mcharge1} and \eqref{mcharge2}
can be performed with the exchange of $\mathcal{F}_{+i}$ and $\mathcal{G}_{+i}$
by the new functions
\begin{equation}
Y_{+i}=\frac{r^5 f^{2}}{\ell^4}\mathcal{F}_{+i},\qquad\qquad
X_{+i}=r^{3}\mathcal{G}_{+i}.
\end{equation}
In terms of these variables, equations \eqref{mcharge2} and \eqref{mcharge3}
become, respectively,
\begin{equation}
\Lambda_{+}Y_{+i}+ip\,\frac{f^2}{\ell^4}\left(1+
\frac{2q_{i}}{p^{2}r}\right)X_{+j}=\frac{r^3 f}{\ell^2}
\left(\mathcal{T}_{\scriptscriptstyle{\cal{B}}}-\frac{iq_i}{p^{\ast}}
\mathcal{J}_{\scriptscriptstyle{\cal{B}}}\right),
\label{mcharge4}
\end{equation}
\begin{equation}
\Lambda_{-}X_{+j}-ip^{\ast}\frac{\ell^2}{f}\left(1+
\frac{q_{j}}{p^{2}r}\right)Y_{+i}=\frac{r^5 f}{\ell^2}\mathcal{T}_{\scriptscriptstyle{\cal{A}}},
\label{mcharge5}
\end{equation}
where the operators $\Lambda_{+}$ and $\Lambda_{-}$ are given by
\begin{equation}
\Lambda_{+}=\frac{d}{dr_{\ast}}+i\varpi=\frac{r^2}{\ell^2}f\,\mathscr{D}_{0}^{\dagger},
\quad\qquad\Lambda_{-}=\frac{d}{dr_{\ast}}-i\varpi=
\frac{r^2}{\ell^2}f\,\mathscr{D}_{0},
\end{equation}
and the tortoise coordinate $r_\ast$ is defined implicitly
by $dr/dr_\ast=r^{2}f/\ell^{2}$.

Finally, we choose to eliminate the $X$'s in favor of the $Y$'s
in \eqref{mcharge4} and \eqref{mcharge5},
and so we obtain the following decoupled equations:
\begin{equation}
\Lambda^{2}Y_{+i}+P_{i}\Lambda_{+}Y_{+i}-Q_{i}Y_{+i}=\mathfrak{S}_{+i},
\label{eq_fund1}
\end{equation}
where
\begin{equation}
\Lambda^2=\Lambda_{+}\Lambda_{-}=\Lambda_{-}\Lambda_{+}=
\frac{d^{\,2}}{dr_{\ast}^{2}}-\varpi^2.
\end{equation}
The functions $P_{i}$ and $Q_{i}$ appearing in equation \eqref{eq_fund1} 
are given by
\begin{equation}
P_{i}=\frac{d}{dr_{\ast}}\ln{\left(\frac{r^{8}}{\mathcal{D}_{i}}\right)},\qquad
Q_{i}=p^{2}\frac{\ell^{2}\,\mathcal{D}_{i}}{r^{8}f}
\left(1+\frac{q_{j}}{p^{2}r}\right),
\end{equation}
with
\begin{equation}
\mathcal{D}_i=\frac{r^{8}f^{2}}{\ell^4}\left(1+\frac{2q_{i}}{p^{2}r}\right),
\end{equation}
and the source terms $\mathfrak{S}_{+i}\,(i=1,2)$ can be written as
\begin{equation}
\mathfrak{S}_{+i}=-ip\,\frac{f}{r^3\ell^2}\,\mathcal{D}_{i}\,
\mathcal{T}_{\scriptscriptstyle{\cal{A}}}+\left(\Lambda_{-}+P_i\right)
\frac{r^3 f}{\ell^2}\left(\mathcal{T}_{\scriptscriptstyle{\cal{B}}}-\frac{iq_i}{p^{\ast}}
\mathcal{J}_{\scriptscriptstyle{\cal{B}}}\right).
\end{equation}

The decoupling of the second group of perturbation equations, \eqref{bcharge5}-\eqref{bcharge8},
can be carried out with the introduction of the functions
\begin{equation}
\begin{aligned}[2]
&\mathcal{F}_{-1}=\Phi_{4}+i\frac{q_{1}}{p}\mathcal{N},\qquad\quad
&\mathcal{G}_{-1}=\Phi_{3}-i\frac{q_{1}}{p^{\ast}}\mathcal{L},\\
&\mathcal{F}_{-2}=\Phi_{4}+i\frac{q_{2}}{p}\mathcal{N},\qquad\quad
&\mathcal{G}_{-2}=\Phi_{3}-i\frac{q_{2}}{p^{\ast}}\mathcal{L},
\end{aligned}
\end{equation}
and
\begin{equation} 
Y_{-i}=\frac{1}{r^3}\mathcal{F}_{-i},\qquad\qquad
X_{-i}=\frac{\ell^2}{rf}\mathcal{G}_{-i}.
\end{equation}

By means of a similar sequence of substitutions and reductions as above,
we obtain a pair of decoupled second-order differential equations for
$Y_{-1}$ and $Y_{-2}$:
\begin{equation}
\Lambda^{2}Y_{-i}+P_{i}\Lambda_{-}Y_{-i}-Q_{i}Y_{-i}=\mathfrak{S}_{-i},
\label{eq_fund2}
\end{equation}
where the source terms $\mathfrak{S}_{-i}\,(i=1,2)$ are given by
\begin{equation}
\mathfrak{S}_{-i}=ip^{\ast}\frac{\ell^2}{r^{11}f}\,\mathcal{D}_{i}\,
\mathcal{T}_{\scriptscriptstyle{\cal{C}}}+\left(\Lambda_{+}+P_i\right)
\frac{f}{r\ell^2}\left(\mathcal{T}_{\scriptscriptstyle{\cal{D}}}-\frac{iq_i}{p}
\mathcal{J}_{\scriptscriptstyle{\cal{D}}}\right).
\end{equation}

Equations~\eqref{eq_fund1} and~\eqref{eq_fund2} are then the uncoupled 
fundamental equations governing the gravitoelectromagnetic perturbations
of the rotating charged black strings.

\section{Chandrasekhar transformations and the Schr\"odinger-like equations}
\label{transformation-theory}

Once we have obtained the fundamental perturbation equations for $Y_{\pm i}$,
we can now use the generalized Chandrasekhar transformation theory
\cite{Chandrasekhar:1985kt,Chandrasekhar:1975,Chandrasekhar:1979iz} of 
\ref{Chandra_transformations}
to reduce a pair of complex equations (those for $Y_{+1}$ and $Y_{+2}$, say) to four
real one-dimensional Schr\"odinger-like equations.
The condition for the equations \eqref{eq_fund1} or \eqref{eq_fund2} to be 
transformed into the form
\begin{equation}
\Lambda^2 Z_{i}^{\scriptscriptstyle{(\pm)}}=V_{i}^{\scriptscriptstyle{(\pm)}}
Z_{i}^{\scriptscriptstyle{(\pm)}}+\mathscr{F}_{i}^{\scriptscriptstyle{(\pm)}},
\label{eq_onda}
\end{equation}
by means of the {\it{dual}} transformations of Heading \cite{Heading:1977},
is the existence of suitable constants $\mbox{\ss}_{i}^{2}$ and $\varkappa_i$, such that
the functions
\begin{equation}
F_i=r^{8}\frac{Q_{i}}{\mathcal{D}_{i}}=\frac{\ell^2}{rf}(p^{2}r+q_{j})
\quad\qquad(i,j=1,2;\; i\neq j)
\label{ctransf2}
\end{equation}
satisfy the nonlinear differential equations
\begin{equation}
\frac{1}{F_{i}}\left(\frac{dF_{i}}{dr_{\ast}}\right)^{2}-
\frac{d^{\,2}F_{i}}{dr_{\ast}^{2}}+
\frac{\mathcal{D}_{i}}{r^{8}}F_{i}^{2}=\frac{\mbox{\ss}_{i}^{2}}{F_{i}}
+\varkappa_{i}.
\label{ctransf3}
\end{equation}
Substituting expression \eqref{ctransf2} for $F_i$ into the last equation,
we obtain the following values for $\mbox{\ss}_{i}^{2}$ and $\varkappa_{i}$:
\begin{equation}
\mbox{\ss}_{i}^{2}=q_{j}^{2}\qquad(i,j=1,2;\; i\neq j)\qquad\quad\mbox{and}
\quad\qquad\varkappa_{i}=p^{4}.
\end{equation}

The fact that $\mbox{\ss}_i$ appears in equation \eqref{ctransf3} as
$\mbox{\ss}_{i}^{2}$ gives rise to a pair of dual transformations for each
value of $i$: a transformation associated with $\mbox{\ss}_{i}^{\scriptscriptstyle{(+)}}=+q_{j}$
and another one corresponding to $\mbox{\ss}_{i}^{\scriptscriptstyle{(-)}}=-q_j$. As it is usual
in theses cases, we use here the superscripts $(\pm)$ to distinguish between transformations
involving $+q_j$ and $-q_j$.

From the generalized transformation theory of 
\ref{Chandra_transformations},
it follows that the frequency-dependent dual potentials $V_{i}^{(\pm)}$ for perturbations
of a black string are such that
\begin{equation}
V_{i}^{(\pm)}=\pm q_{j}\frac{d\mathfrak{f}_{i}}{dr_{\ast}}
+q_{j}^{2}\mathfrak{f}_{i}^{2}+p^4\mathfrak{f}_{i},
\label{potenciais}
\end{equation}
with
\begin{equation}
\mathfrak{f}_i=\frac{1}{F_i}=\frac{rf}{\ell^{2}(p^{2}r+q_{j})}
\qquad(i,j=1,2;\; i\neq j).
\end{equation}
It is not difficult to write the
effective potentials \eqref{potenciais} in an explicit form. For the sector of
perturbations labeled by $(-)$, a direct computation shows that
\begin{equation}
V_{i}^{\scriptscriptstyle{(-)}}=\frac{f}{\ell^2}\left[p^{2}+
4\frac{c^{2}\ell^2}{r^2}-\frac{q_{j}}{r}\right],
\label{potaxial}
\end{equation}
while the effective potentials labeled by $(+)$ are given by
\begin{equation}
V_{i}^{\scriptscriptstyle{(+)}}=\frac{f}{\ell^{2}(p^{2}r+q_{j})^{2}}
\left[q_{j}p^{2}\left(3b\ell-2\frac{c^{2}\ell^2}{r}\right)
+2q_{j}^{2}\left(\frac{r^{2}}{\ell^2}+\frac{b\ell}{2r}-\frac{c^{2}\ell^2}{r^2}\right)
+p^{4}r(p^{2}r+q_{j})\right].
\label{potpolar}
\end{equation}

The source terms $\mathfrak{S}_{\pm i}$ and $\mathscr{F}_{i}^{\scriptscriptstyle{(\pm)}}$,
which appear in the differential equations \eqref{eq_fund1}, \eqref{eq_fund2} and \eqref{eq_onda},
are related by
\begin{equation}
\mathfrak{S}_{+i}=\left(P_i+\Lambda_{-}\right)\left(\Xi_{i}^{\scriptscriptstyle{(\pm)}}
+\Lambda_{-}\right)\mathscr{F}_{i}^{\scriptscriptstyle{(\pm)}},
\label{transform_sources1}
\end{equation}
\begin{equation}
\mathfrak{S}_{-i}=\left(P_i+\Lambda_{+}\right)\left(\Xi_{i}^{\scriptscriptstyle{(\pm)}}
+\Lambda_{+}\right)\mathscr{F}_{i}^{\scriptscriptstyle{(\pm)}},
\label{transform_sources2}
\end{equation}
where the auxiliary functions $\Xi_{i}^{\scriptscriptstyle{(-)}}$
and $\Xi_{i}^{\scriptscriptstyle{(+)}}$ are
\begin{equation}
\Xi_{i}^{\scriptscriptstyle{(\pm)}}=-\frac{d}{dr_{\ast}}\ln \mathfrak{f}_{i}\mp q_{j}\mathfrak{f}_{i},
\label{expression_for_xi}
\end{equation}
or, explicitly,
\begin{equation}
\Xi_{i}^{\scriptscriptstyle{(-)}}=\Xi^{\scriptscriptstyle{(-)}}=
\frac{1}{r^2}\left(-3b\ell+4\frac{c^2\ell^2}{r}\right)
\end{equation}
and
\begin{equation}
\Xi_{i}^{\scriptscriptstyle{(+)}}=\Xi^{\scriptscriptstyle{(-)}}-
2q_j\,\frac{rf}{\ell^2\left(p^2 r+q_j\right)}.
\end{equation}
According to equations \eqref{solution_for_h}-\eqref{inversa_final} of 
\ref{Chandra_transformations}, the inverse
relations of \eqref{transform_sources1} and \eqref{transform_sources2} can be
determined by a pair of consecutive quadratures of the equations
\begin{equation}
\frac{d}{dr_{\ast}}\left[h_{i}^{\scriptscriptstyle{(\pm)}}\frac{d}{dr_{\ast}}
\left(g_{i}^{\scriptscriptstyle{(\pm)}}
\mathscr{F}_{i}^{\scriptscriptstyle{(\pm)}}e^{-i\varpi r_{\ast}}\right)\right]=
h_{i}^{\scriptscriptstyle{(\pm)}}g_{i}^{\scriptscriptstyle{(\pm)}}\mathfrak{S}_{+i}\,
e^{-i\varpi r_{\ast}},
\end{equation}
\begin{equation}
\frac{d}{dr_{\ast}}\left[h_{i}^{\scriptscriptstyle{(\pm)}}\frac{d}{dr_{\ast}}
\left(g_{i}^{\scriptscriptstyle{(\pm)}}
\mathscr{F}_{i}^{\scriptscriptstyle{(\pm)}}e^{+i\varpi r_{\ast}}\right)\right]=
h_{i}^{\scriptscriptstyle{(\pm)}}g_{i}^{\scriptscriptstyle{(\pm)}}\mathfrak{S}_{-i}\,
e^{+i\varpi r_{\ast}}.
\end{equation}

After a convenient choice of integration constant 
($C_{i}^{\scriptscriptstyle{(\pm)}}=0$)
in equations \eqref{solution_for_h} and \eqref{solution_for_g}, the functions
$h_{i}^{\scriptscriptstyle{(-)}}$ and $g_{i}^{\scriptscriptstyle{(-)}}$ are reduced to
\begin{equation}
h_{i}^{\scriptscriptstyle{(-)}}=\frac{r^{6} f}{\ell^{2}\,\mathcal{D}_i}
\left(1+\frac{q_{i}}{p^{2}r}\right)^{2},\qquad\quad
g_{i}^{\scriptscriptstyle{(-)}}=\frac{\ell^{2}r}{f}
\left(1+\frac{q_{i}}{p^{2}r}\right)^{-1},
\end{equation}
while the functions $h_{i}^{\scriptscriptstyle{(+)}}$ and
$g_{i}^{\scriptscriptstyle{(+)}}$ become
\begin{equation}
h_{i}^{\scriptscriptstyle{(+)}}=\frac{\ell^{2}}{f\,\mathcal{D}_{i}F_{i}^{2}}
\left[b\ell r\left(3p^{2}r+4q_{j}\right)-q_{j}\left(2\ell^{2}c^{2}+q_{j}r\right)+p^{4}r^{3}\right]^{2},
\end{equation}
\begin{equation}
g_{i}^{\scriptscriptstyle{(+)}}=F_{i}^{2}\,\frac{r^{4}f}{\ell^2}
\left[b\ell r\left(3p^{2}r+4q_{j}\right)-q_{j}\left(2\ell^{2}c^{2}+q_{j}r\right)+p^{4}r^{3}\right]^{-1}.
\end{equation}

The reductions of the perturbation equations for $Y_{\pm i}$ to the master
equations \eqref{eq_onda} are accomplished by means of the substitutions
\begin{equation}
Y_{+i}=V_{i}^{\scriptscriptstyle{(\pm)}}Z_{i}^{\scriptscriptstyle{(\pm)}}
+(\Xi_{i}^{\scriptscriptstyle{(\pm)}}-2i\varpi)
\Lambda_{-}Z_{i}^{\scriptscriptstyle{(\pm)}}+\mathscr{F}_{i}^{\scriptscriptstyle{(\pm)}},
\label{transf1}
\end{equation}
\begin{equation}
Y_{-i}=V_{i}^{\scriptscriptstyle{(\pm)}}Z_{i}^{\scriptscriptstyle{(\pm)}}
+(\Xi_{i}^{\scriptscriptstyle{(\pm)}}+2i\varpi)
\Lambda_{+}Z_{i}^{\scriptscriptstyle{(\pm)}}+\mathscr{F}_{i}^{\scriptscriptstyle{(\pm)}},
\label{transf2}
\end{equation}
whose inverse transformations assume the form
\begin{equation}
K_{i}^{\scriptscriptstyle{(\mp)}}Z_{i}^{\scriptscriptstyle{(\pm)}}
=\frac{r^{8}}{\mathcal{D}_{i}}Q_{i}Y_{+i}-
\frac{r^{8}}{\mathcal{D}_{i}}(\Xi_{i}^{\scriptscriptstyle{(\pm)}}-2i\varpi)\Lambda_{+}Y_{+i}
-\frac{r^{8}}{\mathcal{D}_{i}}\left[Q_{i}-(\Xi_{i}^{\scriptscriptstyle{(\pm)}}-2i\varpi)
\mathsterling^{\scriptscriptstyle{(\pm)}}_{-i}\right]\mathscr{F}_{i}^{\scriptscriptstyle{(\pm)}},
\label{inv_transf1}
\end{equation}
\begin{equation}
K_{i}^{\scriptscriptstyle{(\pm)}}Z_{i}^{\scriptscriptstyle{(\pm)}}
=\frac{r^{8}}{\mathcal{D}_{i}}Q_{i}Y_{-i}-
\frac{r^{8}}{\mathcal{D}_{i}}(\Xi_{i}^{\scriptscriptstyle{(\pm)}}+2i\varpi)\Lambda_{-}Y_{-i}
-\frac{r^{8}}{\mathcal{D}_{i}}\left[Q_{i}-(\Xi_{i}^{\scriptscriptstyle{(\pm)}}+2i\varpi)
\mathsterling^{\scriptscriptstyle{(\pm)}}_{+i}\right]\mathscr{F}_{i}^{\scriptscriptstyle{(\pm)}},
\label{inv_transf2}
\end{equation}
where $K_{i}^{\scriptscriptstyle{(\pm)}}=p^4 \pm 2i\,\varpi\,q_{j}$.

In the limit of $a\rightarrow 0$, parameters $\varpi=\upgamma\left(\omega-a m/\ell^2\right)$
and $p^{2}=\upgamma^{2}(m-a\omega)^{2}/\ell^2+k^{2}$ tend, respectively, to the frequency and
to the square of the perturbation wave vector.
A lengthy but straightforward calculation shows that equations 
\eqref{eq_onda},
with $\mathscr{F}_{i}^{\scriptscriptstyle{(\pm)}}=0$ and potentials given by \eqref{potaxial}
and \eqref{potpolar}, reduce to the sourceless wave equations for the linear perturbations
of a non-rotating charged black string \cite{Kodama:2003kk} as $a\rightarrow 0$.
The variables $Z_{i}^{\scriptscriptstyle{(+)}}$ ($i=1,2$) correspond to
the even (polar) perturbations under the parity transformation
$\varphi\rightarrow -\varphi$, while $Z_{i}^{\scriptscriptstyle{(-)}}$ ($i=1,2$)
are associated to the odd (axial) sector of perturbations. In the zero charge limit,
the gravitational and electromagnetic perturbations decouple, and the wave equations for
$Z_{1}^{\scriptscriptstyle{(+)}}$ and $Z_{1}^{\scriptscriptstyle{(-)}}$ govern
the electromagnetic perturbations of an electrically neutral black string, while those
for $Z_{2}^{\scriptscriptstyle{(+)}}$ and $Z_{2}^{\scriptscriptstyle{(-)}}$
govern the gravitational perturbations. The source-free equations for
$Z_{1}^{\scriptscriptstyle{(+)}}$ and $Z_{1}^{\scriptscriptstyle{(-)}}$ reduce,
in the $Q\rightarrow 0$ limit, to the master equations for the Regge-Wheeler-Zerilli
variables of the electromagnetic perturbations \cite{Miranda:2008vb,Morgan:2013dv}.

\section{The SUSY quantum mechanics of perturbations}
\label{SUSY_QM}

In addition to allow the reduction of the problem to the four real Schr\"odinger-like
wave equations \eqref{eq_onda}, the transformation theory also reveals important
aspects on the mathematical structure of the
gravitoelectromagnetic perturbations. When properly taken into account,
these aspects will simplify even more the study and application of the
black-string perturbation theory.

A result which follows directly from the transformation theory is the existence of a relation,
pair to pair, between the variables $Z_{i}^{\scriptscriptstyle{(+)}}$ of one sector
of perturbations and the variables $Z_{i}^{\scriptscriptstyle{(-)}}$ of the other
sector (see \ref{Chandra_transformations}). As we will 
show here,
these relations can be viewed as a consequence of an underlying quantum-mechanical
supersymmetry \cite{Witten:1981nf,Cooper:1982dm,Cooper:1994eh} of the black-string
perturbation equations, a typical result for black holes in four
spacetime dimensions \cite{Leung:1999fr,Cardoso:2001bb,Bakas:2008gz}.

To see the emergence of the supersymmetric (SUSY) aspects of the theory,
we first notice that the potentials $V_{i}^{\scriptscriptstyle{(\pm)}}$
can be written as
\begin{equation}
V_{i}^{\scriptscriptstyle{(\pm)}}=\pm \frac{dW_{i}}{dr_{\ast}}
+W_{i}^{2}+\Omega_{i}^{2},
\label{potenciais2}
\end{equation}
with the introduction of the functions
\begin{equation}
W_i=\mbox{\ss}_{i}\mathfrak{f}_i+\frac{\varkappa_i}{2\mbox{\ss}_{i}}=q_j\mathfrak{f}_i+i\Omega_i,
\end{equation}
and the constants
\begin{equation}
\Omega_i=-i\frac{p^4}{2q_j},\qquad(i,j=1,2;\; i\neq j).
\end{equation}

The Riccati form of equation \eqref{potenciais2} shows that $V_{i}^{\scriptscriptstyle{(+)}}$ and
$V_{i}^{\scriptscriptstyle{(-)}}$ are supersymmetric partner potentials generated by
the superpotentials $W_{i}$ ($i=1,2$). In terms of $W_i$ and $\Omega_i$,
the sourceless versions of equations \eqref{eq_onda} take the form
\begin{equation}
\left(-\frac{d^2}{dr_{\ast}^2}+W_{i}^{2}\pm\frac{dW_i}{dr_{\ast}}\right)
Z_{i}^{\scriptscriptstyle{(\pm)}}=\left(\varpi^2-\Omega_{i}^{2}\right)
Z_{i}^{\scriptscriptstyle{(\pm)}},
\label{eq_onda2}
\end{equation}
which shows that $E_{i}=\varpi^2-\Omega_{i}^{2}$ plays the role of the 
energy of the corresponding effective quantum-mechanical problem. From the 
SUSY quantum mechanics formalism,
we know that equations \eqref{eq_onda2} can be simplified with the introduction
of the first-order operators
\begin{equation}
A_i=\frac{d}{dr_\ast}+W_i\qquad\;\mbox{and}\qquad\;
A_{i}^{\dagger}=-\frac{d}{dr_\ast}+W_i\,.
\end{equation}
The effective hamiltonians $H_{i}^{\scriptscriptstyle{(\pm)}}$ are then written as
\begin{equation}
H_{i}^{\scriptscriptstyle{(-)}}=A_{i}^{\dagger}A_{i}\qquad\;\mbox{and}\qquad\;
H_{i}^{\scriptscriptstyle{(+)}}=A_{i}A_{i}^{\dagger},
\end{equation}
so that equations \eqref{eq_onda2} assume the traditional form
\begin{equation}
H_{i}^{\scriptscriptstyle{(\pm)}}Z_{i}^{\scriptscriptstyle{(\pm)}}=
E_{i}Z_{i}^{\scriptscriptstyle{(\pm)}}.
\label{eq_onda3}
\end{equation}

A consequence of the supersymmetric partnership between the effective 
potentials $V_{i}^{\scriptscriptstyle{(+)}}$ and $V_{i}^{\scriptscriptstyle{(-)}}$ is
that wave functions of one sector of perturbations are interconnected to wave functions
of the other sector. In fact, applying the operators $H_{i}^{\scriptscriptstyle{(-)}}$
and $H_{i}^{\scriptscriptstyle{(+)}}$, respectively, to the functions
$A_{i}^{\dagger}  Z_{i}^{\scriptscriptstyle{(+)}}$ and $A_{i}Z_{i}^{\scriptscriptstyle{(-)}}$,
and using equations \eqref{eq_onda3}, we obtain
\begin{equation}
H_{i}^{\scriptscriptstyle{(-)}}\big(A_{i}^{\dagger}Z_{i}^{\scriptscriptstyle{(+)}}\big)=
E_i\big(A_{i}^{\dagger}Z_{i}^{\scriptscriptstyle{(+)}}\big)\qquad\mbox{and}\qquad
H_{i}^{\scriptscriptstyle{(+)}}\big(A_{i} Z_{i}^{\scriptscriptstyle{(-)}}\big)=
E_i\big(A_{i}Z_{i}^{\scriptscriptstyle{(-)}}\big).
\end{equation}
These relations imply that solutions for $Z_{i}^{\scriptscriptstyle{(-)}}$ and
$Z_{i}^{\scriptscriptstyle{(+)}}$ with an eigenvalue $E_i$ are proportional,
respectively, to $A_{i}^{\dagger}  Z_{i}^{\scriptscriptstyle{(+)}}$ and
$A_{i} Z_{i}^{\scriptscriptstyle{(-)}}$; that is,
\begin{equation}
\mathscr{C}_{i}^{\scriptscriptstyle{(+)}}Z_{i}^{\scriptscriptstyle{(-)}}=
A_{i}^{\dagger}Z_{i}^{\scriptscriptstyle{(+)}}
\label{proportional1}
\end{equation}
and
\begin{equation}
\mathscr{C}_{i}^{\scriptscriptstyle{(-)}}Z_{i}^{\scriptscriptstyle{(+)}}=
A_{i} Z_{i}^{\scriptscriptstyle{(-)}},
\label{proportional2}
\end{equation}
where $\mathscr{C}_{i}^{\scriptscriptstyle{(+)}}$ and $\mathscr{C}_{i}^{\scriptscriptstyle{(-)}}$
are proportionality constants. Now applying $A_{i}$ to both sides of \eqref{proportional1},
and using equations \eqref{eq_onda3} and \eqref{proportional2} to simplify the resulting expression,
we find that the constants $\mathscr{C}_{i}^{\scriptscriptstyle{(\pm)}}$ are 
constrained by the following equation:
\begin{equation}
\mathscr{C}_{i}^{\scriptscriptstyle{(+)}}\mathscr{C}_{i}^{\scriptscriptstyle{(-)}}=E_{i}=\varpi^2-\Omega_{i}^{2}\,.
\end{equation}
Therefore, choosing the relative normalization of $Z_{i}^{\scriptscriptstyle{(-)}}$
and $Z_{i}^{\scriptscriptstyle{(+)}}$ such that
\begin{equation}
\mathscr{C}_{i}^{\scriptscriptstyle{(+)}}=\mathscr{C}_{i}^{{\scriptscriptstyle{(-)}}\ast}=
i(\Omega_i+\varpi)=\frac{p^4}{2q_j}+i\varpi,
\end{equation}
relations \eqref{proportional1} and \eqref{proportional2} become
\begin{align}
\left(p^4+2i\varpi q_j\right)Z_{i}^{\scriptscriptstyle{(-)}}=\left(p^4+
\frac{2q_{j}^{2}}{F_i}\right)Z_{i}^{\scriptscriptstyle{(+)}}-2q_{j}\frac{d}{dr_{\ast}}Z_{i}^{\scriptscriptstyle{(+)}}, 
\label{supersymmetric_relation1}
\end{align}
\begin{align}
\left(p^4-2i\varpi q_j\right)Z_{i}^{\scriptscriptstyle{(+)}}=\left(p^4+
\frac{2q_{j}^{2}}{F_i}\right)Z_{i}^{\scriptscriptstyle{(-)}}+2q_{j}\frac{d}{dr_{\ast}}Z_{i}^{\scriptscriptstyle{(-)}}, 
\label{supersymmetric_relation2}
\end{align}
which are identical to the relations \eqref{relation_ZplusZminus} and \eqref{inverse_ZplusZminus}
when we consider that $\mathscr{A}_{i}$ and $\mathscr{F}_{i}^{\scriptscriptstyle{(\pm)}}$ vanish
in the absence of sources, and $\varkappa_{i}=p^4$ and $\mbox{\ss}_{i}=q_j$ for
black-string perturbations.

\section{Summary of results and perspectives}
\label{final}

In this paper a set of four decoupled complex equations for
the variables $Y_{\pm i}$ ($i=1,2$) was obtained by means of a gauge and tetrad
invariant perturbation approach which includes the presence of sources. The variables
$Y_{\pm i}$ are combinations of the spin coefficients, the Weyl and the Maxwell scalars
of the Newman-Penrose formalism, and represent the radiative (non-trivial) parts of
the gravitoelectromagnetic perturbations of rotating charged black strings.

With the aim of obtaining perturbation wave equations of a Schr\"odinger-like form, we have
generalized in \ref{Chandra_transformations} the Chandrasekhar transformation theory
\cite{Chandrasekhar:1985kt, Chandrasekhar:1975,Chandrasekhar:1979iz} to deal with
second-order ordinary differential equations with source terms. It is worth
emphasizing here that the constructed transformation theory is general in
character, and applies to any equation of the form \eqref{eq_fund1} or
\eqref{eq_fund2} with $P_{i}=\partial_{r_{\ast}} \ln\left(r^{8}/\mathcal{D}_{i}\right)$
and arbitrary functions $\mathcal{D}_{i}$, $Q_{i}$ and $\mathfrak{S}_{\pm i}$.
An obvious application is to the study of perturbations of diverse black holes which
admit a formulation in the manner of Teukolsky \cite{Teukolsky:1972my,Teukolsky:1973ha}.
In fact, given its high degree of generality, we can consider the development of this
transformation theory as one of the main results of the present work.

The Chandrasekhar transformation theory with sources was then used to
reduce the black-string perturbation problem to four real decoupled
inhomogeneous Schr\"odinger-like equations for a new set of variables,
$Z_{i}^{\scriptscriptstyle{(-)}}$ and $Z_{i}^{\scriptscriptstyle{(+)}}$.
As a consequence of the way the rotation is implemented in the charged black
strings, the main difference of the resulting wave equations in comparison with
the non-rotating case is the Lorentz transformation of frequency and
wave-vector components:
\begin{equation}
\omega\rightarrow\varpi=\upgamma\left(\omega-\frac{a
m}{\ell^2}\right),\quad\quad
\left(\frac{m}{\ell}\right)^2+k^2 \rightarrow p^{2}=\frac{\upgamma^{2}}{\ell^{2}}(m-a\omega)^{2}+k^{2}.
\end{equation}

The relation between the variables $Z_{i}^{\scriptscriptstyle{(-)}}$ and
$Z_{i}^{\scriptscriptstyle{(+)}}$ reveals an underlying hidden symmetry of
the four-dimensional Einstein-Maxwell theory on a black-hole spacetime. Such
a symmetry can be viewed both as an extension of the electric/magnetic duality
\cite{Witten:2003ya,Herzog:2007ij,Hartnoll:2007ip,deHaro:2007eg,Myers:2010pk}
and of its counterpart in linearized gravity
\cite{Henneaux:2004jw,deHaro:2008gp,Bakas:2008zg,Sadeghi:2010zza}.
When properly taken into account, this result may have important
consequences in connection to the AdS/CFT correspondence, and may also be explored
as a tool in future applications. For instance, it is well known that polar
perturbation equations of static black holes are much more difficult
to be solved (either analytically or numerically) than the axial wave
equations. However, in view of the quantum-mechanical
supersymmetry of the master equations, we can find solutions in the sector
labeled by $(-)$ and use relation \eqref{supersymmetric_relation2} to
obtain the wave functions in the $(+)$ sector of perturbations.

Other aspect of the black-string perturbation theory, which is relevant
for future applications, is the invariance of the set of 
variables $\{\Phi_{0},\,\Phi_{1},\,\mathcal{K},\,\mathcal{S}\}$ and
$\{\Phi_{3},\,\Phi_{4},\,\mathcal{N},\,\mathcal{L}\}$ under infinitesimal
tetrad transformations. As shown by Wald \cite{Wald:1978vm,Wald:1979},
a tetrad-invariant approach for  perturbations is an essential
ingredient to allow the use of the Chrzanowski-Cohen-Kegeles procedure
\cite{Chrzanowski:1975wv,Kegeles:1979an} to reconstruct
the metric and vector potentials from the perturbed NP scalars.
Aditionally, the explicit presence of source terms in the wave equations opens
the possibility of a series of important applications, including the study of
the influence of a hypothetical astrophysical environment on a rotating
charged black string.

\appendix

\renewcommand{\theequation}{\Alph{section}.\arabic{equation}} 
\setcounter{equation}{0} 

\section{The Newman-Penrose formalism}
\label{apenA}

The method of spin coefficients of Newman and Penrose \cite{Newman:1961qr}
is a tetrad formalism with a basis of null vectors
$\{\vec{l},\, \vec{n},\, \vec{m}, \vec{m}^{\ast}\}$,
where $\vec{l}$ and $\vec{n}$ are real and $\vec{m}$ and $\vec{m}^{\ast}$ are
complex conjugates of each other. In order to introduce the notation
and sign conventions used in this work and to make the paper self-contained,
we present below the main equations of the Newman-Penrose formalism with
the explicit inclusion of source
terms due to an electromagnetic field and other nongravitational fields.

\subsection{The null basis and the spin coefficients}
\label{Base-Nula}

The tetrad system of vectors $\{\vec{l},\, \vec{n},\, \vec{m}, \vec{m}^{\ast}\}$
are required to satisfy the orthogonality conditions,
\begin{equation}
l_{\mu}m^{\mu}=l_{\mu}m^{\ast\mu}=n_{\mu}m^{\mu}=n_{\mu}m^{\ast\mu}=0,
\end{equation}
and the normalization conditions,
\begin{equation}
l_{\mu}n^{\mu}=-m_{\mu}m^{\ast\mu}=-1,
\end{equation}
besides, of course, the null-vector conditions,
\begin{equation}
l_{\mu}l^{\mu}=n_{\mu}n^{\mu}=m_{\mu}m^{\mu}=m_{\mu}^{\ast}m^{\ast\mu}=0.
\end{equation}
As a result of these conditions, the metric takes the form of a flat-space metric
in a null basis,
\begin{equation}
\left[\eta_{ab}\right]=\left[\eta^{ab}\right]=
\left[
\begin{array}{ccrr}
0 & -1 & 0 & 0 \\
-1 & 0 & 0 & 0 \\
0 & 0 & 0 & 1 \\
0 & 0 & 1 & 0
\end{array}
\right],
\label{metrica_basenula}
\end{equation}
with the identifications $\vec{e}_{1}=\vec{l}$,
$\vec{e}_{2}=\vec{n}$, $\vec{e}_{3}=\vec{m}$
and $\vec{e}_{4}=\vec{m}^{\ast}$.

The connection coefficients of the usual tensor calculus
are substituted by 12 complex functions, which are called
the spin coefficients and designated by the symbols
\begin{equation}
\begin{split}
-\kappa & =\gamma_{mll}=l_{\mu;\nu}m^{\mu}l^{\nu},
\quad\qquad\,-\rho =\gamma_{mlm^{\ast}}=l_{\mu;\nu}m^{\mu}m^{\ast\nu},\\
-\sigma & =\gamma_{mlm}=l_{\mu;\nu}m^{\mu}m^{\nu},
\quad\quad\,-\mu=\gamma_{nm^{\ast}m}=m_{\mu;\nu}^{\ast}n^{\mu}m^{\nu},\\
-\lambda & =\gamma_{nm^{\ast}m^{\ast}}=m_{\mu;\nu}^{\ast}n^{\mu}m^{\ast\nu},
\;\;-\tau=\gamma_{mln}=l_{\mu;\nu}m^{\mu}n^{\nu},\\
-\nu & =\gamma_{nm^{\ast}n}=m_{\mu;\nu}^{\ast}n^{\mu}n^{\nu},
\quad\;\;\,-\pi=\gamma_{nm^{\ast}l}=m_{\mu;\nu}^{\ast}n^{\mu}l^{\nu},\\
-\varepsilon & =\tfrac{1}{2}\left[\gamma_{nll}
+\gamma_{mm^{\ast}l}\right]
=\tfrac{1}{2}[l_{\mu;\nu}n^{\mu}l^{\nu}+m_{\mu;\nu}^{\ast}m^{\mu}l^{\nu}],\\
-\gamma & =\tfrac{1}{2}\left[\gamma_{nln}
+\gamma_{mm^{\ast}n}\right]=
\tfrac{1}{2}[l_{\mu;\nu}n^{\mu}n^{\nu}+m_{\mu;\nu}^{\ast}m^{\mu}n^{\nu}],\\
-\alpha & =\tfrac{1}{2}\left[\gamma_{nlm^{\ast}}
+\gamma_{mm^{\ast}m^{\ast}}\right]=\tfrac{1}{2}[l_{\mu;\nu}n^{\mu}m^{\ast\nu}
+m_{\mu;\nu}^{\ast}m^{\mu}m^{\ast\nu}],\\
-\beta &=\tfrac{1}{2}\left[\gamma_{nlm}
+\gamma_{mm^{\ast}m}\right]=\tfrac{1}{2}[l_{\mu;\nu}n^{\mu}m^{\nu}
+m_{\mu;\nu}^{\ast}m^{\mu}m^{\nu}],
\end{split}
\label{coef-spin}
\end{equation}
where the quantities $\gamma_{abc}=e_{a}^{\;\;\mu}\,e_{b\,\mu;\nu}\,
e_{c}^{\;\;\nu}$ are the so-called Ricci rotation coefficients
of the general tetrad formalism.
The basis vectors, when acting as directional derivatives,
are also named by special symbols:
\begin{equation}
D=l^{\mu}\frac{\partial}{\partial x^{\mu}},\qquad
\Delta=n^{\mu}\frac{\partial}{\partial x^{\mu}},
\qquad\delta=m^{\mu}\frac{\partial}{\partial x^{\mu}},
\qquad\delta^{\ast}=m^{\ast\mu}\frac{\partial}{\partial x^{\mu}}.
\label{NP-operators}
\end{equation}
These operators do not commute with each other and the Lie brackets among 
them give rise to the following commutation relations,
\begin{align}
[\Delta,\,D] & =\Delta D-D\Delta=(\gamma+\gamma^{\ast})D+
(\varepsilon+\varepsilon^{\ast})\Delta-
(\tau^{\ast}+\pi)\delta-(\tau+\pi^{\ast})
\delta^{\ast},\label{NP-comut1}\\
[\delta,\,D] & =\delta D-D\delta=(\alpha^{\ast}+\beta-\pi^{\ast})D+
\kappa\Delta-(\rho^{\ast}+\varepsilon-\varepsilon^{\ast})
\delta-\sigma\delta^{\ast},
\label{NP-comut2}\\
[\delta,\,\Delta] & =\delta\Delta-\Delta\delta=-\nu^{\ast}D+
(\tau-\alpha^{\ast}-\beta)\Delta+(\mu-\gamma+\gamma^{\ast})
\delta+\lambda^{\ast}\delta^{\ast},
\label{NP-comut3}\\
[\delta^{\ast},\,\delta] & =\delta^{\ast}\delta-\delta\delta^{\ast}=(\mu^{\ast}-\mu)D+
(\rho^{\ast}-\rho)\Delta+(\alpha-\beta^{\ast})
\delta+(\beta-\alpha^{\ast})\delta^{\ast},
\label{NP-comut4}
\end{align}
which are part of the basic set of equations of the NP formalism.

\subsection{Weyl, Ricci and Maxwell scalars}
\label{tensores}

In a Riemannian four-dimensional manifold, half of the independent
components of the Riemann curvature tensor are given by the Ricci
tensor $R_{ac}=\eta^{ad}
R_{abcd}$, and the other half are
given by the Weyl tensor
\begin{equation}
\begin{split}
C_{abcd}=&R_{abcd}-\frac{1}{2}\left[\eta_{ac}R_{bd}+\eta_{bd}R_{ac}
-\eta_{bc}R_{ad}-\eta_{ad}R_{bc}\right]\\
&+\frac{1}{6}\left[\eta_{ac}\eta_{bd}-\eta_{ad}\eta_{bc}\right]R,
\end{split}
\label{Weyl-tensor}
\end{equation}
where $R=\eta^{ab}R_{ab}=-2\left[R_{ln}-R_{mm^{\ast}}\right]$
is the curvature (Ricci) scalar.

The ten independent components of the Weyl tensor in a null basis are completely specified
by the five complex scalars
\begin{equation}
\begin{aligned}
\Psi_{0}&=C_{lmlm}=C_{\mu\nu\rho\sigma}l^{\mu}m^{\nu}l^{\rho}m^{\sigma},\\
\Psi_{1}&=C_{lnlm}=C_{\mu\nu\rho\sigma}l^{\mu}n^{\nu}l^{\rho}m^{\sigma},\\
\Psi_{2}&=C_{lmm^{\ast}n}=C_{\mu\nu\rho\sigma}l^{\mu}m^{\nu}m^{\ast\rho}n^{\sigma},\\
\Psi_{3}&=C_{lnm^{\ast}n}=C_{\mu\nu\rho\sigma}l^{\mu}n^{\nu}m^{\ast\rho}n^{\sigma},\\
\Psi_{4}&=C_{nm^{\ast}nm^{\ast}}=C_{\mu\nu\rho\sigma}n^{\mu}m^{\ast\nu}n^{\rho}m^{\ast\sigma},\\
\label{Weyl-scalars}
\end{aligned}
\vspace{-0.7cm}
\end{equation}
and the components of the Ricci tensor are represented by the
quantities
\begin{gather}
\begin{aligned}
&\Phi_{00}=\tfrac{1}{2}R_{ll}=\tfrac{1}{2}
R_{\mu\nu}l^{\mu}l^{\nu},\; & \;\Phi_{22}& =\tfrac{1}{2}R_{nn}
=\tfrac{1}{2}R_{\mu\nu}n^{\mu}n^{\nu},\\
&\Phi_{01}=\tfrac{1}{2}R_{lm}=\tfrac{1}{2}
R_{\mu\nu}l^{\mu}m^{\nu},\; & \;\Phi_{10}& =\tfrac{1}{2}R_{lm^{\ast}}=
\tfrac{1}{2}R_{\mu\nu}l^{\mu}m^{\ast\nu},\\
&\Phi_{02}=\tfrac{1}{2}R_{mm}=
\tfrac{1}{2}R_{\mu\nu}m^{\mu}m^{\nu},\; & \;\Phi_{20}&=
\tfrac{1}{2}R_{m^{\ast}m^{\ast}}=\tfrac{1}{2}R_{\mu\nu}m^{\ast\mu}m^{\ast\nu},\\
&\Phi_{12}=\tfrac{1}{2}R_{nm}=
\tfrac{1}{2}R_{\mu\nu}n^{\mu}m^{\nu},\; & \;\Phi_{21}&=\tfrac{1}{2}R_{nm^{\ast}}=
\tfrac{1}{2}R_{\mu\nu}n^{\mu}m^{\ast\nu},
\label{Ricci-scalars}
\end{aligned}\\
\Phi_{11}=\tfrac{1}{4}[R_{ln}+R_{mm^{\ast}}]=
\tfrac{1}{4}[R_{\mu\nu}l^{\mu}n^{\nu}+R_{\mu\nu}m^{\mu}m^{\ast\nu}],\nonumber\\
\Pi=\tfrac{1}{24}R=-\tfrac{1}{12}[R_{ln}-R_{mm^{\ast}}]=
-\tfrac{1}{12}[R_{\mu\nu}l^{\mu}n^{\nu}-R_{\mu\nu}m^{\mu}m^{\ast\nu}].\nonumber
\end{gather}
On basis of equations \eqref{Weyl-tensor}-\eqref{Ricci-scalars} and the symmetry properties
of curvature tensors, one finds that the tetrad components of the Riemman 
tensor are
related to the Weyl and Ricci scalars by
\begin{equation}
\begin{aligned}
R_{lnmm^{\ast}}&=-\Psi_{2}+\Psi_{2}^{\ast},
\;\;\;R_{lmlm}=\Psi_{0}, & \, 
R_{lmnm^{\ast}}&=-\Psi_{2}+2\Pi, &
R_{nmnm}&=\Psi_{4}^{\ast},\\
R_{lnln}&=\Psi_{2}+\Psi_{2}^{\ast}+2\Phi_{11}-2\Pi,
&\, R_{lnlm}&=\Psi_{1}+\Phi_{01},
&\, R_{lmlm^{\ast}}&=\Phi_{00},\\
R_{mm^{\ast}mm^{\ast}}&=\Psi_{2}+\Psi_{2}^{\ast}-2\Phi_{11}-2\Pi,
&\, R_{lmmm^{\ast}}&=-\Psi_{1}+\Phi_{01},
& \, R_{nmnm^{\ast}}&=\Phi_{22},\\
R_{lnnm}&=-\Psi_{3}^{\ast}-\Phi_{12},
&\, R_{nmmm^{\ast}}&=-\Psi_{3}^{\ast}+\Phi_{12},
&\, R_{mlmn}&=\Phi_{02}.
\end{aligned}
\label{Tcurvature}
\end{equation}

In a spacetime with vanishing cosmological constant, the Ricci tensor is identically
zero in the vacuum and the spacetime curvature is given only by the Weyl scalars.
In the general case, however, the Ricci scalars must be considered
and their values are fixed by the Einstein field equations,
\begin{equation}
R_{ab}=8\uppi G\left[T_{ab}
-\tfrac{1}{2}T\eta_{ab}\right]+\Lambda_{c}\, \eta_{ab},
\label{Einstein}
\end{equation}
where $\Lambda_{c}$ is the cosmological constant and $T_{ab}$
is the energy-momentum tensor of all non-gravitational fields (including matter).
In the present work, the cosmological constant takes the value
$\Lambda_{c}=-3/\ell^2$ and the tensor $T_ab$ is conveniently separated
into an electromagnetic component,
\begin{equation}
T_{ab}^{\mbox{\tiny{(EM)}}}=\frac{\ell^2}{4\uppi G}
\left[F^{c}_{\,\;\;a}F_{cb}
-\frac{1}{4}\eta_{ab}F_{cd}
F^{cd}\right],
\label{EM_energy}
\end{equation}
and an energy-momentum tensor $T_{ab}^{\mbox{\tiny{(MAT)}}}$
for the matter and the remaining fields.

The six non-vanishing components of the Maxwell tensor $F_{\mu\nu}$ are replaced,
in the Newman-Penrose formalism, by three complex scalars defined as follows
\begin{equation}
\begin{aligned}
&\phi_{0}=F_{lm}=F_{\mu\nu}l^{\mu}m^{\nu},\\
&\phi_{1}=\tfrac{1}{2}\left[F_{ln}+F_{m^{\ast}m}\right]
=\tfrac{1}{2}F_{\mu\nu}\left[l^{\mu}n^{\nu}+m^{\ast\mu}m^{\nu}\right],\\
&\phi_{2}=F_{m^{\ast}n}=F_{\mu\nu}m^{\ast\mu}n^{\nu}.
\end{aligned}
\label{Maxwell-scalars}
\end{equation}
Using equations \eqref{Einstein}-\eqref{Maxwell-scalars}, we can write the 
Ricci scalars
\eqref{Ricci-scalars} in terms of the Maxwell scalars and the tetrad components of
the tensor $T_{\mu\nu}^{\mbox{\tiny{(MAT)}}}$:
\begin{equation}
\begin{aligned}
&\Phi_{00}=2\ell^2\phi_{0}\phi_{0}^{\ast}+4\uppi G\, T_{ll}^{\mbox{\tiny{(MAT)}}};\;
& \;\Phi_{22}& =2\ell^2\phi_{2}\phi_{2}^{\ast}+4\uppi G\, T_{nn}^{\mbox{\tiny{(MAT)}}};\\
&\Phi_{01}=2\ell^2\phi_{0}\phi_{1}^{\ast}+4\uppi G\, T_{lm}^{\mbox{\tiny{(MAT)}}};
\; & \;\Phi_{10}& =2\ell^2\phi_{1}\phi_{0}^{\ast}+4\uppi G\, T_{lm^{\ast}}^{\mbox{\tiny{(MAT)}}};\\
&\Phi_{02}=2\ell^2\phi_{0}\phi_{2}^{\ast}+4\uppi G\, T_{mm}^{\mbox{\tiny{(MAT)}}};
\; & \;\Phi_{20}&=2\ell^2\phi_{2}\phi_{0}^{\ast}+4\uppi G\, T_{m^{\ast}m^{\ast}}^{\mbox{\tiny{(MAT)}}};\\
&\Phi_{12}=2\ell^2\phi_{1}\phi_{2}^{\ast}+4\uppi G\, T_{nm}^{\mbox{\tiny{(MAT)}}};
\; & \;\Phi_{21}&=2\ell^2\phi_{2}\phi_{1}^{\ast}+4\uppi G\, T_{nm^{\ast}}^{\mbox{\tiny{(MAT)}}};\\
&\Phi_{11}=2\ell^2\phi_{1}\phi_{1}^{\ast}+2\uppi G\, \left[T_{ln}^{\mbox{\tiny{(MAT)}}}
+T_{mm^{\ast}}^{\mbox{\tiny{(MAT)}}}\right];\;&\;
\Pi &=-\tfrac{1}{3}\uppi G\,T^{\mbox{\tiny{(MAT)}}}-\tfrac{1}{2}\ell^{-2}.
\label{Ricci-scalars2}
\end{aligned}
\end{equation}
%

\subsection{The Ricci equations}
\label{Ricci-equations}

In the Newman-Penrose formalism, the Ricci identities
\begin{equation}
R_{abcd}=\gamma_{abd,c}-\gamma_{abc,d}+\gamma_{fac}\gamma^{\;\;f}_{b\;\;\;d}
-\gamma_{fad}\gamma^{\;\;f}_{b\;\;\;c}+\gamma_{baf}
\left(\gamma^{\;\;f}_{c\;\;\;d}-\gamma^{\;\;f}_{d\;\;\;c}\right)
\label{I-Ricci}
\end{equation}
comprise a set of 18 complex equations involving the spin coefficients
and the Weyl and the Ricci scalars. For completeness, we present the Ricci
equations below, following the notation and sign conventions adopted
in this paper:
\begin{alignat}{2}
&D\rho-\delta^{\ast}\kappa=(\rho^{2}+\sigma\sigma^{\ast})+
\rho(\varepsilon+\varepsilon^{\ast})&&\notag \\
&\;\quad\qquad\qquad-\kappa^{\ast}\tau-\kappa(3\alpha+
\beta^{\ast}-\pi)+\Phi_{00};\tag{a}\\
&D\sigma-\delta\kappa=\sigma(3\varepsilon-
\varepsilon^{\ast}+\rho+\rho^{\ast})&&\notag\\
&\;\;\;\qquad\qquad+\kappa(\pi^{\ast}-\tau-3\beta-
\alpha^{\ast})+\Psi_{0};\tag{b}\\
&D\tau-\Delta\kappa=\rho(\tau+\pi^{\ast})
+\sigma(\tau^{\ast}+\pi)+\tau(\varepsilon-
\varepsilon^{\ast})&&\notag\\
&\quad\qquad\qquad-\kappa(3\gamma+\gamma^{\ast})
+\Psi_{1}+\Phi_{01};\tag{c}\\
&D\alpha-\delta^{\ast}\varepsilon=
\alpha(\rho+\varepsilon^{\ast}-
2\varepsilon)+\beta\sigma^{\ast}-\beta^{\ast}\varepsilon-
\kappa\lambda&&\notag\\
&\quad\qquad\qquad-\kappa^{\ast}\gamma+\pi(\varepsilon+\rho)
+\Phi_{10};\tag{d}\\
&D\beta-\delta\varepsilon=\sigma(\alpha+\pi)+
\beta(\rho^{\ast}-
\varepsilon^{\ast})-\kappa(\mu+\gamma)&&\notag\\
&\;\;\qquad\qquad-\varepsilon(\alpha^{\ast}-\pi^{\ast})+\Psi_{1};\tag{e}\\
\notag
&D\gamma-\Delta\varepsilon=\alpha(\tau+
\pi^{\ast})+\beta(\tau^{\ast}+\pi)-\gamma(\varepsilon+
\varepsilon^{\ast})&&\notag\\
&\quad\qquad\qquad-\varepsilon(\gamma+\gamma^{\ast})+\tau\pi-
\nu\kappa+\Psi_{2}+\Phi_{11}-\Pi;\tag{f}\\
&D\lambda-\delta^{\ast}\pi=(\rho\lambda+
\sigma^{\ast}\mu)+
\pi(\pi+\alpha-\beta)-\nu\kappa^{\ast}&&\notag\\
&\;\quad\qquad\qquad-\lambda(3\varepsilon-\varepsilon^{\ast})
+\Phi_{20};\tag{g}\\
&D\mu-\delta\pi=(\rho^{\ast}\mu+\sigma\lambda)+
\pi(\pi^{\ast}-\alpha^{\ast}+\beta)&&\notag\\
&\quad\qquad\qquad-\mu(\varepsilon+\varepsilon^{\ast})-
\nu\kappa+\Psi_{2}+2\Pi;\tag{h}\\
&D\nu-\Delta\pi=\mu(\pi+\tau^{\ast})+
\lambda(\pi^{\ast}+\tau)+\pi(\gamma-\gamma^{\ast})&&\notag\\
&\:\quad\qquad\qquad-\nu(3\varepsilon+\varepsilon^{\ast})
+\Psi_{3}+\Phi_{21};\tag{i}\\
&\Delta\lambda-\delta^{\ast}\nu=
-\lambda(\mu+\mu^{\ast}+
3\gamma-\gamma^{\ast})&&\notag\\
&\:\,\quad\qquad\qquad+\nu(3\alpha+\beta^{\ast}+\pi-\tau^{\ast})
-\Psi_{4};\tag{j}\\
&\delta\rho-\delta^{\ast}\sigma=
\rho(\alpha^{\ast}+\beta)-
\sigma(3\alpha-\beta^{\ast})+\tau(\rho-\rho^{\ast})&&\notag\\
&\;\:\:\qquad\qquad+\kappa(\mu-\mu^{\ast})-\Psi_{1}+\Phi_{01};\tag{k}\\
&\delta\alpha-\delta^{\ast}\beta=
(\mu\rho-\lambda\sigma)+
\alpha\alpha^{\ast}+\beta\beta^{\ast}-2\alpha\beta&&\notag\\
&\quad\qquad\qquad+\gamma(\rho-\rho^{\ast})+\varepsilon
(\mu-\mu^{\ast})-\Psi_{2}+\Phi_{11}+\Pi;\tag{l}\\
&\delta\lambda-\delta^{\ast}\mu=\nu(\rho-
\rho^{\ast})+\pi(\mu-\mu^{\ast})+\mu(\alpha+
\beta^{\ast})&&\notag\\
&\;\;\:\qquad\qquad+\lambda(\alpha^{\ast}-3\beta)-\Psi_{3}+
\Phi_{21};\tag{m}\\
&\delta\nu-\Delta\mu=(\mu^{2}+
\lambda\lambda^{\ast})+
\mu(\gamma+\gamma^{\ast})-\nu^{\ast}\pi&&\notag\\
&\quad\qquad\qquad+\nu(\tau-3\beta-\alpha^{\ast})+\Phi_{22};\tag{n}\\
&\delta\gamma-\Delta\beta=\gamma(\tau-
\alpha^{\ast}-\beta)+
\mu\tau-\sigma\nu-\varepsilon\nu^{\ast}&&\notag\\
&\quad\qquad\qquad-\beta(\gamma-\gamma^{\ast}-\mu)+
\alpha\lambda^{\ast}+\Phi_{12};\tag{o}\\
\notag
\end{alignat}
\begin{alignat}{2}
&\delta\tau-\Delta\sigma=(\mu\sigma+
\lambda^{\ast}\rho)+
\tau(\tau+\beta-\alpha^{\ast})&&\notag\\
&\;\;\:\:\qquad\qquad-\sigma(3\gamma-\gamma^{\ast})-
\kappa\nu^{\ast}+\Phi_{02};\tag{p}\\
&\Delta\rho-\delta^{\ast}\tau=-
(\rho\mu^{\ast}+
\sigma\lambda)+\tau(\beta^{\ast}-\alpha-\tau^{\ast})
&&\notag\\
&\;\quad\qquad\qquad+\rho(\gamma+\gamma^{\ast})+
\nu\kappa-\Psi_{2}-2\Pi;\tag{q}\\
&\Delta\alpha-\delta^{\ast}\gamma=
\nu(\rho+\varepsilon)-\lambda(\tau+\beta)+\alpha(\gamma^{\ast}-
\mu^{\ast})&&\notag\\
&\;\quad\qquad\qquad+\gamma(\beta^{\ast}-\tau^{\ast})-\Psi_{3}.\tag{r}\\
\label{NP-Ricci}
\end{alignat}

From the complete set of Ricci equations, we are mainly interested here
in the equations (\ref{NP-Ricci}b), (\ref{NP-Ricci}c), (\ref{NP-Ricci}i), (\ref{NP-Ricci}j),
(\ref{NP-Ricci}k), (\ref{NP-Ricci}m), which are important in the
study of perturbations of rotating charged black strings. By using relations
\eqref{Ricci-scalars2} it is possible to rewrite such equations in the following form:
\begin{equation}
D\sigma-\delta\kappa=\sigma(3\varepsilon-\varepsilon^{\ast}+\rho+\rho^{\ast})
+\kappa(\pi^{\ast}-\tau-3\beta-\alpha^{\ast})+\Psi_{0};
\label{Ricci-1}
\end{equation}
\begin{equation}
\begin{split}
D\tau-\Delta\kappa=\rho(\tau+\pi^{\ast})
&+\sigma(\tau^{\ast}+\pi)+\tau(\varepsilon-\varepsilon^{\ast})\\
&-\kappa(3\gamma+\gamma^{\ast})
+\Psi_{1}+2\ell^2\phi_{0}\phi_{1}^{\ast}+4\uppi G\, T_{lm}^{\mbox{\tiny{(MAT)}}};
\label{Ricci-3}
\end{split}
\end{equation}
\begin{equation}
\begin{split}
D\nu-\Delta\pi=\mu(\pi+\tau^{\ast})& +\lambda(\pi^{\ast}
+\tau)+\pi(\gamma-\gamma^{\ast})\\
&-\nu(3\varepsilon+\varepsilon^{\ast})+\Psi_{3}+2\ell^2\phi_{2}\phi_{1}^{\ast}
+4\uppi G\, T_{nm^{\ast}}^{\mbox{\tiny{(MAT)}}};
\label{Ricci-4}
\end{split} 
\end{equation}
\begin{equation}
\Delta\lambda-\delta^{\ast}\nu=-\lambda(\mu+\mu^{\ast}+3\gamma-\gamma^{\ast})
+\nu(3\alpha+\beta^{\ast}+\pi-\tau^{\ast})-\Psi_{4};
\label{Ricci-2}
\end{equation}
\begin{equation}
\begin{split}
\delta\rho-\delta^{\ast}\sigma=\rho(\alpha^{\ast}+\beta)
&-\sigma(3\alpha-\beta^{\ast})+\tau(\rho-\rho^{\ast})\\
&+\kappa(\mu-\mu^{\ast})-\Psi_{1}+2\ell^2\phi_{0}\phi_{1}^{\ast}
+4\uppi G\, T_{lm}^{\mbox{\tiny{(MAT)}}};
\label{Ricci-5}
\end{split}
\end{equation}
\begin{equation}
\begin{split}
\delta\lambda-\delta^{\ast}\mu=\nu(\rho-\rho^{\ast})
&+\pi(\mu-\mu^{\ast})+\mu(\alpha+\beta^{\ast})\\
&+\lambda(\alpha^{\ast}-3\beta)-
\Psi_{3}+2\ell^2\phi_{2}\phi_{1}^{\ast}+4\uppi G\, T_{nm^{\ast}}^{\mbox{\tiny{(MAT)}}}. 
\label{Ricci-6}
\end{split}
\end{equation}
%

\subsection{The Bianchi identities}
\label{Bianchi}

The twenty linearly independent Bianchi identities are
given by eight complex equations,
\begin{equation}
\begin{aligned}
R_{lm[lm|m^{\ast}]}&=0,&\quad R_{lm[nl|m^{\ast}]}&=0,&\quad
R_{lm[lm|n]}&=0,&\quad R_{lm[m^{\ast}m|n]}&=0,\\
R_{m^{\ast}n[lm|m^{\ast}]}&=0,&\quad R_{m^{\ast}n[nl|m^{\ast}]}&=0,
&\quad R_{m^{\ast}n[lm|n]}&=0,&\quad R_{m^{\ast}n[m^{\ast}m|n]}&=0,
\end{aligned}
\label{B1-complexa}
\end{equation}
and by four real equations that are derived from
\begin{equation}
\eta^{bc}(R_{ab}-\tfrac{1}{2}\eta_{ab}R)_{|c}=0,
\label{B2-complexa}
\end{equation}
where above we have used square brackets to denote anti-symmetrization
and a vertical bar to indicate the intrinsic derivative,
$\Omega_{a_{1}\ldots a_{n}|b}=\Omega_{\mu_{1}\ldots\mu_{n};\nu}
\,e_{a_{1}}^{\;\;\;\;\mu_{1}}\ldots e_{a_{n}}^{\;\;\;\;\mu_{n}}\,e_{b}^{\;\;\;\nu}$.

In terms of the Newman-Penrose quantities, the identities \eqref{B1-complexa} become
\begin{align}
&(\delta^{\ast}-4\alpha+\pi)\Psi_{0}-(D-4\rho-2\varepsilon)\Psi_{1}
-3\kappa\Psi_{2}=-(D-2\varepsilon-2\rho^{\ast})\Phi_{01}\notag\\
&\qquad\qquad\qquad\quad\;\;\;+(\delta+\pi^{\ast}-2\alpha^{\ast}-2\beta)\Phi_{00}
+2\sigma\Phi_{10}-2\kappa\Phi_{11}-\kappa^{\ast}\Phi_{02},\tag{a}\\
&(\delta^{\ast}+2\pi-2\alpha)\Psi_{1}-(D-3\rho)\Psi_{2}
-\lambda\Psi_{0}-2\kappa\Psi_{3}=-(\delta^{\ast}-2\alpha-2\tau^{\ast})\Phi_{01}\notag\\
&\qquad\qquad\qquad\quad\;\;+(\Delta+\mu^{\ast}
-2\gamma-2\gamma^{\ast})\Phi_{00}-2\rho\Phi_{11}-\sigma^{\ast}\Phi_{02}
+2\tau\Phi_{10}+2D\Pi,\tag{b}\\
&(\delta^{\ast}+3\pi)\Psi_{2}-(D+2\varepsilon-2\rho)\Psi_{3}
-2\lambda\Psi_{1}-\kappa\Psi_{4}=-(D-2\rho^{\ast}+2\varepsilon)\Phi_{21}\notag\\
&\qquad\qquad\qquad+(\delta-2\alpha^{\ast}+2\beta+\pi^{\ast})
\Phi_{20}-2\mu\Phi_{10}+2\pi\Phi_{11}-\kappa^{\ast}
\Phi_{22}-2\delta^{\ast}\Pi,\tag{c}\\
&(\delta^{\ast}+4\pi+2\alpha)\Psi_{3}-(D+4\varepsilon-\rho)\Psi_{4}-3\lambda\Psi_{2}
=(\Delta+\mu^{\ast}+2\gamma-2\gamma^{\ast})\Phi_{20}\notag\\
&\qquad\qquad\qquad\quad\;\;-(\delta^{\ast}+2\alpha-2\tau^{\ast})\Phi_{21}-2\nu\Phi_{10}-\sigma^{\ast}
\Phi_{22}+2\lambda\Phi_{11},\tag{d}\\
&(\Delta-4\gamma+\mu)\Psi_{0}-(\delta-4\tau-2\beta)\Psi_{1}-3\sigma\Psi_{2}=
-(D-\rho^{\ast}-2\varepsilon+2\varepsilon^{\ast})\Phi_{02}\notag\\
&\qquad\qquad\qquad\quad\;\;\,+(\delta+
2\pi^{\ast}-2\beta)\Phi_{01}-2\kappa\Phi_{12}-\lambda^{\ast}\Phi_{00}+2\sigma
\Phi_{11},\tag{e}\\
&(\Delta-2\gamma+2\mu)\Psi_{1}-(\delta-3\tau)\Psi_{2}-\nu\Psi_{0}-2\sigma\Psi_{3}=
(\Delta+2\mu^{\ast}-2\gamma)\Phi_{01}\notag\\
&\qquad\qquad\qquad\qquad\;-(\delta^{\ast}-\tau^{\ast}+2\beta^{\ast}
-2\alpha)\Phi_{02}-2\rho\Phi_{12}-\nu^{\ast}\Phi_{00}
+2\tau\Phi_{11}+2\delta\Pi,\tag{f}\\
&(\Delta+3\mu)\Psi_{2}-(\delta+2\beta-2\tau)\Psi_{3}-2\nu\Psi_{1}-\sigma\Psi_{4}=
-(D-\rho^{\ast}+2\varepsilon+2\varepsilon^{\ast})\Phi_{22}\notag\\
&\qquad\qquad\quad\;\;\,\,+(\delta+2\pi^{\ast}+2\beta)
\Phi_{21}-2\mu\Phi_{11}-\lambda^{\ast}\Phi_{20}+2\pi
\Phi_{12}-2\Delta\Pi,\tag{g}\\
&(\Delta+2\gamma+4\mu)\Psi_{3}-(\delta-\tau+4\beta)\Psi_{4}
-3\nu\Psi_{2}=(\Delta+2\mu^{\ast}+2\gamma)\Phi_{21}\notag\\
&\qquad\qquad\qquad\qquad\quad\;\,-(\delta^{\ast}-\tau^{\ast}+2\alpha+2\beta^{\ast})\Phi_{22}
-2\nu\Phi_{11}-\nu^{\ast}\Phi_{20}+2\lambda\Phi_{12}.\tag{h}\\
\label{NP-Bianchi1}
\end{align}
The contracted Bianchi identities \eqref{B2-complexa}, in turn,
may be written as
\begin{align}
\delta^{\ast}&\Phi_{01}+\delta\Phi_{10}-D(\Phi_{11}+3\Pi)-
\Delta\Phi_{00}=\kappa^{\ast}\Phi_{12}+\kappa\Phi_{21}+
(2\alpha+2\tau^{\ast}-\pi)\Phi_{01}\notag\\
&+(2\alpha^{\ast}+2\tau-\pi^{\ast})\Phi_{10}-2(\rho+
\rho^{\ast})\Phi_{11}-\sigma^{\ast}\Phi_{02}-\sigma
\Phi_{20}+(\mu+\mu^{\ast}-2\gamma-2\gamma^{\ast})
\Phi_{00},\tag{i}\\
\delta^{\ast}&\Phi_{12}+\delta\Phi_{21}-\Delta(\Phi_{11}
+3\Pi)-D\Phi_{22}=-\nu\Phi_{01}-\nu^{\ast}\Phi_{10}+
(\tau^{\ast}-2\beta^{\ast}-2\pi)\Phi_{12}\notag\\
&+(\tau-2\beta-2\pi^{\ast})\Phi_{21}+2(\mu+
\mu^{\ast})\Phi_{11}-(\rho+\rho^{\ast}-2\varepsilon-
2\varepsilon^{\ast})\Phi_{22}+\lambda
\Phi_{02}+\lambda^{\ast}\Phi_{20},\tag{j}\\
\delta&(\Phi_{11}-3\Pi)-D\Phi_{12}-\Delta\Phi_{01}+
\delta^{\ast}\Phi_{02}=\kappa\Phi_{22}-\nu^{\ast}\Phi_{00}+
(\tau^{\ast}-\pi+2\alpha-2\beta^{\ast})\Phi_{02}\notag\\
&-\sigma\Phi_{21}+\lambda^{\ast}\Phi_{10}+2(\tau-\pi^{\ast})
\Phi_{11}-(2\rho+\rho^{\ast}-2\varepsilon^{\ast})\Phi_{12}+
(2\mu^{\ast}+\mu-2\gamma)\Phi_{01}.\tag{k}\\
\label{NP-Bianchi2}
\end{align}

The commutation relations \eqref{NP-comut1}-\eqref{NP-comut4},
the Ricci equations \eqref{NP-Ricci} and the Bianchi identities
\eqref{NP-Bianchi1} and \eqref{NP-Bianchi2} constitute the basic
set of equations of the NP formalism.

\subsection{Maxwell equations and the Ricci terms}
\label{NPmaxwell}

In spacetimes with electromagnetic fields,
the NP equations are supplemented by the Maxwell equations,
\begin{equation}
F_{[\mu\nu;\rho]}=0,\qquad\quad\qquad \eta^{\nu\rho}F_{\mu\nu;\rho}=4\uppi J_{\mu}.
\label{Maxwell-1}
\end{equation}
Projecting these equations on the complex null vectors
$\{\vec{l},\, \vec{n},\, \vec{m}, \vec{m}^{\ast}\}$ and combining
the resulting equations in a suitable form, one gets
\begin{align}
&(\delta^{\ast}+\pi-2\alpha)\phi_{0}-(D-2\rho)\phi_{1}-\kappa\phi_{2}=2\uppi J_l,
\label{Max1}\\
&(\Delta+2\mu)\phi_{1}-(\delta-\tau+2\beta)\phi_{2}-\nu\phi_{0}=2\uppi J_n,
\label{Max2}\\
&(\Delta+\mu-2\gamma)\phi_{0}-(\delta-2\tau)\phi_{1}-\sigma\phi_{2}=2\uppi J_m,
\label{Max3}\\
&(\delta^{\ast}+2\pi)\phi_{1}-(D-\rho+2\varepsilon)\phi_{2}
-\lambda\phi_{0}=2\uppi J_{m^{\ast}}.
\label{Max4}
\end{align}

The electromagnetic contribution to the Ricci (source) terms,
which appear on the right hand side of equations \eqref{NP-Bianchi1},
can be simplified with the use of the Maxwell equations 
\eqref{Max1}-\eqref{Max4}.
For instance, considering the expressions \eqref{Ricci-scalars2} for the Ricci scalars
and using \eqref{Max1} to eliminate $D\phi_{1}^{\ast}-\delta\phi_{0}^{\ast}$,
we have
\begin{equation}
\begin{split}
-D\Phi_{01}+\delta\Phi_{00} & =2\ell^2\left[-\phi_{0}(D\phi_{1}^{\ast}-\delta\phi_{0}^{\ast})-
\phi_{1}^{\ast}D\phi_{0}+\phi_{0}^{\ast}\delta\phi_{0}\right]+\cdots\\
&=2\ell^{2}\left[-\phi_{0}\{(\pi^{\ast}-2\alpha^{\ast})\phi_{0}^{\ast}
+2\rho^{\ast}\phi_{1}^{\ast}-\kappa^{\ast}\phi_{2}^{\ast}-2\uppi J_{l}\}\right. \\
&\qquad\;\;\; \left. -\phi_{1}^{\ast}D\phi_{0}+\phi_{0}^{\ast}\delta\phi_{0}\right]+\cdots,
\end{split}
\end{equation}
and substituting this expression in equation (\ref{NP-Bianchi1}a), we obtain
\begin{equation}
2\ell^{2}\left[-\phi_{1}^{\ast}D\phi_{0}+\phi_{0}^{\ast}\delta\phi_{0}+2(\varepsilon
\phi_{0}\phi_{1}^{\ast}+\sigma\phi_{1}\phi_{0}^{\ast}-\kappa\phi_{1}\phi_{1}^{\ast}
-\beta\phi_{0}\phi_{0}^{\ast})\right]+4\uppi\ell^{2}\phi_{0}J_{l}+\cdots,
\end{equation}
where the ellipses stand for matter, cosmological constant and other non-electromagnetic
field contributions to the Ricci term in equation (\ref{NP-Bianchi1}a).

With a similar sequence of replacements and simplifications,
we find that the Ricci terms of equations \eqref{NP-Bianchi1} for
a spacetime with electromagnetic field take the form
\begin{align}
&2\ell^{2}\left[-\phi_{1}^{\ast}D\phi_{0}+\phi_{0}^{\ast}\delta\phi_{0}+2(\varepsilon
\phi_{0}\phi_{1}^{\ast}+\sigma\phi_{1}\phi_{0}^{\ast}-\kappa\phi_{1}\phi_{1}^{\ast}
-\beta\phi_{0}\phi_{0}^{\ast})\right]+4\uppi\ell^{2}\phi_{0}J_{l}+\cdots,\tag{a}\\
&2\ell^{2}\left[-\phi_{1}^{\ast}\delta^{\ast}\phi_{0}+\phi_{0}^{\ast}\Delta\phi_{0}+2(\alpha
\phi_{0}\phi_{1}^{\ast}-\rho\phi_{1}\phi_{1}^{\ast}-\gamma\phi_{0}\phi_{0}^{\ast}
+\tau\phi_{1}\phi_{0}^{\ast})\right]+4\uppi\ell^{2}\phi_{0}J_{m^{\ast}}+\cdots,\tag{b}\\
&2\ell^{2}\left[-\phi_{1}^{\ast}D\phi_{2}+\phi_{0}^{\ast}\delta\phi_{2}-2(\varepsilon
\phi_{2}\phi_{1}^{\ast}+\mu\phi_{1}\phi_{0}^{\ast}-\beta\phi_{2}\phi_{0}^{\ast}
-\pi\phi_{1}\phi_{1}^{\ast})\right]+4\uppi\ell^{2}\phi_{2}J_{l}+\cdots,\tag{c}\\
&2\ell^{2}\left[-\phi_{1}^{\ast}\delta^{\ast}\phi_{2}+\phi_{0}^{\ast}\Delta\phi_{2}-2(\alpha
\phi_{2}\phi_{1}^{\ast}+\nu\phi_{1}\phi_{0}^{\ast}-\gamma\phi_{2}\phi_{0}^{\ast}
-\lambda\phi_{1}\phi_{1}^{\ast})\right]+4\uppi\ell^{2}\phi_{2}J_{m^{\ast}}+\cdots,\tag{d}\\
&2\ell^{2}\left[-\phi_{2}^{\ast}D\phi_{0}+\phi_{1}^{\ast}\delta\phi_{0}-2(\kappa
\phi_{1}\phi_{2}^{\ast}+\beta\phi_{0}\phi_{1}^{\ast}-\sigma\phi_{1}\phi_{1}^{\ast}
-\varepsilon\phi_{0}\phi_{2}^{\ast})\right]+4\uppi\ell^{2}\phi_{0}J_{m}+\cdots,\tag{e}\\
&2\ell^{2}\left[-\phi_{2}^{\ast}\delta^{\ast}\phi_{0}+\phi_{1}^{\ast}\Delta\phi_{0}-2(\rho
\phi_{1}\phi_{2}^{\ast}+\gamma\phi_{0}\phi_{1}^{\ast}-\tau\phi_{1}\phi_{1}^{\ast}
-\alpha\phi_{0}\phi_{2}^{\ast})\right]+4\uppi\ell^{2}\phi_{0}J_{n}+\cdots,\tag{f}\\
&2\ell^{2}\left[-\phi_{2}^{\ast}D\phi_{2}+\phi_{1}^{\ast}\delta\phi_{2}-2(\varepsilon
\phi_{2}\phi_{2}^{\ast}+\mu\phi_{1}\phi_{1}^{\ast}-\beta\phi_{2}\phi_{1}^{\ast}
-\pi\phi_{1}\phi_{2}^{\ast})\right]+4\uppi\ell^{2}\phi_{2}J_{m}+\cdots,\tag{g}\\
&2\ell^{2}\left[-\phi_{2}^{\ast}\delta^{\ast}\phi_{2}+\phi_{1}^{\ast}\Delta\phi_{2}-2(\alpha
\phi_{2}\phi_{2}^{\ast}+\nu\phi_{1}\phi_{1}^{\ast}-\gamma\phi_{2}\phi_{1}^{\ast}
-\lambda\phi_{1}\phi_{2}^{\ast})\right]+4\uppi\ell^{2}\phi_{2}J_{n}+\cdots.\tag{h}\\
\end{align}

The Bianchi identities (\ref{NP-Bianchi1}a), (\ref{NP-Bianchi1}d), (\ref{NP-Bianchi1}e),
(\ref{NP-Bianchi1}h) are particularly important
for the study of the perturbations of rotating charged black strings developed in
the present work. In view of that, we take into account the foregoing
Ricci terms and rewrite these equations in full form:
\begin{equation}
\begin{split}
(\delta^{\ast} & -4\alpha+\pi)\Psi_{0}-(D-4\rho-2\varepsilon)\Psi_{1}
-3\kappa\Psi_{2}=2\ell^{2}\left[-\phi_{1}^{\ast}D\phi_{0}+\phi_{0}^{\ast}\delta\phi_{0}
+2(\varepsilon\phi_{0}\phi_{1}^{\ast}\right.\\
&\left.+\sigma\phi_{1}\phi_{0}^{\ast}-\kappa\phi_{1}\phi_{1}^{\ast}-\beta\phi_{0}\phi_{0}^{\ast})\right]
+4\uppi\ell^{2}\phi_{0}J_{l}+4\uppi G\left[-(D-2\varepsilon-2\rho^{\ast})
T_{lm}^{\mbox{\tiny{(MAT)}}}\right.\\
&\left.+(\delta+\pi^{\ast}-2\alpha^{\ast}-2\beta)T_{ll}^{\mbox{\tiny{(MAT)}}}
+2\sigma T_{lm^{\ast}}^{\mbox{\tiny{(MAT)}}}-\kappa\left(T_{ln}^{\mbox{\tiny{(MAT)}}}+
T_{mm^{\ast}}^{\mbox{\tiny{(MAT)}}}\right)-\kappa^{\ast}T_{mm}^{\mbox{\tiny{(MAT)}}}\right];
\end{split}
\label{Bianchi-full1}
\end{equation}
\vspace{0.1cm}
\begin{equation}
\begin{split}
(\delta^{\ast} & +4\pi+2\alpha)\Psi_{3}-(D+4\varepsilon-\rho)\Psi_{4}-3\lambda\Psi_{2}
=2\ell^{2}\left[-\phi_{1}^{\ast}\delta^{\ast}\phi_{2}+\phi_{0}^{\ast}\Delta\phi_{2}-2(\alpha
\phi_{2}\phi_{1}^{\ast}\right.\\
&\left.+\nu\phi_{1}\phi_{0}^{\ast}-\gamma\phi_{2}\phi_{0}^{\ast}
-\lambda\phi_{1}\phi_{1}^{\ast})\right]+4\uppi\ell^{2}\phi_{2}J_{m^{\ast}}
+4\uppi G\left[(\Delta+\mu^{\ast}+2\gamma-2\gamma^{\ast})
T_{m^{\ast}m^{\ast}}^{\mbox{\tiny{(MAT)}}}\right.\\
&\left.-(\delta^{\ast}+2\alpha-2\tau^{\ast})T_{nm^{\ast}}^{\mbox{\tiny{(MAT)}}}
-2\nu T_{lm^{\ast}}^{\mbox{\tiny{(MAT)}}}-\sigma^{\ast}
T_{nn}^{\mbox{\tiny{(MAT)}}}+\lambda\left(T_{ln}^{\mbox{\tiny{(MAT)}}}+
T_{mm^{\ast}}^{\mbox{\tiny{(MAT)}}}\right)\right]; 
\end{split}
\label{Bianchi-full2}
\end{equation}
\vspace{0.2cm}
\begin{equation}
\begin{split}
(\Delta & -4\gamma+\mu)\Psi_{0}-(\delta-4\tau-2\beta)\Psi_{1}-3\sigma\Psi_{2}=
2\ell^{2}\left[-\phi_{2}^{\ast}D\phi_{0}+\phi_{1}^{\ast}\delta\phi_{0}-2(\kappa
\phi_{1}\phi_{2}^{\ast}\right.\\
&\left.+\beta\phi_{0}\phi_{1}^{\ast}-\sigma\phi_{1}\phi_{1}^{\ast}
-\varepsilon\phi_{0}\phi_{2}^{\ast})\right]+4\uppi\ell^{2}\phi_{0}J_{m}+4\uppi G
\left[-(D-\rho^{\ast}-2\varepsilon+2\varepsilon^{\ast})T_{mm}^{\mbox{\tiny{(MAT)}}}\right.\\
&\left.+(\delta+2\pi^{\ast}-2\beta)T_{lm}^{\mbox{\tiny{(MAT)}}}
-2\kappa T_{nm}^{\mbox{\tiny{(MAT)}}}-\lambda^{\ast}T_{ll}^{\mbox{\tiny{(MAT)}}}
+\sigma\left(T_{ln}^{\mbox{\tiny{(MAT)}}}+T_{mm^{\ast}}^{\mbox{\tiny{(MAT)}}}\right)\right]; 
\end{split}
\label{Bianchi-full3}
\end{equation}
\vspace{0.2cm}
\begin{equation}
\begin{split}
(\Delta & +2\gamma+4\mu)\Psi_{3}-(\delta-\tau+4\beta)\Psi_{4}
-3\nu\Psi_{2}=2\ell^{2}\left[-\phi_{2}^{\ast}\delta^{\ast}\phi_{2}
+\phi_{1}^{\ast}\Delta\phi_{2}-2(\alpha\phi_{2}\phi_{2}^{\ast}\right.\\
&\left.+\nu\phi_{1}\phi_{1}^{\ast}-\gamma\phi_{2}\phi_{1}^{\ast}
-\lambda\phi_{1}\phi_{2}^{\ast})\right]+4\uppi\ell^{2}\phi_{2}J_{n}+4\uppi G
\left[(\Delta+2\mu^{\ast}+2\gamma)T_{nm^{\ast}}^{\mbox{\tiny{(MAT)}}}\right.\\
&\left.-(\delta^{\ast}-\tau^{\ast}+2\alpha+2\beta^{\ast})
T_{nn}^{\mbox{\tiny{(MAT)}}}-\nu\left(T_{ln}^{\mbox{\tiny{(MAT)}}}+
T_{mm^{\ast}}^{\mbox{\tiny{(MAT)}}}\right)-\nu^{\ast}T_{m^{\ast}m^{\ast}}^{\mbox{\tiny{(MAT)}}}
+2\lambda T_{nm}^{\mbox{\tiny{(MAT)}}}\right]. 
\end{split}
\label{Bianchi-full4} 
\end{equation}

\section{Gauge invariance and tetrad transformations}
\label{gauge_invariance}

The physical quantities in the NP formalism are obtained from
the projection of tensor fields on a basis of null vectors.
As a result, the equations of the formalism involve only
scalar functions, and so these quantities are independent
of the choice of the coordinate system or, equivalently, they
are invariant under {\textit{gauge transformations of the first kind}},
as defined by Sachs \cite{Sachs:1964}. In a general-relativistic perturbation
theory, there is also the gauge freedom associated to the choice of
the point identification map between the physical spacetime and the
unperturbed (background) spacetime \cite{Stewart:1974uz}. The so-called
{\it{gauge transformation of the second kind}} is a change of identification map
that can be represented by
\begin{equation}
x^{\mu}_{\mbox{\scriptsize{new}}}=x^{\mu}_{\mbox{\scriptsize{old}}}+\xi^{\mu},
\label{change_coordinates}
\end{equation}
where $\xi^{\mu}$ is an infinitesimal arbitrary vector field. As a result
of this transformation, a certain NP quantity $T$ change as
\begin{equation}
T^{\mbox{\scriptsize{new}}}=T^{\mbox{\scriptsize{old}}}
-T_{,\mu}^{\mbox{\scriptsize{old}}}\,\xi^{\mu},
\label{change_scalar1}
\end{equation}
or, considering the first-order perturbation $T^{\scriptscriptstyle{(1)}}$
of the scalar $T$, we have
\begin{equation}
T^{\scriptscriptstyle{(1)}\mbox{\scriptsize{new}}}=T^{\scriptscriptstyle{(1)}\mbox{\scriptsize{old}}}
-T_{,\mu}^{\mbox{\scriptsize{old}}}\,\xi^{\mu}.
\label{change_scalar2}
\end{equation}

In the case of a rotating charged black string in the Kinnersley frame
\eqref{vec-l}-\eqref{vec-m}, the only  non-vanishing Weyl and Maxwell scalars
are respectively $\Psi_{2}$ and  $\phi_1$, and the only non-vanishing spin
coefficients are $\rho$, $\mu$ and $\gamma$.
According to the transformation \eqref{change_scalar2}, these quantities with
non-vanishing background values depend on the point identification map,
while the remaining NP scalars are invariant under the gauge
transformations \eqref{change_coordinates}.
However, once we have chosen a specific null basis, it is also
possible to perform a Lorentz transformation on this tetrad
so that the metric \eqref{metrica_basenula} remains unchanged.
Then, another relevant question to the study of linear perturbations
in the Newman-Penrose formalism is what kind of quantities are invariant
under infinitesimal changes in the tetrad.

Associated to the six parameters of the Lorentz group of transformations,
there are six degrees of freedom related to rotations on the
specific chosen null basis. For the analysis of the changes induced by
these rotations on the various NP quantities, it
is convenient to separate the rotations in three classes, based on
their effect on the vectors
$\{\vec{l},\, \vec{n},\, \vec{m}, \vec{m}^{\ast}\}$:
(a) null rotations which leave $\vec{l}$ unchanged are of class I; (b) null rotations
which leave $\vec{n}$ unchanged are of class II; and (c) transformations of class III
consist of boosts and rotations which leave the directions of $\vec{l}$ and $\vec{n}$
unchanged, while rotate $\vec{m}$ and $\vec{m}^{\ast}$ by an angle $\theta$.
These rotations induce the following changes in the null tetrad basis:
\vspace{0.15cm}
\begin{enumerate}
\item[I:] $\vec{l}\rightarrow\vec{l}$,\;
$\vec{m}\rightarrow\vec{m}+
a\vec{l}$,\; $\vec{m}^{\ast}
\rightarrow\vec{m}^{\ast}+a^{\ast}
\vec{l}$,\; $\vec{n}
\rightarrow\vec{n}+a^{\ast}\vec{m}+
a\vec{m}^{\ast}+aa^{\ast}\vec{l}$;
\item[II:] $\vec{n}\rightarrow\vec{n}$,\;
$\vec{m}\rightarrow\vec{m}+
b\vec{n}$,\; $\vec{m}^{\ast}
\rightarrow\vec{m}^{\ast}+b^{\ast}
\vec{n}$,\; $\vec{l}\rightarrow
\vec{l}+b^{\ast}\vec{m}+
b\vec{m}^{\ast}+bb^{\ast}\vec{n}$;
\item[III:] $\vec{l}\rightarrow A
\vec{l}$,\; $\vec{n}\rightarrow
A^{-1}\vec{n}$,\; $\vec{m}\rightarrow
e^{i\theta}\vec{m}$,\; $\vec{m}^{\ast}
\rightarrow e^{-i\theta}\vec{m}^{\ast}$;
\end{enumerate}
\vspace{0.15cm}
where $a$ and $b$ are complex functions and $A$ and $\theta$ are real functions.

For a rotation of class I, the spin coefficients \eqref{coef-spin}
transform as
\begin{align}
&\kappa\rightarrow\kappa,\;\sigma\rightarrow\sigma+
a\kappa,\;\rho\rightarrow\rho+a^{\ast}\kappa,\;\varepsilon
\rightarrow\varepsilon+a^{\ast}\kappa,\notag\\
&\tau\rightarrow\tau+a\rho+a^{\ast}\sigma+aa^{\ast}
\kappa,\;\pi\rightarrow\pi+2a^{\ast}\varepsilon+
(a^{\ast})^{2}\kappa+Da^{\ast},\notag\\
&\alpha\rightarrow\alpha+a^{\ast}(\rho+\varepsilon)+
(a^{\ast})^{2}\kappa,\;\beta\rightarrow\beta+a
\varepsilon+a^{\ast}\sigma+aa^{\ast}\kappa,\notag\\
&\gamma\rightarrow\gamma+a\alpha+a^{\ast}(\beta+\tau)+
aa^{\ast}(\rho+\varepsilon)+(a^{\ast})^{2}\sigma+
a(a^{\ast})^{2}\kappa,\notag\\
&\lambda\rightarrow\lambda+a^{\ast}(2\alpha+\pi)+
(a^{\ast})^{2}(\rho+2\varepsilon)+(a^{\ast})^{3}\kappa+
\delta^{\ast}a^{\ast}+a^{\ast}Da^{\ast},\notag\\
&\mu\rightarrow\mu+a\pi+2a^{\ast}\beta+2aa^{\ast}\varepsilon+
(a^{\ast})^{2}\sigma+a(a^{\ast})^{2}\kappa+\delta
a^{\ast}+aDa^{\ast},\notag\\
&\nu\rightarrow\nu+a\lambda+a^{\ast}(\mu+2\gamma)+(a^{\ast})^{2}
(\tau+2\beta)+(a^{\ast})^{3}\sigma+aa^{\ast}(\pi+2\alpha)\notag\\
&\;\;\:\quad+a(a^{\ast})^{2}(\rho+2\varepsilon)+
a(a^{\ast})^{3}\kappa+(\Delta+a^{\ast}\delta+
a\delta^{\ast}+aa^{\ast}D)a^{\ast},
\label{spin-I}
\end{align}
and the Weyl \eqref{Weyl-scalars} and Maxwell \eqref{Maxwell-scalars}
scalars become
\begin{equation}
\begin{aligned}
&\Psi_{0}\rightarrow\Psi_{0},\;\Psi_{1}\rightarrow\Psi_{1}+
a^{\ast}\Psi_{0},\;\Psi_{2}\rightarrow\Psi_{2}+2a^{\ast}\Psi_{1}+
(a^{\ast})^{2}\Psi_{0},\\
&\Psi_{3}\rightarrow\Psi_{3}+3a^{\ast}\Psi_{2}+3(a^{\ast})^{2}\Psi_{1}+
(a^{\ast})^{3}\Psi_{0},\\
&\Psi_{4}\rightarrow\Psi_{4}+4a^{\ast}\Psi_{3}+6(a^{\ast})^{2}\Psi_{2}+
4(a^{\ast})^{3}\Psi_{1}+(a^{\ast})^{4}\Psi_{0},\\
&\phi_{0}\rightarrow\phi_{0},\;\;\phi_{1}\rightarrow\phi_{1}+
a^{\ast}\phi_{0},\;\;\phi_{2}\rightarrow\phi_{2}+2a^{\ast}\phi_{1}+
(a^{\ast})^{2}\phi_{0}.
\end{aligned}
\label{Weyl-Maxwell-I}
\end{equation}
Considering that $\rho$, $\mu$, $\gamma$, $\Psi_{2}$ and $\phi_{1}$ 
are the only non-vanishing NP quantities for the black-string background
in a Kinnersley null frame, the effect of an infinitesimal rotation by a parameter
$a^{\scriptscriptstyle{(1)}}$ on the spin coefficients $\kappa$, $\sigma$,
$\lambda$ and $\nu$ is
\begin{equation}
\kappa^{\scriptscriptstyle{(1)}}\rightarrow\kappa^{\scriptscriptstyle{(1)}},
\quad\sigma^{\scriptscriptstyle{(1)}}\rightarrow\sigma^{\scriptscriptstyle{(1)}},
\quad\lambda^{\scriptscriptstyle{(1)}}\rightarrow\lambda^{\scriptscriptstyle{(1)}}
+\delta^{\ast}a^{{\scriptscriptstyle{(1)}}\ast},\quad
\nu^{\scriptscriptstyle{(1)}}\rightarrow\nu^{\scriptscriptstyle{(1)}}
+a^{{\scriptscriptstyle{(1)}}\ast}(\mu+2\gamma)+
\Delta a^{{\scriptscriptstyle{(1)}}\ast}.
\label{spin-I2}
\end{equation}
and the effect on the first-order perturbations of the Weyl and Maxwell scalars is
\begin{equation}
\begin{aligned}
&\Psi_{0}^{\scriptscriptstyle{(1)}}\rightarrow\Psi_{0}^{\scriptscriptstyle{(1)}},
\quad\Psi_{1}^{\scriptscriptstyle{(1)}}\rightarrow\Psi_{1}^{\scriptscriptstyle{(1)}},\quad
\Psi_{2}^{\scriptscriptstyle{(1)}}\rightarrow\Psi_{2}^{\scriptscriptstyle{(1)}},\quad
\Psi_{3}^{\scriptscriptstyle{(1)}}\rightarrow\Psi_{3}^{\scriptscriptstyle{(1)}}+
3a^{{\scriptscriptstyle{(1)}}\ast}\Psi_{2},\\
&\Psi_{4}^{\scriptscriptstyle{(1)}}\rightarrow\Psi_{4}^{\scriptscriptstyle{(1)}},\quad
\phi_{0}^{\scriptscriptstyle{(1)}}\rightarrow\phi_{0}^{\scriptscriptstyle{(1)}},\quad\;\,
\phi_{1}^{\scriptscriptstyle{(1)}}\rightarrow\phi_{1}^{\scriptscriptstyle{(1)}},\quad\;
\phi_{2}^{\scriptscriptstyle{(1)}}\rightarrow\phi_{2}^{\scriptscriptstyle{(1)}}
+2a^{{\scriptscriptstyle{(1)}}\ast}\phi_{1}.
\end{aligned}
\label{Weyl-Maxwell-I2}
\end{equation}

The equations of transformation for a rotation of class II can be directly
obtained from the formulas \eqref{spin-I} and \eqref{Weyl-Maxwell-I},
simply by replacing $a$ by $b$ and considering that the interchange of
$\vec{l}$ and $\vec{n}$ results in the transformation
\begin{equation}
\begin{aligned}
&\Psi_{0}\rightleftarrows\Psi_{4}^{\ast},\quad
\Psi_{1}\rightleftarrows\Psi_{3}^{\ast},\quad
\Psi_{2}\rightleftarrows\Psi_{2}^{\ast},\quad
\phi_{0}\rightleftarrows-\phi_{2}^{\ast},\quad
\phi_{1}\rightleftarrows-\phi_{1}^{\ast},\\
&\kappa\rightleftarrows-\nu^{\ast},\;\;
\sigma\rightleftarrows-\lambda^{\ast},\;\;\rho\rightleftarrows-\mu^{\ast},\;\;
\tau\rightleftarrows-\pi^{\ast},\;\;
\varepsilon\rightleftarrows-\gamma^{\ast},
\;\;\alpha\rightleftarrows-\beta^{\ast}.
\end{aligned}
\label{equivalencia}
\end{equation}
In particular, 
for the perturbations of a charged black string in a Kinnersley null 
frame, an infinitesimal rotation of class II by a parameter
$b^{\scriptscriptstyle{(1)}}$ leads to the changes
\begin{equation}
\begin{aligned}
&\kappa^{\scriptscriptstyle{(1)}}\rightarrow\kappa^{\scriptscriptstyle{(1)}}+
b^{\scriptscriptstyle{(1)}}\rho-Db^{\scriptscriptstyle{(1)}},\;
\sigma^{\scriptscriptstyle{(1)}}\rightarrow\sigma^{\scriptscriptstyle{(1)}}-
\delta b^{\scriptscriptstyle{(1)}},\;
\lambda^{\scriptscriptstyle{(1)}}\rightarrow\lambda^{\scriptscriptstyle{(1)}},\;
\nu^{\scriptscriptstyle{(1)}}\rightarrow\nu^{\scriptscriptstyle{(1)}},\\
&\Psi_{0}^{\scriptscriptstyle{(1)}}\rightarrow\Psi_{0}^{\scriptscriptstyle{(1)}},
\quad\Psi_{1}^{\scriptscriptstyle{(1)}}\rightarrow\Psi_{1}^{\scriptscriptstyle{(1)}}
+3b^{\scriptscriptstyle{(1)}}\Psi_{2},\quad
\Psi_{2}^{\scriptscriptstyle{(1)}}\rightarrow\Psi_{2}^{\scriptscriptstyle{(1)}},\quad
\Psi_{3}^{\scriptscriptstyle{(1)}}\rightarrow\Psi_{3}^{\scriptscriptstyle{(1)}},\\
&\Psi_{4}^{\scriptscriptstyle{(1)}}\rightarrow\Psi_{4}^{\scriptscriptstyle{(1)}},\quad
\phi_{0}^{\scriptscriptstyle{(1)}}\rightarrow\phi_{0}^{\scriptscriptstyle{(1)}}
+2b^{\scriptscriptstyle{(1)}}\phi_{1},\quad\;\,
\phi_{1}^{\scriptscriptstyle{(1)}}\rightarrow\phi_{1}^{\scriptscriptstyle{(1)}},\quad\;
\phi_{2}^{\scriptscriptstyle{(1)}}\rightarrow\phi_{2}^{\scriptscriptstyle{(1)}}.
\end{aligned}
\label{Spin-Weyl-Maxwell-II}
\end{equation}

The effect of a rotation of class III on the Newman-Penrose quantities is the following:
\begin{equation}
\begin{aligned}
&\Psi_{0}\rightarrow A^{2}e^{2i\theta}\Psi_{0};\;
\Psi_{1}\rightarrow Ae^{i\theta}\Psi_{1};\;
\Psi_{2}\rightarrow\Psi_{2};\\
&\Psi_{3}\rightarrow
A^{-1}e^{-i\theta}\Psi_{3};\;\Psi_{4}\rightarrow A^{-2}
e^{-2i\theta}\Psi_{4};\\
&\phi_{0}\rightarrow Ae^{i\theta}\phi_{0};\;
\phi_{1}\rightarrow \phi_{1};\;
\phi_{2}\rightarrow A^{-1}e^{-i\theta}\phi_{2};\\
&\kappa\rightarrow A^{2}e^{i\theta}\kappa;\;
\sigma\rightarrow Ae^{2i\theta}\sigma;\;
\rho\rightarrow A\rho;\\
&\pi\rightarrow e^{-i\theta}\pi;\;
\lambda\rightarrow A^{-1}e^{-2i\theta}\lambda;\;
\mu\rightarrow A^{-1}\mu;\\
&\gamma\rightarrow A^{-1}\gamma+\tfrac{1}{2}A^{-2}\Delta A+
\tfrac{1}{2}iA^{-1}\Delta\theta;\;\nu\rightarrow
A^{-2}e^{-i\theta}\nu;\\
&\varepsilon\rightarrow A\varepsilon+
\tfrac{1}{2}DA+\tfrac{1}{2}iAD\theta;\;
\tau\rightarrow e^{i\theta}\tau;\\
&\alpha\rightarrow e^{-i\theta}\alpha+
\tfrac{1}{2}A^{-1}e^{-i\theta}\delta^{\ast}A+
\tfrac{1}{2}ie^{-i\theta}\delta^{\ast}\theta;\\
&\beta\rightarrow e^{-i\theta}\beta+
\tfrac{1}{2}A^{-1}e^{i\theta}\delta A+
\tfrac{1}{2}ie^{i\theta}\delta\theta.
\end{aligned}
\label{NP-III}
\end{equation}
Also in relation to a rotation of class III, a scalar $T$ is said to have a conformal
weight $C$ and a spin weight $S$ if it transforms as
\begin{equation}
T\rightarrow A^{C}e^{iS\theta}T.
\end{equation}
For a rotating charged black string, an infinitesimal rotation of class III leaves the
spin coefficients $\kappa$, $\sigma$, $\lambda$, $\nu$ and the Weyl and Maxwell scalars
unchanged, since the transformed quantities are proportional to the original ones
and the quantities are either vanishing in the background spacetime (the case of
$\kappa$, $\sigma$, $\lambda$, $\nu$, $\Psi_0$, $\Psi_1$, $\Psi_3$, $\Psi_4$,
$\phi_0$ and $\phi_2$) or they are invariant under class-III rotations (the case of
$\Psi_2$ and $\phi_1$).

\section{The transformation theory with source terms}
\label{Chandra_transformations}

In this appendix, we extend the Chandrasekhar transformation
theory \cite{Chandrasekhar:1985kt,Chandrasekhar:1975,Chandrasekhar:1979iz} to take
into account the presence of source terms
in the fundamental wave equations. The aim is to find
transformations that relate the solutions of equations of the form
\begin{equation}
\Lambda^{2}Y_{+ i}+P_{i}\Lambda_{+}Y_{+ i}-Q_{i}Y_{+ i}=\mathfrak{S}_{+ i}
\label{ap-desejada}
\end{equation}
to solutions of the one-dimensional Schr\"odinger-like equations
\begin{equation}
\Lambda^{2}Z_{i}=V_{i}Z_{i}+\mathscr{F}_{i},
\label{ap-onda}
\end{equation}
where
\begin{equation}
P_{i}=\frac{d}{dr_{\ast}}\ln\left(\frac{r^{8}}{\mathcal{D}_{i}}\right),
\label{relation_PD}
\end{equation}
and $Q_i$ and $V_i$ are, for the moment, unspecified functions of the radial 
coordinate, and the operators $\Lambda^2$ and $\Lambda_{\pm}$ are defined in
section \ref{Decoupling_eqs}.

The only difference between the form of equations \eqref{eq_fund1} for 
$Y_{+i}$ (with source terms $\mathfrak{S}_{+i}$) and the form of equations
\eqref{eq_fund2} for $Y_{-i}$ (with source terms $\mathfrak{S}_{-i}$) is the
presence of the operator $\Lambda_{+}$ in one case and the operator
$\Lambda_{-}$ in the other. Then, the equations of a transformation
theory developed for $Y_{+i}$ can be obtained from those for $Y_{-i}$ by the
exchange  $+\varpi \rightarrow-\varpi$, and vice-versa. In order to simplify
the notation, we restrict the study to equation \eqref{ap-desejada}
and write $Y_{i}$ and $\mathfrak{S}_{i}$ in place of $Y_{+i}$ and $\mathfrak{S}_{+i}$.

As usual in the transformation theory, let's assume that
$Y_{i}$ is given by a combination of $Z_{i}$ and its
derivative of the form
\begin{equation}
Y_{i}=\rqm_{i}\Lambda_{-}\Lambda_{-}Z_{i}+\Xi_{i}\,\Lambda_{-}Z_{i},
\end{equation}
or, considering that $\Lambda_{-}=\Lambda_{+}- 2i\varpi$
and substituting $\,\Lambda^2 Z_{i}$ from \eqref{ap-onda}, we have
\begin{equation}
Y_{i}=\rqm_{i}V_{i}Z_{i}+(\Xi_{i}- 2i\varpi\,\rqm_{i})\Lambda_{-}Z_{i}
+\rqm_{i}\,\mathscr{F}_{i},
\label{relac-yz}
\end{equation}
where $\rqm_{i}$ and $\Xi_{i}$ are functions of $r$ (or $r_{\ast}$) to be determined.

Applying the operator $\Lambda_{+}$ to equation \eqref{relac-yz}
and making use of equation \eqref{ap-onda}, we find 
\begin{equation}
\begin{split}
\Lambda_{+}Y_{i} &=\left[\frac{d}{dr_{\ast}}(\rqm_{i}\,V_{i})+\Xi_{i}\,V_{i}\right]Z_{i}+
\left[\rqm_{i}\,V_{i}+\frac{d}{dr_{\ast}}(\Xi_{i}- 2i\varpi\,\rqm_{i})\right]\Lambda_{-}Z_{i}
+\mathsterling_{-i}\mathscr{F}_{i}\\
&=-\mbox{\ss}_{i}\frac{\mathcal{D}_{i}}{r^{8}}Z_{i}+R_{i}\Lambda_{-}Z_{i}
+\mathsterling_{-i}\,\mathscr{F}_{i},
\end{split}
\label{der-first}
\end{equation}
where the operators $\mathsterling_{+i}$ and $\mathsterling_{-i}$ are defined by
\begin{equation} 
\mathsterling_{\pm i}=\Xi_{i}+\frac{d\,\rqm_{i}}{dr_{\ast}}+\rqm_{i}\,\Lambda_{\pm}\,,
\label{operador-L} 
\end{equation}
and the functions $\mbox{\ss}_{i}$ and $R_{i}$ are given by
\begin{gather}
\mbox{\ss}_{i}=-\frac{r^{8}}{\mathcal{D}_{i}}\left[\frac{d}{dr_{\ast}}(\rqm_{i}\,V_{i})
+\Xi_{i}\,V_{i}\right],
\label{fund1}\\
R_{i}=\rqm_{i}\,V_{i}+\frac{d}{dr_{\ast}}(\Xi_{i}-2i\varpi\,\rqm_{i}).
\label{fund2}
\end{gather}

A similar procedure, with $\Lambda_{+}$ now applied to equation \eqref{der-first} and the use
of \eqref{ap-onda} to simplify the resulting expression, yields
\begin{equation}
\begin{split}
\Lambda_{+}\Lambda_{+}Y_{i}=&-\mbox{\ss}_{i}\frac{\mathcal{D}_{i}}{r^{8}}(\Lambda_{-}
+ 2i\varpi)Z_{i}-\mbox{\ss}_{i} Z_{i}\frac{d}{dr_{\ast}}\left(\frac{\mathcal{D}_{i}}{r^{8}}\right)\\
&-\frac{d\mbox{\ss}_{i}}{dr_{\ast}}\frac{\mathcal{D}_{i}}{r^{8}}Z_{i}+R_{i}V_{i}Z_{i}
+\frac{dR_{i}}{dr_{\ast}}\Lambda_{-}Z_{i}+\left(R_{i}+\Lambda_{+}
\mathsterling_{- i}\right)\mathscr{F}_{i}.
\end{split}
\label{der-second1}
\end{equation}
On the other hand, it follows from equation \eqref{ap-desejada} that
\begin{equation}
\Lambda_{+}\Lambda_{+}Y_{i}=\Lambda^{2}Y_{i}+ 2i\varpi \Lambda_{+}Y_{i}
=-(P_{i}- 2i\varpi)\Lambda_{+}Y_{i}+Q_{i}Y_{i}+\mathfrak{S}_{i},
\end{equation}
or, substituting $Y_{i}$ and $\Lambda_{+}Y_{i}$ from equations \eqref{relac-yz}
and \eqref{der-first}, we find
\begin{equation}
\begin{split}
\Lambda_{+}\Lambda_{+}Y_{i}=& -(P_{i}- 2i\varpi)\left[-\mbox{\ss}_{i}
\frac{\mathcal{D}_{i}}{r^{8}}Z_{i}
+R_{i}\Lambda_{-}Z_{i}\right]+Q_{i}\left[\rqm_{i}V_{i}Z_{i}
+(\Xi_{i}- 2i\varpi\,\rqm_{i})\Lambda_{-}Z_{i}\right]\\
&-(P_{i}- 2i\varpi)\mathsterling_{- i}\mathscr{F}_{i}
+Q_{i}\,\rqm_{i}\,\mathscr{F}_{i}+\mathfrak{S}_{i}.
\end{split}
\label{der-second2}
\end{equation}

Considering that equations \eqref{der-second1} and \eqref{der-second2}
must be equal to each other, one can compare the terms which do not contain 
$Z_{i}$, as well as the coefficients of $Z_{i}$ and $\Lambda_{-}Z_{i}$, in 
both equations. After some algebra, we obtain
\begin{gather}
-\frac{\mathcal{D}_{i}}{r^{8}}\frac{d\mbox{\ss}_{i}}{dr_{\ast}}=(Q_{i}\,\rqm_{i}-R_{i})V_{i},
\label{fund3}\\
\frac{d}{dr_{\ast}}\left(\frac{r^{8}}{\mathcal{D}_{i}}R_{i}\right)=\frac{r^{8}}{\mathcal{D}_{i}}
\left[Q_{i}(\Xi_{i}- 2i\varpi\,\rqm_{i})+ 2i\varpi R_{i}\right]+\mbox{\ss}_{i},
\label{fund4}\\
\mathfrak{S}_{i}=(P_{i}+\Lambda_{-})\mathsterling_{- i}\,\mathscr{F}_{i}
-(Q_{i}\,\rqm_{i}-R_{i})\mathscr{F}_{i}. 
\label{fund5}
\end{gather}

As shown by Chandrasekhar in \cite{Chandrasekhar:1985kt,Chandrasekhar:1975,Chandrasekhar:1979iz},
the system of equations \eqref{fund1}, \eqref{fund2}, \eqref{fund3} and
\eqref{fund4} admits the integral
\begin{equation}
\frac{r^{8}}{\mathcal{D}_{i}}R_{i}\,\rqm_{i}\,V_{i}+\mbox{\ss}_{i}
(\Xi_{i}- 2i\varpi \,\rqm_{i})=K_{i}=\mbox{constant}.
\label{fund6}
\end{equation}
This equation, in turn, allows to write the inverse of relations
\eqref{relac-yz} and \eqref{der-first} as
\begin{equation}
K_{i}Z_{i}=\frac{r^{8}}{\mathcal{D}_{i}}R_{i}Y_{i}
-\frac{r^{8}}{\mathcal{D}_{i}}(\Xi_{i}- 2i\varpi\,\rqm_{i})
\Lambda_{+}Y_{i}-\frac{r^{8}}{\mathcal{D}_{i}}
\left[R_{i}\,\rqm_{i}-(\Xi_{i}- 2i\varpi\,\rqm_{i})
\mathsterling_{- i}\right]\mathscr{F}_{i}\,,
\end{equation}
\begin{equation}
K_{i}\Lambda_{-}Z_{i}=\mbox{\ss}_{i} Y_{i}
+\frac{r^{8}}{\mathcal{D}_{i}}\rqm_{i}V_{i}\Lambda_{+}Y_{i}
-\mbox{\ss}_{i}\rqm_{i}\mathscr{F}_{i}-\frac{r^{8}}{\mathcal{D}_{i}}\rqm_{i}V_{i}
\mathsterling_{- i}\,\mathscr{F}_{i}\,.
\end{equation}

In the study of perturbations of Schwarzschild and Reissner-Nordstr\"om black holes
\cite{Chandrasekhar:1985kt,Chandrasekhar:1979iz} (without sources),
it is found that equations \eqref{fund1}, \eqref{fund2},
\eqref{fund3}, \eqref{fund4} and \eqref{fund6} are satisfied by a special set of transformations
for which
\begin{equation}
\mbox{\ss}_{i}=\mbox{constant}\qquad\mbox{and}\qquad\rqm_{i}=1.
\label{ap-hipoteses}
\end{equation}
Since these equations do not change with
the inclusion of $\mathfrak{S}_{i}$ and $\mathscr{F}_{i}$ in the
wave equations \eqref{ap-desejada} and \eqref{ap-onda}, the same assumptions
can be considered here and will lead to consistent solutions.
For a given function $Q_{i}$, there are conditions to be satisfied
by $R_{i}$, $V_{i}$, $\Xi_{i}$, $\mbox{\ss}_{i}$ and $K_{i}$ for the existence of
transformations compatible with \eqref{ap-hipoteses}. Such conditions
are presented below.

Imposing $\mbox{\ss}_{i}=\mbox{constant}$ and $\rqm_{i}=1$ on \eqref{fund3},
we find that $R_{i}=Q_{i}$. In conjunction with \eqref{fund4} and \eqref{fund5},
these conditions yield
\begin{gather}
\frac{d}{dr_{\ast}}\left(\frac{r^{8}}{\mathcal{D}_{i}}Q_{i}\right)=
\frac{r^{8}}{\mathcal{D}_{i}}Q_{i}\Xi_{i}+\mbox{\ss}_{i}
\label{ap-sub1}\\
\notag\\[-1.0cm]
\intertext{and}
\notag\\[-1.0cm]
\mathfrak{S}_{i}=(P_{i}+\Lambda_{-})(\Xi_{i}+\Lambda_{-})\mathscr{F}_{i}=
(P_{i}+\Lambda_{-})\mathsterling_{- i}\,\mathscr{F}_{i}. 
\label{ap-sub2}
\end{gather}
The basic set of equations are completed by \eqref{fund2} and \eqref{fund6},
which can be put into the form
\begin{gather}
V_{i}=Q_{i}-\frac{d}{dr_{\ast}}\Xi_{i},
\label{ap-sub3}\\
\left(\frac{r^{8}}{\mathcal{D}_{i}}Q_{i}\right)V_{i}+\mbox{\ss}_{i}\Xi_{i}
=K_{i}+ 2i\varpi\,\mbox{\ss}_{i}=\mbox{constant}\equiv\varkappa_{i}.
\label{ap-sub4}
\end{gather}

Introducing the function $F_{i}=(r^{8}/\mathcal{D}_{i})Q_{i}$ and using the expression
\eqref{ap-sub3} for $V_{i}$, we can rewrite equations \eqref{ap-sub1} and \eqref{ap-sub4} as
\begin{equation}
\Xi_{i}=\frac{1}{F_{i}}\left(\frac{dF_{i}}{dr_{\ast}}-\mbox{\ss}_{i}\right)
\quad\qquad\mbox{and}\qquad\quad
F_{i}\left(Q_{i}-\frac{d\Xi_{i}}{dr_{\ast}}\right)+\mbox{\ss}_{i}\Xi_{i}=\varkappa_{i}.
\label{ap-sub5}
\end{equation}
Therefore, eliminating $\Xi_{i}$ from the above equations and performing 
some simplifications, we obtain
\begin{equation}
\frac{1}{F_{i}}\left(\frac{dF_{i}}{dr_{\ast}}\right)^{2}-\frac{d^{\,2}F_{i}}{dr_{\ast}^{2}}+
\frac{\mathcal{D}_{i}}{r^{8}}F_{i}^{2}=\frac{\mbox{\ss}_{i}^{2}}{F_{i}}+\varkappa_{i}.
\label{ap-fundamental}
\end{equation}
The last equation shows that a necessary and sufficient condition
for the compatibility of the transformation equations with
\eqref{ap-hipoteses} is the existence of constants $\mbox{\ss}_{i}$ and
$\varkappa_{i}$ such that equation \eqref{ap-fundamental} is satisfied
by the given function $Q_{i}=(\mathcal{D}_{i}/r^{8})F_{i}$. Considering that
$\mbox{\ss}_{i}$ appears in equation \eqref{ap-fundamental} as $\mbox{\ss}_{i}^{2}$,
we have a pair of the so-called dual transformations, one of them generated 
by $\mbox{\ss}_{i}^{\scriptscriptstyle{(+)}}=
+\mbox{\ss}_{i}$ and the other one by 
$\mbox{\ss}_{i}^{\scriptscriptstyle{(-)}}=-\mbox{\ss}_{i}$.

Using the superscripts $(\pm)$ to distinguish between the transformations
with $+\mbox{\ss}_{i}$ and $-\mbox{\ss}_{i}$, we have
\begin{gather}
\Xi_{i}^{\scriptscriptstyle{(\pm)}}=\frac{1}{F_{i}}\left(\frac{dF_{i}}{dr_{\ast}}
\mp\mbox{\ss}_{i}\right)\label{weq}\\
\notag\\[-1.3cm]
\intertext{and}
\notag\\[-1.3cm]
V_{i}^{\scriptscriptstyle{(\pm)}}=Q_{i}-\frac{d\Xi_{i}^{\scriptscriptstyle{(\pm)}}}{dr_{\ast}}.
\label{veq}
\end{gather}
Substituting $\Xi_{i}^{\scriptscriptstyle{(\pm)}}$ from equation \eqref{weq} in \eqref{veq}
and making use of the second of equations \eqref{ap-sub5}, we find
\begin{equation}
V_{i}^{\scriptscriptstyle{(\pm)}}=Q_{i}-\frac{d}{dr_{\ast}}\left[\frac{1}{F_{i}}
\left(\frac{dF_{i}}{dr_{\ast}}\mp\mbox{\ss}_{i}\right)\right]=
\frac{\varkappa_{i}}{F_{i}}\mp\frac{\mbox{\ss}_{i}}{F_{i}^{\,2}}
\left(\frac{dF_{i}}{dr_{\ast}}\mp\mbox{\ss}_{i}\right).
\end{equation}
So, introducing the function $\mathfrak{f}_{i}=1/F_{i}$,
we obtain the following formulas for the potentials:
\begin{equation}
V_{i}^{\scriptscriptstyle{(\pm)}}=\pm\mbox{\ss}_{i}
\frac{d\mathfrak{f}_{i}}{dr_{\ast}}+\mbox{\ss}_{i}^2\,
\mathfrak{f}_{i}^{2}+\varkappa_{i}\,\mathfrak{f}_{i}.
\label{potential_transf}
\end{equation}

The associated transformations relating $Y_{i}$ to $Z_{i}^{\scriptscriptstyle{(\pm)}}$
are given, in explicit form, by
\begin{equation}
\begin{gathered}
Y_{i}=V_{i}^{\scriptscriptstyle{(\pm)}}Z_{i}^{\scriptscriptstyle{(\pm)}}
+(\Xi_{i}^{\scriptscriptstyle{(\pm)}}- 2i\varpi)
\Lambda_{-}Z_{i}^{\scriptscriptstyle{(\pm)}}+\mathscr{F}_{i}^{\scriptscriptstyle{(\pm)}},\\
\Lambda_{+}Y_{i}=\mp\mbox{\ss}_{i}
\frac{\mathcal{D}_{i}}{r^{8}}Z_{i}^{\scriptscriptstyle{(\pm)}}+
Q_{i}\Lambda_{-}Z_{i}^{\scriptscriptstyle{(\pm)}}+
\mathsterling^{\scriptscriptstyle{(\pm)}}_{- i}\mathscr{F}_{i}^{\scriptscriptstyle{(\pm)}},
\end{gathered}
\label{ap-transf1}
\end{equation}
and
\begin{equation}
\begin{gathered}
K_{i}^{\scriptscriptstyle{(\mp)}}Z_{i}^{\scriptscriptstyle{(\pm)}}
=\frac{r^{8}}{\mathcal{D}_{i}}Q_{i}Y_{i}-
\frac{r^{8}}{\mathcal{D}_{i}}(\Xi_{i}^{\scriptscriptstyle{(\pm)}}- 2i\varpi)\Lambda_{+}Y_{i}
-\frac{r^{8}}{\mathcal{D}_{i}}\left[Q_{i}-(\Xi_{i}^{\scriptscriptstyle{(\pm)}}- 2i\varpi)
\mathsterling^{\scriptscriptstyle{(\pm)}}_{- i}\right]\mathscr{F}_{i}^{\scriptscriptstyle{(\pm)}},\\
K_{i}^{\scriptscriptstyle{(\mp)}}\Lambda_{-}Z_{i}^{\scriptscriptstyle{(\pm)}}=
\pm\mbox{\ss}_{i}\,Y_{i}+\frac{r^{8}}{\mathcal{D}_{i}}
V_{i}^{\scriptscriptstyle{(\pm)}}\Lambda_{+}Y_{i}\mp\mbox{\ss}_{i}
\,\mathscr{F}_{i}^{\scriptscriptstyle{(\pm)}}-\frac{r^{8}}{\mathcal{D}_{i}}V_{i}^{\scriptscriptstyle{(\pm)}}
\mathsterling^{\scriptscriptstyle{(\pm)}}_{- i}\mathscr{F}_{i}^{\scriptscriptstyle{(\pm)}},
\end{gathered}
\label{ap-transf2}
\end{equation}
where $K_{i}^{\scriptscriptstyle{(\pm)}}=\varkappa_{i}\pm 2i\,\varpi\,
\mbox{\ss}_{i}$, and the operators 
$\mathsterling^{\scriptscriptstyle{(\pm)}}_{-i}$ are given by
\begin{equation}
\mathsterling^{\scriptscriptstyle{(\pm)}}_{-i}
=\Xi_{i}^{\scriptscriptstyle{(\pm)}}+\Lambda_{-}.
\end{equation}

The source terms appearing in equations \eqref{ap-desejada} and \eqref{ap-onda}
are related by
\begin{equation}
\mathfrak{S}_{i}=(P_{i}+\Lambda_{-})\mathsterling^{\scriptscriptstyle{(\pm)}}_{- i}
\mathscr{F}_{i}^{\scriptscriptstyle{(\pm)}}.
\label{sources1}
\end{equation}
The present form of such equations is not appropriate for a transformation theory,
since they give us a prescription to compute the known terms $\mathfrak{S}_{i}$
from the unknown source terms $\mathscr{F}_{i}^{\scriptscriptstyle{(\pm)}}$, 
while the opposite would be expected. In order to invert these equations, we follow
Sasaki and Nakamura \cite{Sasaki:1981} and look for a set of functions
$h_{i}^{\scriptscriptstyle{(\pm)}}(r)$ and $g_{i}^{\scriptscriptstyle{(\pm)}}(r)$ such that
\begin{equation}
\mathfrak{S}_{i}=\frac{1}{h_{i}^{\scriptscriptstyle{(\pm)}}g_{i}^{\scriptscriptstyle{(\pm)}}}
\Lambda_{-}h_{i}^{\scriptscriptstyle{(\pm)}}\Lambda_{-}\left(g_{i}^{\scriptscriptstyle{(\pm)}}
\mathscr{F}_{i}^{\scriptscriptstyle{(\pm)}}\right). 
\label{sources2}
\end{equation}
Comparing \eqref{sources2} to \eqref{sources1}, we find the functions
$h_{i}^{\scriptscriptstyle{(\pm)}}(r)$ and $g_{i}^{\scriptscriptstyle{(\pm)}}(r)$
satisfy the following set of inhomogeneous coupled equations:
\begin{gather}
2\frac{d}{dr_{\ast}}\mbox{ln}g_{i}^{\scriptscriptstyle{(\pm)}}+\frac{d}{dr_{\ast}}
\mbox{ln}h_{i}^{\scriptscriptstyle{(\pm)}}=P_{i}+\Xi_{i}^{\scriptscriptstyle{(\pm)}};\\
\frac{d^2}{dr_{\ast}^{2}}\mbox{ln}g_{i}^{\scriptscriptstyle{(\pm)}}+
\left(\frac{d}{dr_{\ast}}\mbox{ln}g_{i}^{\scriptscriptstyle{(\pm)}}+
\frac{d}{dr_{\ast}}\mbox{ln}h_{i}^{\scriptscriptstyle{(\pm)}}\right)
\frac{d}{dr_{\ast}}\mbox{ln}g_{i}^{\scriptscriptstyle{(\pm)}}
=P_{i}\,\Xi_{i}^{\scriptscriptstyle{(\pm)}}
+\frac{d}{dr_{\ast}}\Xi_{i}^{\scriptscriptstyle{(\pm)}}.
\end{gather}
This system of equations can be analytically solved and the solutions are given by
\begin{gather}
h_{i}^{\scriptscriptstyle{(\pm)}}(r)=\frac{\varUpsilon_{i}^{\scriptscriptstyle{(\pm)}}}{Q_{i}}
\left(C_{i}^{\scriptscriptstyle{(\pm)}}-\int^{r_{\ast}}
\frac{Q_{i}(r_{\ast}^{\prime})}{\varUpsilon_{i}^{\scriptscriptstyle{(\pm)}}(r_{\ast}^{\prime})}
dr_{\ast}^{\prime}\right)^2,
\label{solution_for_h}\\
g_{i}^{\scriptscriptstyle{(\pm)}}(r)=\frac{F_{i}}{\varUpsilon_{i}^{\scriptscriptstyle{(\pm)}}}
\left(C_{i}^{\scriptscriptstyle{(\pm)}}-\int^{r_{\ast}}
\frac{Q_{i}(r_{\ast}^{\prime})}{\varUpsilon_{i}^{\scriptscriptstyle{(\pm)}}(r_{\ast}^{\prime})}
dr_{\ast}^{\prime}\right)^{-1},
\label{solution_for_g}
\end{gather}
where the functions $\varUpsilon_{i}^{\scriptscriptstyle{(\pm)}}$ are defined as
\begin{equation}
\varUpsilon_{i}^{\scriptscriptstyle{(\pm)}}=\mbox{exp}
\left(\pm\mbox{\ss}_{i}
\int^{r_{\ast}}\mathfrak{f}_{i}(r_{\ast}^{\prime})dr_{\ast}^{\prime}\right)
\label{definition_of_Y}
\end{equation}
and $C_{i}^{\scriptscriptstyle{(\pm)}}$ are integration constants, which can 
be chosen as zero by
convenience.

Equation \eqref{sources2} can be further simplified with the
introduction of the new functions
\begin{equation}
\mathscr{W}_{i}^{\scriptscriptstyle{(\pm)}}=g_{i}^{\scriptscriptstyle{(\pm)}}
\mathscr{F}_{i}^{\scriptscriptstyle{(\pm)}}e^{-i\varpi r_{\ast}}.
\end{equation}
In terms of $\mathscr{W}_{i}^{\scriptscriptstyle{(\pm)}}$,
equation \eqref{sources2} becomes
\begin{equation}
\frac{d}{dr_{\ast}}\left[h_{i}^{\scriptscriptstyle{(\pm)}}\frac{d}{dr_{\ast}}
\mathscr{W}_{i}^{\scriptscriptstyle{(\pm)}}\right]=
h_{i}^{\scriptscriptstyle{(\pm)}}g_{i}^{\scriptscriptstyle{(\pm)}}\mathfrak{S}_{i}\,
e^{-i\varpi r_{\ast}},
\label{inversa_final}
\end{equation}
so that the problem of expressing $\mathscr{F}_{i}^{\scriptscriptstyle{(\pm)}}$
in terms of $\mathfrak{S}_{i}$ is reduced to quadratures.

To complete this analysis, we notice that, as in the
standard transformation theory, the solutions for $Z_{i}^{\scriptscriptstyle{(+)}}$
and $Z_{i}^{\scriptscriptstyle{(-)}}$ can be related to each other. In fact,
taking from \eqref{ap-transf1} the expressions for $Y_{i}$ and
$\Lambda_{+}Y_{i}$ in terms of $Z_{i}^{\scriptscriptstyle{(+)}}$,
$\Lambda_{-}Z_{i}^{\scriptscriptstyle{(+)}}$, and
$\mathscr{F}_{i}^{\scriptscriptstyle{(+)}}$ and substituting them into the 
first of equations \eqref{ap-transf2} for 
$Z_{i}^{\scriptscriptstyle{(-)}}$, we find
\begin{equation}
\begin{split}
K_{i}^{\scriptscriptstyle{(+)}}Z_{i}^{\scriptscriptstyle{(-)}}=&
\left[F_{i}V_{i}^{\scriptscriptstyle{(+)}}+\mbox{\ss}_{i}
(\Xi_{i}^{\scriptscriptstyle{(+)}}- 2i\varpi)
-\mbox{\ss}_{i}(\Xi_{i}^{\scriptscriptstyle{(+)}}-\Xi_{i}^{\scriptscriptstyle{(-)}})\right]
Z_{i}^{\scriptscriptstyle{(+)}}+F_{i}(\Xi_{i}^{\scriptscriptstyle{(+)}}
-\Xi_{i}^{\scriptscriptstyle{(-)}})\Lambda_{-}Z_{i}^{\scriptscriptstyle{(+)}}\\
&+\frac{r^8}{\mathcal{D}_{i}}\left(\Xi_{i}^{\scriptscriptstyle{(-)}}- 2i\varpi\right)
\left(\mathsterling^{\scriptscriptstyle{(-)}}_{- i}
\mathscr{F}_{i}^{\scriptscriptstyle{(-)}}-\mathsterling^{\scriptscriptstyle{(+)}}_{- i}
\mathscr{F}_{i}^{\scriptscriptstyle{(+)}}\right)+F_{i}
\left(\mathscr{F}_{i}^{\scriptscriptstyle{(+)}}-\mathscr{F}_{i}^{\scriptscriptstyle{(-)}}\right).
\end{split}
\label{intermediaria1}
\end{equation}
However, since
\begin{equation}
\mathfrak{S}_{i}=(P_{i}+\Lambda_{-})\mathsterling^{\scriptscriptstyle{(-)}}_{- i}
\mathscr{F}_{i}^{\scriptscriptstyle{(-)}}=(P_{i}+\Lambda_{-})
\mathsterling^{\scriptscriptstyle{(+)}}_{- i}
\mathscr{F}_{i}^{\scriptscriptstyle{(+)}},
\end{equation}
we have
\begin{equation}
\mathsterling^{\scriptscriptstyle{(-)}}_{- i}
\mathscr{F}_{i}^{\scriptscriptstyle{(-)}}-\mathsterling^{\scriptscriptstyle{(+)}}_{- i}
\mathscr{F}_{i}^{\scriptscriptstyle{(+)}}=\mathscr{A}_{i}\,\frac{\mathcal{D}_{i}}{r^8}e^{i\varpi r_{\ast}},
\end{equation}
where $\mathscr{A}_{i}$ are non-vanishing arbitrary integration constants.
From equations \eqref{ap-sub4} and \eqref{weq}, it also follows that
\begin{equation}
F_{i}V_{i}^{\scriptscriptstyle{(+)}}+\mbox{\ss}_{i}
(\Xi_{i}^{\scriptscriptstyle{(+)}}- 2i\varpi)=K_{i}^{\scriptscriptstyle{(+)}}
\qquad\mbox{and}\qquad \Xi_{i}^{\scriptscriptstyle{(+)}}-\Xi_{i}^{\scriptscriptstyle{(-)}}
=-\frac{2\mbox{\ss}_{i}}{F_{i}}.
\end{equation}
Substituting these relations into \eqref{intermediaria1}, we obtain the 
desired expression
\begin{equation}
K_{i}^{\scriptscriptstyle{(+)}}Z_{i}^{\scriptscriptstyle{(-)}}=
\left(\varkappa_{i}+2\frac{\mbox{\ss}_{i}^{2}}{F_{i}}\right)
Z_{i}^{\scriptscriptstyle{(+)}}-2\,\mbox{\ss}_{i}\frac{dZ_{i}^{\scriptscriptstyle{(+)}}}{dr_{\ast}}
+\mathscr{A}_{i}\left(\Xi_{i}^{\scriptscriptstyle{(-)}}- 2i\varpi\right)e^{i\varpi r_{\ast}}
+F_{i}\left(\mathscr{F}_{i}^{\scriptscriptstyle{(+)}}-\mathscr{F}_{i}^{\scriptscriptstyle{(-)}}\right).
\label{relation_ZplusZminus}
\end{equation}
The inverse of this relation is given by
\begin{equation}
K_{i}^{\scriptscriptstyle{(-)}}Z_{i}^{\scriptscriptstyle{(+)}}
=\left(\varkappa_{i}+2\frac{\mbox{\ss}_{i}^{2}}{F_{i}}\right)
Z_{i}^{\scriptscriptstyle{(-)}}+2\,\mbox{\ss}_{i}\frac{dZ_{i}^{\scriptscriptstyle{(-)}}}{dr_{\ast}}
-\mathscr{A}_{i}\left(\Xi_{i}^{\scriptscriptstyle{(+)}}- 2i\varpi\right)e^{i\varpi r_{\ast}}
+F_{i}\left(\mathscr{F}_{i}^{\scriptscriptstyle{(-)}}-\mathscr{F}_{i}^{\scriptscriptstyle{(+)}}\right).
\label{inverse_ZplusZminus}
\end{equation}

This completes an important result generalizing the
Chandrasekhar transformation theory including source terms, useful for
applications to first order perturbations of nonempty spacetimes.


\section*{Acknowledgments}
We thank Vitor Cardoso for stimulating conversations encouraging us to 
complete this work.
VTZ would like to thank Conselho Nacional de
Desenvolvimento Cient\'ifico e Tecnol\'ogico - CNPq, Brazil, for
grants, and Funda\c{c}\~ao de Amparo \`a Pesquisa do Estado de S\~ao
Paulo, Brazil, for a grant, Processo 2011/18729-1.


\section*{References}

\begin{thebibliography}{99}

   
\bibitem{Banados:1992wn} 
  M.~Banados, C.~Teitelboim and J.~Zanelli,
{\it{The Black hole in three-dimensional space-time}},
  Phys.\ Rev.\ Lett.\  {\bf 69}, 1849 (1992).

\bibitem{Martinez:1999qi} 
  C.~Martinez, C.~Teitelboim and J.~Zanelli,
{\it Charged rotating black hole in three space-time dimensions},
  Phys.\ Rev.\ D {\bf 61}, 104013 (2000).

\bibitem{Zanchin:2003nu} 
  V.~T.~Zanchin and A.~S.~Miranda,
{\it Spherical and planar three-dimensional anti-de Sitter black holes},
  Class.\ Quant.\ Grav.\  {\bf 21}, 875 (2004).
  
\bibitem{Carlip:1995qv} 
  S.~Carlip, {\it{The (2+1)-Dimensional black hole}},
  Class.\ Quant.\ Grav.\  {\bf 12}, 2853 (1995).  
 
\bibitem{Carlip:2005zn} 
  S.~Carlip, {\it{Conformal field theory, (2+1)-dimensional gravity,
and the BTZ black hole}},
  Class.\ Quant.\ Grav.\  {\bf 22}, R85 (2005).  
 
\bibitem{Witten:2007kt} 
  E.~Witten, {\it{Three-Dimensional Gravity Revisited}},
  arXiv:0706.3359 [hep-th].

\bibitem{Maldacena:2013xja} 
  J.~Maldacena and L.~Susskind,
{\it{Cool horizons for entangled black holes}},
  Fortsch.\ Phys.\  {\bf 61}, 781 (2013).
  
\bibitem{Jensen:2013ora} 
  K.~Jensen and A.~Karch,
{\it{Holographic Dual of an Einstein-Podolsky-Rosen Pair has a Wormhole}},
  Phys.\ Rev.\ Lett.\  {\bf 111}, no. 21, 211602 (2013).
  
\bibitem{Sonner:2013mba} 
  J.~Sonner,
{\it{Holographic Schwinger Effect and the Geometry of Entanglement}},
  Phys.\ Rev.\ Lett.\  {\bf 111}, no. 21, 211603 (2013).
 
\bibitem{Gharibyan:2013aha} 
  H.~Gharibyan and R.~F.~Penna,
{\it{Are entangled particles connected by wormholes? Evidence for the
ER=EPR conjecture from entropy inequalities}},
  Phys.\ Rev.\ D {\bf 89}, no. 6, 066001 (2014).
  
\bibitem{Chernicoff:2013iga} 
  M.~Chernicoff, A.~Güijosa and J.~F.~Pedraza,
{\it{Holographic EPR Pairs, Wormholes and Radiation}},
  JHEP {\bf 1310}, 211 (2013).
  
\bibitem{JohnCBaez:2014sra} 
  J.~C.~Baez and J.~Vicary,
{\it{Wormholes and entanglement}},
  Class.\ Quant.\ Grav.\  {\bf 31}, no. 21, 214007 (2014).
  
  
\bibitem{Maldacena:1997re}
  J.~M.~Maldacena, {\it{The large N limit of superconformal field theories and
supergravity}},
  Adv.\ Theor.\ Math.\ Phys.\  {\bf 2}, 231 (1998)
  [Int.\ J.\ Theor.\ Phys.\  {\bf 38} (1999) 1113].

\bibitem{Witten:1998qj}
  E.~Witten, {\it{Anti-de Sitter space and holography}},
  Adv.\ Theor.\ Math.\ Phys.\  {\bf 2}, 253 (1998).

\bibitem{Gubser:1998bc}
  S.~S.~Gubser, I.~R.~Klebanov and A.~M.~Polyakov,
  {\it{Gauge theory correlators from non-critical string theory}},
  Phys.\ Lett.\  {\bf {B 428}}, 105 (1998).

\bibitem{Aharony:1999ti}
  O.~Aharony, S.~S.~Gubser, J.~M.~Maldacena, H.~Ooguri and Y.~Oz,
  {\it{Large N field theories, string theory and gravity}},
  Phys.\ Rept.\  {\bf 323}, 183 (2000).

\bibitem{Son:2007vk}
  D.~T.~Son and A.~O.~Starinets,
  {\it{Viscosity, black holes, and quantum field theory}},
  Ann.\ Rev.\ Nucl.\ Part.\ Sci.\  {\bf 57}, 95 (2007).

\bibitem{Gubser:2007zz}
  S.~S.~Gubser,
  {\it{Heavy ion collisions and black hole dynamics,}}
  Gen.\ Rel.\ Grav.\  {\bf 39}, 1533 (2007)
  [{{Int.\ J.\ Mod.\ Phys.}} {\bf D 17} (2008) 673].

\bibitem{Myers:2008fv}
  R.~C.~Myers and S.~E.~Vazquez,
  {\it{Quark Soup al dente: Applied Superstring Theory}},
  Class.\ Quant.\ Grav.\ {\bf 25}, 114008 (2008).
  
\bibitem{Berti:2009kk} 
  E.~Berti, V.~Cardoso and A.~O.~Starinets,
{\it Quasinormal modes of black holes and black branes},
  Class.\ Quant.\ Grav.\  {\bf 26}, 163001 (2009).
    
\bibitem{Herzog:2009xv}
  C.~P.~Herzog,
  {\it{Lectures on holographic superfluidity and superconductivity}},
  J.\ Phys.\ {\bf{A 42}}, 343001 (2009).

\bibitem{Hartnoll:2009sz}
  S.~A.~Hartnoll,
  {\it{Lectures on holographic methods for condensed matter physics}},
  Class.\ Quant.\ Grav.\ {\bf 26}, 224002 (2009).

\bibitem{McGreevy:2009xe}
  J.~McGreevy,
  {\it{Holographic duality with a view toward many-body physics}},
  Adv.\ High Energy Phys.\  {\bf 2010}, 723105 (2010).
  
\bibitem{Hubeny:2010ry}
  V.~E.~Hubeny and M.~Rangamani,
  {\it{A holographic view on physics out of equilibrium}},
  Adv.\ High Energy Phys.\ {\bf 2010}, 297916 (2010).
    
\bibitem{Hawking:1971vc} 
  S.~W.~Hawking,
{\it{Black holes in general relativity}},
Commun.\ Math.\ Phys.\  {\bf 25}, 152 (1972).

\bibitem{Hawking:1973uf} 
  S.~W.~Hawking and G.~F.~R.~Ellis,
{\it The Large scale structure of space-time},
(Cambridge University Press, Cambridge, 1973).

\bibitem{Lemos:1994fn}
J.P.S.~Lemos, {\it{Two-dimensional black holes and planar general
relativity}}, Class. Quant. Grav., {\bf{12}}, 1081 (1995).

\bibitem{cai:1996} R. -G. Cai and Y. -Z. Zhang, {\it{Black plane solutions in
four-dimensional spacetimes}}, Phys. Rev. {\bf{D 54}}, 4891 (1996).

\bibitem{huang:1995} C. G. Huang and C. B. Liang, {\it{A torus-like black hole}},
Phys. Lett. {\bf{A 201}}, 27 (1995).

\bibitem{Stachel:1981fg} 
  J. Stachel, {\it{Globally stationary but locally static space-times: A
gravitational analog of the Aharonov-Bohm effect}}, Phys. Rev. {\bf{D 26}},
1281 (1982).

\bibitem{Lemos:1994xp}
  J.~P.~S.~Lemos, {\it{Cylindrical black hole in general relativity}},
  Phys.\ Lett.\ {\bf B 353}, 46 (1995).

\bibitem{Lemos:1995cm}
  J.~P.~S.~Lemos and V.~T.~Zanchin, {\it{Rotating charged black string and
three-dimensional black holes}},
  Phys.\ Rev.\ {\bf {D 54}}, 3840 (1996).


\bibitem{Edalati:2010hk} 
  M.~Edalati, J.~I.~Jottar and R.~G.~Leigh,
{\it{Shear Modes, Criticality and Extremal Black Holes}},
  JHEP {\bf 1004}, 075 (2010).

\bibitem{Edalati:2010pn} 
  M.~Edalati, J.~I.~Jottar and R.~G.~Leigh,
{\it{Holography and the sound of criticality}},
  JHEP {\bf 1010}, 058 (2010).
  
\bibitem{Brattan:2010bw} 
  D.~K.~Brattan,
{\it{Charged, conformal non-relativistic hydrodynamics}},
 JHEP {\bf 1010}, 015 (2010).
  
\bibitem{Brattan:2010pq} 
  D.~K.~Brattan and S.~A.~Gentle,
{\it{Shear channel correlators from hot charged black holes}},
  JHEP {\bf 1104}, 082 (2011).

\bibitem{Davison:2011uk} 
  R.~A.~Davison and N.~K.~Kaplis,
{\it{Bosonic excitations of the $AdS_4$ Reissner-Nordstrom black hole}},
  JHEP {\bf 1112}, 037 (2011).

\bibitem{Ge:2010yc} 
  X.~H.~Ge, K.~Jo and S.~J.~Sin,
{\it{Hydrodynamics of RN AdS$_4$ black hole and Holographic Optics}},
  JHEP {\bf 1103}, 104 (2011).
  
\bibitem{Davison:2013bxa} 
  R.~A.~Davison and A.~Parnachev,
{\it{Hydrodynamics of cold holographic matter}},
  JHEP {\bf 1306}, 100 (2013).

\bibitem{Phukon:2013tda} 
  P.~Phukon and T.~Sarkar,
{\it{R-Charged Black Holes and Holographic Optics}},
JHEP {\bf 1309}, 102 (2013).
  
\bibitem{Kim:2014bza} 
  K.~Y.~Kim, K.~K.~Kim, Y.~Seo and S.~J.~Sin,
{\it{Coherent/incoherent metal transition in a holographic model}},
  JHEP {\bf 1412}, 170 (2014).
  
\bibitem{Blake:2014lva} 
  M.~Blake, A.~Donos and D.~Tong,
{\it{Holographic Charge Oscillations}},
  JHEP {\bf 1504}, 019 (2015).
  
\bibitem{Guica:2008mu} 
  M.~Guica, T.~Hartman, W.~Song and A.~Strominger,
{\it{The Kerr/CFT Correspondence}},
  Phys.\ Rev.\ D {\bf 80}, 124008 (2009).

\bibitem{Lu:2008jk} 
  H.~Lu, J.~Mei and C.~N.~Pope,
{\it{Kerr/CFT Correspondence in Diverse Dimensions}},
  JHEP {\bf 0904}, 054 (2009).
  
\bibitem{Dias:2009ex} 
  O.~J.~C.~Dias, H.~S.~Reall and J.~E.~Santos,
{\it{Kerr-CFT and gravitational perturbations}},
  JHEP {\bf 0908}, 101 (2009).
  
\bibitem{Guica:2010ej} 
  M.~Guica and A.~Strominger,
{\it{Microscopic Realization of the Kerr/CFT Correspondence}},
  JHEP {\bf 1102}, 010 (2011).

\bibitem{Mei:2012wd} 
  J.~Mei,
{\it{On the General Kerr/CFT Correspondence in Arbitrary Dimensions}},
  JHEP {\bf 1204}, 113 (2012).

\bibitem{Compere:2012jk} 
  G.~Compere,
{\it{The Kerr/CFT correspondence and its extensions: a comprehensive review}},
  Living Rev.\ Rel.\  {\bf 15}, 11 (2012).
  
\bibitem{Newman:1965my} 
  E.~T.~Newman, R.~Couch, K.~Chinnapared, A.~Exton, A.~Prakash and R.~Torrence,
{\it{Metric of a Rotating, Charged Mass}},
  J.\ Math.\ Phys.\  {\bf 6}, 918 (1965).

\bibitem{Lee:1976}
  C.~A.~Lee,
{\it Coupled gravitational and electromagnetic perturbations around a charged
black hole},
  J.\ Math.\ Phys.\  {\bf 17}, 1226 (1976).
  
\bibitem{Chitre:1976bb} 
  D.~M.~Chitre,
{\it{Perturbations of Charged Black Holes}},
  Phys.\ Rev.\ D {\bf 13}, 2713 (1976).

\bibitem{Chandrasekhar:1978ab} 
  S.~Chandrasekhar,
{\it The Gravitational Perturbations of the Kerr Black Hole. I. The Perturbations
in the Quantities which Vanish in the Stationary State},
  Proc.\ Roy.\ Soc.\ Lond.\ A {\bf 358}, 421 (1978).

\bibitem{Hartman:2008pb} 
  T.~Hartman, K.~Murata, T.~Nishioka and A.~Strominger,
{\it{CFT Duals for Extreme Black Holes}},
  JHEP {\bf 0904}, 019 (2009).
  
\bibitem{Hartman:2009nz} 
  T.~Hartman, W.~Song and A.~Strominger,
{\it{Holographic Derivation of Kerr-Newman Scattering Amplitudes for General Charge and Spin}},
  JHEP {\bf 1003}, 118 (2010).

\bibitem{Mark:2014aja} 
  Z.~Mark, H.~Yang, A.~Zimmerman and Y.~Chen,
{\it{Quasinormal modes of weakly charged Kerr-Newman spacetimes}},
  Phys.\ Rev.\ D {\bf 91}, no. 4, 044025 (2015).
  
\bibitem{Pani:2013ija} 
  P.~Pani, E.~Berti and L.~Gualtieri,
{\it{Gravitoelectromagnetic Perturbations of Kerr-Newman Black Holes:
Stability and Isospectrality in the Slow-Rotation Limit}},
  Phys.\ Rev.\ Lett.\  {\bf 110}, no. 24, 241103 (2013).
  
\bibitem{Pani:2013wsa} 
  P.~Pani, E.~Berti and L.~Gualtieri,
{\it{Scalar, Electromagnetic and Gravitational Perturbations of Kerr-Newman
Black Holes in the Slow-Rotation Limit}},
  Phys.\ Rev.\ D {\bf 88}, 064048 (2013).
  
\bibitem{Dias:2015wqa} 
  O.~J.~C.~Dias, M.~Godazgar and J.~E.~Santos,
{\it{Linear Mode Stability of the Kerr-Newman Black Hole and Its Quasinormal Modes}},
  Phys.\ Rev.\ Lett.\  {\bf 114}, no. 15, 151101 (2015).

\bibitem{Zilhao:2014wqa} 
  M.~Zilhão, V.~Cardoso, C.~Herdeiro, L.~Lehner and U.~Sperhake,
{\it{Testing the nonlinear stability of Kerr-Newman black holes}},
  Phys.\ Rev.\ D {\bf 90}, no. 12, 124088 (2014).

\bibitem{Chandrasekhar:1975} 
  S.~Chandrasekhar and S. Detweiler,
{\it Equations governing axisymmetric perturbations of the Kerr black-hole}   
Proc.\ Roy.\ Soc.\ Lond.\ A {\bf 345}, 145 (1975).


\bibitem{Chandrasekhar:1979iz} 
  S.~Chandrasekhar,
{\it On The Equations Governing The Perturbations Of The Reissner-nordstrom Black Hole},
  Proc.\ Roy.\ Soc.\ Lond.\ A {\bf 365}, 453 (1979).
  
  
\bibitem{Chandrasekhar:1985kt}
  S.~Chandrasekhar,
  {\it The Mathematical Theory of Black Holes}
(Oxford University Press, New York, 1983).


\bibitem{Newman:1961qr}
  E.~Newman and R.~Penrose,
{\it An Approach to gravitational radiation by a method of spin coefficients},
  J.\ Math.\ Phys.\  {\bf 3}, 566 (1962).


\bibitem{Dehghani:2002rr}
  M.~H.~Dehghani,
{\it Thermodynamics of rotating charged black strings and (A)dS/CFT
cor\-res\-pon\-den\-ce},
Phys.\ Rev.\  D {\bf 66}, 044006 (2002).
  
  
\bibitem{Kinnersley:1969zza}
  W.~Kinnersley,
{\it Type D Vacuum Metrics},
  J.\ Math.\ Phys.\  {\bf 10}, 1195 (1969).
  
\bibitem{goldberg1962}
J. N. Goldberg and R. K. Sachs,
{\it A theorem on Petrov types},
Acta Phys. Polon. {\bf 22}, 13 (1962).
   
\bibitem{Bardeen:1973xb} 
  J.~M.~Bardeen and W.~H.~Press,
{\it Radiation fields in the schwarzschild background},
  J.\ Math.\ Phys.\  {\bf 14}, 7 (1973).
  
\bibitem{Teukolsky:1973ha} 
  S.~A.~Teukolsky,
{\it Perturbations of a rotating black hole.
1. Fundamental equations for gravitational electromagnetic and neutrino field perturbations},
  Astrophys.\ J.\  {\bf 185}, 635 (1973).
 
 
\bibitem{Sachs:1964}
  R.~K.~Sachs, ``Gravitational radiation'',
  {\it Relativity, Groups and Topology}, ed. C. DeWitt and B. DeWitt
(Gordon and Breach, New York, 1964).

\bibitem{Heading:1977}
  J.~Heading,
{\it Resolution of the mystery behind Chandrasekhar's black hole transformations},
  J. Phys. A: Math. Gen. {\bf 10}, 885 (1977).

\bibitem{Kodama:2003kk}
  H.~Kodama and A.~Ishibashi,
{\it Master equations for perturbations of generalized static black holes with
charge in higher dimensions},
  Prog.\ Theor.\ Phys.\  {\bf 111}, 29 (2004).

\bibitem{Miranda:2008vb} 
  A.~S.~Miranda, J.~Morgan and V.~T.~Zanchin,
{\it Quasinormal modes of plane-symmetric black holes according to the AdS/CFT correspondence},
  JHEP {\bf 0811}, 030 (2008).
 
\bibitem{Morgan:2013dv} 
  J.~Morgan, A.~S.~Miranda and V.~T.~Zanchin,
{\it Electromagnetic quasinormal modes of rotating black strings and the AdS/CFT correspondence},
  JHEP {\bf 1303}, 169 (2013).
  
\bibitem{Witten:1981nf} 
  E.~Witten,
{\it Dynamical Breaking of Supersymmetry}
  Nucl.\ Phys.\ B {\bf 188}, 513 (1981).

\bibitem{Cooper:1982dm} 
  F.~Cooper and B.~Freedman,
{\it Aspects of Supersymmetric Quantum Mechanics},
  Annals Phys.\  {\bf 146}, 262 (1983).

\bibitem{Cooper:1994eh} 
  F.~Cooper, A.~Khare and U.~Sukhatme,
{\it Supersymmetry and quantum mechanics},
  Phys.\ Rept.\  {\bf 251}, 267 (1995).

\bibitem{Leung:1999fr} 
  P.~T.~Leung, A. M. van den Brink, W.~M.~Suen, C.~W.~Wong and K.~Young,
{\it SUSY transformations for quasinormal and total transmission modes of open sys\-tems},
[math-ph/9909030].

\bibitem{Cardoso:2001bb} 
  V.~Cardoso and J.~P.~S.~Lemos,
{\it Quasinormal modes of Schwarzschild anti-de Sitter black holes: Electromagnetic and gravitational perturbations},
  Phys.\ Rev.\ D {\bf 64}, 084017 (2001).

\bibitem{Bakas:2008gz} 
  I.~Bakas,
{\it Energy-momentum/Cotton tensor duality for AdS(4) black holes},
  JHEP {\bf 0901}, 003 (2009).

\bibitem{Teukolsky:1972my} 
  S.~A.~Teukolsky,
{\it{Rotating black holes: separable wave equations for gravitational and electro\-magnetic perturbations}},
  Phys.\ Rev.\ Lett.\  {\bf 29}, 1114 (1972).

  
\bibitem{Witten:2003ya} 
  E.~Witten, {\it SL(2,Z) action on three-dimensional conformal field theories with Abelian symmetry},
  In *Shifman, M. (ed.) et al.: From fields to strings, vol. 2* 1173-1200
  [hep-th/0307041].

\bibitem{Herzog:2007ij} 
  C.~P.~Herzog, P.~Kovtun, S.~Sachdev and D.~T.~Son,
{\it Quantum critical transport, duality, and M-theory},
  Phys.\ Rev.\ D {\bf 75}, 085020 (2007).

  
\bibitem{Hartnoll:2007ip} 
  S.~A.~Hartnoll and C.~P.~Herzog,
{\it Ohm's Law at strong coupling: S duality and the cyclotron resonance},
  Phys.\ Rev.\ D {\bf 76}, 106012 (2007).

\bibitem{deHaro:2007eg} 
  S.~de Haro and P.~Gao,
{\it{Electric-magnetic duality and deformations of three-dimensional CFT's}},
  Phys.\ Rev.\ D {\bf 76}, 106008 (2007).
  
\bibitem{Myers:2010pk} 
  R.~C.~Myers, S.~Sachdev and A.~Singh,
{\it Holographic Quantum Critical Transport without Self-Duality},
  Phys.\ Rev.\ D {\bf 83}, 066017 (2011).

\bibitem{Henneaux:2004jw} 
  M.~Henneaux and C.~Teitelboim,
{\it Duality in linearized gravity},
  Phys.\ Rev.\ D {\bf 71}, 024018 (2005).
  
\bibitem{deHaro:2008gp} 
  S.~de Haro,
{\it Dual Gravitons in AdS(4) / CFT(3) and the Holographic Cotton Tensor},
  JHEP {\bf 0901}, 042 (2009).
  
\bibitem{Bakas:2008zg} 
  I.~Bakas,
{\it Duality in linearized gravity and holography},
  Class.\ Quant.\ Grav.\  {\bf 26}, 065013 (2009).
    
\bibitem{Sadeghi:2010zza} 
  J.~Sadeghi, M.~R.~Pahlavani and H.~Farahani,
{\it The AdS(4) gravitational perturbation and supersymmetry},
  Int.\ J.\ Theor.\ Phys.\  {\bf 49}, 914 (2010).


\bibitem{Wald:1978vm} 
  R.~M.~Wald,
{\it Construction of Solutions of Gravitational, Electromagnetic, Or Other Perturbation
Equations from Solutions of Decoupled Equations},
  Phys.\ Rev.\ Lett.\  {\bf 41}, 203 (1978).

\bibitem{Wald:1979}  
R. M. Wald, {\it Construction of metric and vector potential perturbations of
a Reissner-Nordstr\"om black hole},
Proc.\ Roy.\ Soc.\ Lond.\ A {\bf 369}, 67 (1979).

\bibitem{Chrzanowski:1975wv} 
  P.~L.~Chrzanowski,
{\it Vector Potential and Metric Perturbations of a Rotating Black Hole},
  Phys.\ Rev.\ D {\bf 11}, 2042 (1975).

\bibitem{Kegeles:1979an} 
  L.~S.~Kegeles and J.~M.~Cohen,
{\it Constructive Procedure For Perturbations Of Space-times},
  Phys.\ Rev.\ D {\bf 19}, 1641 (1979).
  
\bibitem{Stewart:1974uz} 
  J.~M.~Stewart and M.~Walker,
{\it Perturbations of spacetimes in general relativity},
  Proc.\ Roy.\ Soc.\ Lond.\ A {\bf 341}, 49 (1974).
  
\bibitem{Sasaki:1981}
  M.~Sasaki and T.~Nakamura,
{\it The Regge-Wheeler equation with sources for both even and odd parity perturbations
of the Schwarzschild geometry},
  Phys.\ Lett.\ A {\bf 87}, 85 (1981).
 
  
\end{thebibliography}
\end{document}